\documentclass[12pt,english,onehalfspacing]{report}
\usepackage{amssymb}
\usepackage{amsmath}
\usepackage{xcolor}
\usepackage{textcomp}
\usepackage{setspace}
\usepackage{url}
\usepackage[utf8]{inputenc}
\usepackage[toc,page,title]{appendix}
\newcommand\ra{\rangle}
\newcommand\la{\langle}
\newcommand\nn{\nonumber}
\newcommand\f{\frac}
\newcommand\p{\partial}
\usepackage{graphicx}
\graphicspath {{./Figures/}}
\usepackage{textcomp}
\usepackage{comment}
\usepackage[
colorlinks=true, 
pdfstartview=FitV, 
linkcolor=magenta, 
citecolor=blue, 
urlcolor=black, 
bookmarks=true,
bookmarksnumbered=true,
pdftitle={},
pdfauthor={}
]{hyperref}

\usepackage{sidecap}
\usepackage[top=2.5cm, bottom=2.5cm, left=2.5cm, right=2.5cm,a4paper]{geometry}
\DeclareMathOperator\erf{erf}

\begin{document}

\section*{\centering{\textcolor{magenta}{Thermalization, Chaos and Hydrodynamics \\ in Classical Hamiltonian Systems}}}
\vspace{31mm} 
 \thispagestyle{empty}
\begin{center}
	A thesis

\vspace{31mm} 
Submitted to the

\vspace{1mm} 
Tata Institute of Fundamental Research, Mumbai

\vspace{1mm} 
for the degree of Doctor of Philosophy

\vspace{1mm} 
in Physics

\vspace{31mm} 

by

\vspace{1mm} 
Santhosh Ganapa

\vspace{1mm} 
International Centre for Theoretical Sciences

\vspace{1mm} 
Tata Institute of Fundamental Research

\vspace{1mm} 
Bangalore, India

\vspace{31mm} 
September 13 2022
\end{center}
\newpage
\thispagestyle{empty}
\mbox{}
\newpage
\pagenumbering{roman}
\section*{\centering{Declaration}}
This thesis is a presentation of my original research work. Wherever contributions of
others are involved, every effort is made to indicate this clearly, with due reference to the
literature, and acknowledgement of collaborative research and discussions. 

\begin{flushleft}
	The work was done under the guidance of Prof. Abhishek Dhar and Prof. Amit Apte, at International Centre for Theoretical
	Sciences, Tata Institute of Fundamental Research (ICTS-TIFR), Bangalore.
\end{flushleft}
\vspace{5mm} 
\begin{flushleft}
Santhosh Ganapa
\end{flushleft}
\vspace{5mm} 
\begin{flushleft}
In my capacity as supervisor of the candidate’s thesis, I certify
that the above statements are true to the best of my knowledge.
\end{flushleft}
\vspace{7mm} 
\begin{flushleft}
Abhishek Dhar \hspace{50mm} Amit Apte\\
\vspace{5mm} 
Date: September 13 2022
\end{flushleft}
\newpage

\begin{center}
	\textit{To my parents, sister, Dhidhi and Mamma.}
\end{center}

\newpage
\section*{\centering{Acknowledgements}}
I thank God Almighty for giving me the strength and the endurance to pursue my passion: physics. In particular, i would like to thank God for giving me an opportunity to meet and work with my master Abhishek Dhar. I am very fortunate to have him as my role model and my advisor. He must be credited for converting a nobody to a somebody. 

I would like to thank my co-advisor Amit Apte for introducing me to the beautiful area of Hamiltonian chaos, which eventually lead me to work in thermalization and hydrodynamics. I benefited from having Sriram Ramaswamy and Samriddhi Sankar Ray in my thesis monitoring committee, who clarified my confusions and also allowed me to express myself freely. I thank them for their friendliness. I acknowledge Anupam Kundu and Vishal Vasan for teaching me some of the tricks of the trade. I will forever be indebted to my almae matres BITS Pilani and ICTS for helping me realize what i want to do in life.

I rejoice the time spent with my friends Kohinoor Ghosh and Hemanta Kumar, who made my time during PhD even more memorable. I enjoyed discussions with Varun Dubey and many of the other talented scholars of ICTS, from whom i learnt how to do physics. My time in ICTS was made unforgettable by my kung fu and violin teachers. I thank my football friends for helping me control my weight and stay healthy during my PhD. Some of this work was done during the Covid-19 pandemic and i must credit the cooks, security guards, maintenance workers and the cleaning ladies of ICTS for doing their part during the lockdown, which made it easy for me to concentrate only on my work.  I thank my friend Bhanu Teja for always challenging me to achieve higher in every aspect of my life. 

And finally, I dedicate this work to my family. I marvel at the hard work and dedication of my parents. I also thank them for always giving me the freedom to choose what i want to do in life, and my sister Pavani for always counseling me after i fall into some sort of trouble by making some stupid choice. I thank my grandmothers Dhidhi and Mamma for their sacrifices, and looking after me when i needed them the most.
\newpage

\textit{``The mystery of life is not a problem to be solved but a reality to be experienced."}
\rightline{...Alan Watts}
\newpage
\section*{\centering{List of Publications}}
\begin{enumerate}
	\item Ganapa, S., Apte, A. and Dhar, A. Thermalization of Local Observables in the $\alpha$-FPUT Chain. \textit{J Stat Phys}	180 (1), 1010–1030 Springer (2020). \url{https://doi.org/10.1007/s10955-020-02576-2}
	
	\item Ganapa, S., Chakraborti, S., Krapivsky, P., L., and Dhar, A. (2021). Blast in the One-Dimensional Cold Gas: Comparison of Microscopic Simulations with Hydrodynamic Predictions. \textit{Phys. Fluids} 33 (8).
	\url{https://aip.scitation.org/doi/10.1063/5.0058152}
	
	\item Chakraborti, S., Ganapa, S., Krapivsky, P., L., and Dhar, A. (2021). Blast in a One-Dimensional Cold Gas: From Newtonian Dynamics to Hydrodynamics. \textit{Phy. Rev. Lett.} 126, 244503.
	\url{https://journals.aps.org/prl/pdf/10.1103/PhysRevLett.126.244503}
	
\end{enumerate}
\vspace{7mm} 
\section*{\centering{List of Collaborators}}
\begin{enumerate}
	\item[--] Abhishek Dhar, Professor, ICTS-TIFR, Ind.
	\item[--] Amit Apte, Professor, ICTS-TIFR and IISER Pune, Ind.
	\item[--] Pavel. L.  Krapivsky, Professor, Boston University, USA and Skolkovo Institute of Science and Technology, Moscow, Rus.	
	\item[--] Subhadip Chakraborti, postdoctoral fellow, ICTS-TIFR (former) and Friedrich–Alexander University Erlangen-Nuremberg, Deu (current).
\end{enumerate}
\newpage
\section*{\centering{List of Abbreviations}}
\begin{enumerate}
	\item[--] FPUT chain: Fermi-Pasta-Ulam-Tsingou chain (formerly Fermi-Pasta-Ulam chain).
	\item[--] TvNS solution: Taylor-von Neumann-Sedov solution.	
	\item[--] 1D: one dimensional.
	\item[--] AHP gas: Alternate mass hard particle gas.
	\item[--] NSF equations: Navier-Stokes-Fourier equations.
	\item[--] SLE: Space localized excitations.	
	\item[--] NMLE: Normal mode localized excitations.
	\item[--] BBGKY hierarchy: Bogoliubov–Born–Green–Kirkwood–Yvon hierarchy.
	\item[--] KdV equation: Korteweg–de Vries equation.	
	\item[--] ODE: Ordinary differential equations.
	\item[--] PDE: Partial differential equations.
	\item[--] LE: Local equilibrium.
	\item[--] MD simulations: Molecular dynamics simulations.
\end{enumerate}

\textcolor{magenta}{\tableofcontents}

\chapter{Introduction} 
\pagenumbering{arabic}
\label{chap:intro}Hamiltonian systems are ubiquitous in physics and attempts to understand their behaviour in a better way have been made for a long time. Consider a one dimensional system of $N$ particles with positions and momenta given by $\{q_i,p_i\}$, for $i=0,1,\ldots,N-1$.  Let $H(\{p_i,q_i\})$ be the Hamiltonian of the system. The Hamilton's equations of motion relate the positions and momenta as:
\begin{equation}\label{eq:hamiton equations}
\begin{split}
\dot{p}_i = -\frac{\partial H}{\partial q_i},\ \
\dot{q}_i = \frac{\partial H}{\partial p_i}.
\end{split}
\end{equation}
Solving the above $2N$ first order differential equations for a given initial condition describes the behaviour of the system at subsequent times. So, describing a Hamiltonian system is now reduced to solving a set of differential equations with some initial condition. A Hamiltonian system with $2N$ degrees of freedom is said to be integrable if there exists $N$ independent constants of motion $I_1, I_2,...I_N$ which Poison commute with each other: $\{I_i,I_j \} = 0\ $, $\forall\ i, j$. Then we have a class of systems whose description is possible in principle, by solving the equations of motion analytically. But integrable systems are rare, and we encounter non-integrable systems that only have a few conserved quantities in general, in which case it is not possible to solve the equations of motion Eq.~ \eqref{eq:hamiton equations} analytically, and we have
to resort to numerical methods. Now let's say we are interested in describing the behaviour of a macroscopic non-integrable system. If we try to solve Eq.~\eqref{eq:hamiton equations}, then there would be a huge number of equations to solve, which may hide the necessary information that would have otherwise been sufficient to describe the system. So, we are not interested in the behaviour of individual particles. We have to move on from solving the set of Hamilton's equations to something else. We use the methods of statistical physics to describe emergent behaviour in physical systems when we are dealing with such macroscopic systems that have a huge number of particles. We describe the system as a whole now instead of its constituent particles.

\section{A General Introduction to this Work}
\subsection{Thermalization and its Importance in Statistical Physics}

There are many interesting problems in statistical physics that one can study in the context of classical Hamiltonian systems. One of the fundamental questions is to understand why statistical physics works, which has been a long-standing problem. The theory of equilibrium statistical physics says that the observed values of physical observables, for a system in thermal equilibrium, can be obtained by taking an average over an appropriate equilibrium statistical ensemble. A justification of why this works is usually made by invoking the ergodic hypothesis, which says that generic Hamiltonian systems would have few conserved quantities, for example the total energy, and the motion of the system in phase space would fill the energy surface densely. Given that Hamiltonian dynamics is area preserving, this means that the infinite time average of a physical observable would equal the microcanonical ensemble average. However, we note that typical measurements involve time averages over a period which is minute compared to the time required to sample the energy surface of a typical many-body system. In other words, we observe thermal equilibrium at time scales much shorter than the time required for the ergodic hypothesis to hold. This is especially true for macroscopic systems, where the time taken for an initial point in the phase space to cover  most of it is astronomical. Hence, ergodic hypothesis cannot be a sufficient basis for the foundations of statistical physics. Then the question is to understand thermal equilibrium itself and to know what leads a system to thermal equilibrium so that we are able to use statistical methods after much smaller times. Some causes that can be explored are chaos, non-integrability, nonlinearity and mixing. One can also study the role of coarse-graining, the role of initial conditions and the choice of macrovariables used to estimate equilibration. One of the first studies on this problem was done by Fermi, Pasta, Ulam and Tsingou (FPUT) \cite{Fermi1955,dauxois2008fermi}. They expected that nonlinearity should be sufficient for a system to thermalize but their results did not indicate that. They instead observed a quasiperiodic behaviour in the system. Since the original work of FPUT, there have been a number of studies aimed at understanding their results, which led to significant developments in statistical physics, nonlinear dynamics and mathematical physics. To this day, there remain some unanswered questions. Thus, understanding the requirements for the validity of equilibrium statistical physics remains a challenge. At the same time, with the drastic improvement of computational performance, we are now in a better position to probe these questions even further.
\subsection{Evolution Before Thermalization - Description in Terms of Conserved Fields}
Another interesting problem would be to study how a system thermalizes. This question is different from what causes the system to thermalize, which is addressed in the previous paragraph. Here, starting from a typical initial condition, one aims to understand how the system evolves to the thermal state. This can be done by studying emergent phenomena in the system - we would like to find ways to actually simplify the description of a macroscopic Hamiltonian system. One of the standard ways to do this is to consider a coarse-grained description of the system in terms of the conserved fields, which are the slow degrees of freedom. This is the purview of hydrodynamics. This is the realm of nonequilibrium statistical physics, in contrast to equilibrium statistical physics, where there are steady state values of the conserved quantities. 
One of the interesting ways to study hydrodynamics is to initially excite the system to a large amount of energy concentrated within a small region and then study its evolution. This leads to the formation of shocks in the system.  
This so called blast problem was first studied during the Second World War by Taylor \cite{Taylor19501,Taylor19502}, von Neumann \cite{VonNeumann1963} and Sedov \cite{Sedov1946,Sedov2014} (TvNS) in order to understand atomic explosions. The study of the nonequilibrium regime of the system is now reduced to studying the dynamics of shock propagation in the system. Because a large amount of energy is initially excited in a small region, this is an extreme example where we can test the predictions of hydrodynamics. One such prediction is a self-similar structure of the shock wave (in time). It is then possible to derive the exact solution of the evolution of conserved quantities, which helps us understand the nonequilibrium regime. It will be very interesting if one can observe these scaling solutions in the microscopic simulations of a many body Hamiltonian system. However, an agreement between molecular dynamics simulations of a macroscopic Hamiltonian system and hydrodynamics remains an open question. 
\subsection{Structure of the Thesis}
This work is aimed to address the problems motivated by the above-mentioned line of thinking. As the title suggests, we study various aspects of thermalization, chaos and hydrodynamics in one dimensional Hamiltonian systems. We study two problems. In the first problem, we revisit the problem of equilibration in the $\alpha-$Fermi-Pasta-Ulam-Tsingou (FPUT) system. In this system we try to understand what causes a system to thermalize. The second problem deals with the evolution of a blast wave in an alternate mass hard particle (AHP) gas. Here we study the nonequilibrium regime in terms of shock propagation in the system. We aim to describe the system using hydrodynamic equations up to times the shock wave reaches the boundary of the system. This thesis is organized as follows:

In this chapter we next go on to explain the problem statements and also the methods that are used in this study. Chapters \ref{chap:fput1} and \ref{chap:fput2} discuss the first problem mentioned in the previous paragraph. In chapter \ref{chap:fput1}, we give a brief literature review of the Fermi-Pasta-Ulam-Tsingou (FPUT) problem. There is a lot of literature on this problem, some of which are mentioned in \cite{weissert1997,Berman2005, Gallavotti2008}. We emphasize only on those aspects which are relevant to our study. Our work is motivated by the results of \cite{Onorato2015}, and so we will explain the results of that work in detail. Then in chapter \ref{chap:fput2}, we present the results of our work \cite{Ganapa2020}, and also compare with the results of \cite{Onorato2015}. Chapters \ref{chap:tvns1} and \ref{chap:tvns2} discuss the second problem. In the chapter \ref{chap:tvns1}, we derive the Taylor-von Neumann-Sedov (TvNS) solution of the Euler equations in one dimension, which will be used later to compare our results \cite{Ganapa2021, Chakraborti2021} of the molecular dynamics simulations of the alternate mass hard particle (AHP) gas. We also give a literature review of the works which aimed at observing the TvNS scaling in molecular dynamics simulations. In chapter \ref{chap:tvns2}, we present our results of the molecular dynamics simulations, which we compare with the exact TvNS scaling solution and also with the numerical solution of the Navier-Stokes-Fourier (NSF) equations. 
In chapter \ref{chap:conc}, we conclude our work and present some interesting extensions. For completeness sake, we also mention the limitations of our study.

\section{Problem Statements}
\subsection{Thermalization in the $\alpha-$FPUT Chain}
Consider a harmonic chain of $N$ particles, each of mass $m$ connected by a spring of spring constant $\mu$.  The Hamiltonian of this system is then described as a function of the position $q_i$ and momentum $p_i$ of each particle as:

\begin{equation}\label{eq:hamharm}
H(\boldsymbol{p},\boldsymbol{q})=\sum_{i=0}^{N-1}\left[\frac{p_i^2}{2m}+\frac{\mu(q_{i+1}-q_{i})^2}{2}\right]~, 
\end{equation}
where we have considered periodic boundary conditions with $q_N \equiv q_0$ and $q_{-1} \equiv q_{N-1}$, and from now on we take $m = \mu = 1$ unless otherwise mentioned. One can first make a canonical transformation from $p,q$ variables to the normal mode variables $P,Q$ defined by:

\begin{equation}\label{eq:nm}
Q_k =\frac{1}{N} \sum_{j=0}^{N-1}q_je^{-i2\pi kj/N},~~ P_k =\frac{1}{N} \sum_{j=0}^{N-1}p_je^{-i2\pi kj/N}.  
\end{equation}
The  Hamiltonian then becomes
\begin{align}
H=N\sum_{k=0}^{N-1} E_k,~~{\rm where}~ E_k=\Bigg[\frac{|P_k|^2}{2}+\frac{\Omega_k^2 |Q_{k}|^2}{2}\Bigg]~ \label{HNM}
\end{align}
is the energy of each mode and $\Omega_k = 2 \sin(k\pi/N), k = 0,1,2,...,N-1$ is the normal mode frequency of that particular mode. The zero mode corresponding to $k=0$ does not participate in the dynamics, and we only deal with initial conditions that have $E_0 = 0$ (zero initial momentum). From Hamilton's equations of motion in these canonical coordinates we can observe that these normal modes decouple. The evolution of each normal mode can then be found independently of the others. So, energy exchange does not take place between different normal modes. So, most of the phase space is not accessible to a harmonic chain and hence the system does not thermalize. So, we would like the normal modes to somehow interact with each other in order to expect thermalization. For that we add a cubic interaction term to the Hamiltonian. If we now ask whether the system thermalizes, then the answer is not very obvious. We then get the $\alpha-$FPUT system described by the Hamiltonian:

\begin{equation}\label{eq:hamFPU}
H(\boldsymbol{p},\boldsymbol{q})=\sum_{i=0}^{N-1}\left[\frac{p_i^2}{2m}+\frac{\mu (q_{i+1}-q_{i})^2}{2}+\frac{\alpha(q_{i+1}-q_{i})^3}{3}\right]~.
\end{equation}
The origin of this Hamiltonian can be understood as arising from a spring that has restoring force $F$ dependent on the displacement of the particle $x$ connected to it as $F = - \mu x -\alpha x^2$. Thus, $\alpha$ is the origin of nonlinearity in the system. A heuristic estimate of the strength of the nonlinearity  can be obtained by comparing the contribution of the nonlinear interaction part to the total energy. Roughly, if $r$ is the average spacing between particles, we expect $\mu r^2\sim E/(N-1)=e$ which gives a length scale $r\sim\sqrt{e/\mu}$. An estimate of the ratio of the nonlinear and harmonic energies is then given by the parameter
\begin{equation}\label{eq:eps1}
\epsilon = \frac{\alpha r^3}{\mu r^2}=\frac{\alpha e^{1/2}}{\mu^{3/2}}~.
\end{equation}
This dimensionless number, $\epsilon$, and the system size, $N$, are the only relevant parameters. In our subsequent discussions we assume that $\mu=1$ and one can change $\epsilon=\alpha e^{1/2}$ by either changing the nonlinearity strength $\alpha$ or equivalently, by changing the energy density $e$.
Note that the cubic potential of the $\alpha$-FPUT system implies that the system stays bounded only if the total energy  is sufficiently small and the precise condition is $E< \mu^3/(6 \alpha^2)$, corresponding to all energy contributing to the potential energy of a single particle. In practice this is highly improbable and one can work with energies slightly higher than this bound.

\begin{figure}[ht]
	\centering
	\includegraphics[width=0.6\textwidth]{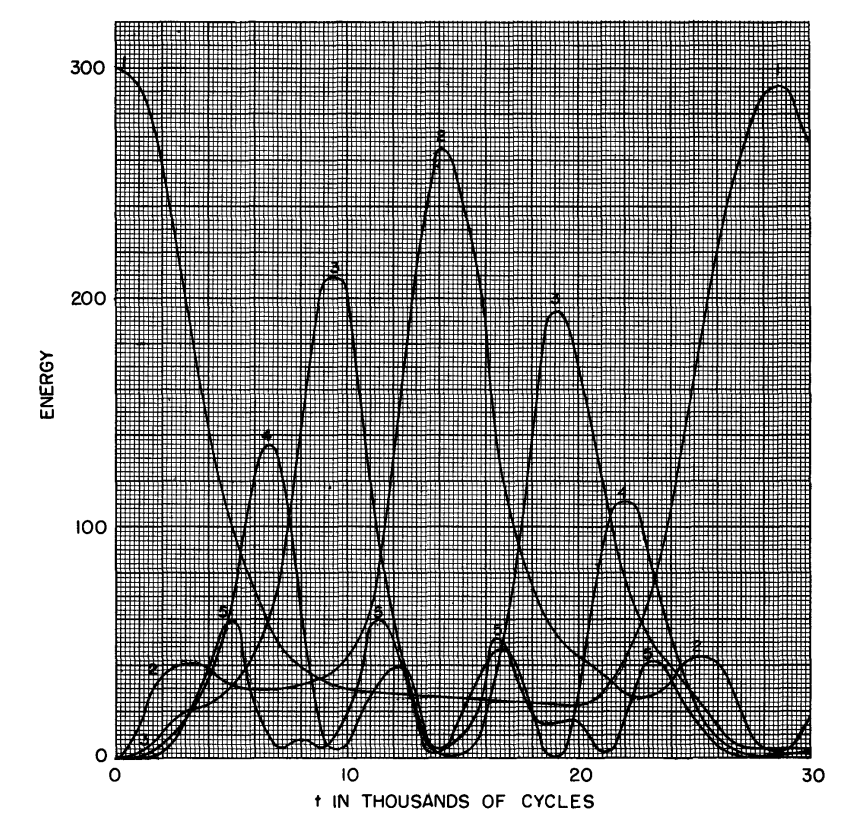}
	\caption{$\alpha-$FPUT chain: Graph from the original paper \cite{Fermi1955} showing the quasiperiodic evolution of the system with near recurrences to the initial condition when the first mode is excited. The units for energy are arbitrary.}
	\label{FPUoriginal}
\end{figure}

Because of the presence of the cubic term, the normal modes do not decouple and energy exchange between different normal modes takes place and one might expect that the entire constant energy surface $H(\boldsymbol{p},\boldsymbol{q})=E$ becomes accessible. In the original problem \cite{Fermi1955}, the system was started in a highly atypical initial condition (all energy in the first normal mode, $E_1=0.08$) with $\alpha=0.25$ (where  
however the fixed boundary condition case was studied).
It was asked if the system evolves to a state where all the normal modes have the same energy. To the surprise of the authors, they did not find such a state in the time scales of their observations. Instead, they found quasi-periodic behaviour and near-recurrences to the initial state as shown in Fig.~\ref{FPUoriginal}.

The original paper also tried with quartic nonlinearity (the $\beta-$FPUT system) but no thermalization was observed. The problem of whether the FPUT system thermalizes has been studied ever since and even today there remain some open problems in this field. Recently, \cite{Onorato2015} was able to answer positively that the $\alpha-$FPUT system thermalizes for an arbitrary (nonzero) value of $\alpha$. Using the methods of wave turbulence \cite{Zakharov1992,Nazarenko2011}, they derived that the time scale for equipartition $\tau_{eq}$ in the $\alpha-$FPUT system scaled with the dimensionless nonlinearity parameter $\epsilon$ as: $ \tau_{eq} \sim 1/\epsilon^8$. They also verified this scaling numerically for the system by initially exciting the first mode and then checking that the energy indeed gets divided between all the normal modes equally when some sort of averaging is performed.

The current work, motivated by the results of \cite{Onorato2015}, looks at some aspects of thermalization in the $\alpha-$FPUT system from a somewhat different perspective. The problems that are addressed, and the differences in the approaches from the previous studies are listed below:

\begin{enumerate}
	
	\item[--] \textbf{Dependence of equilibration on the initial conditions:} Most studies look at initial conditions specified in the normal mode space - they excite the first or the first few normal modes and then study the evolution of the $\alpha-$FPUT chain. However, we study the evolution by using initial conditions that are localized in the real space - we excite a few particles initially and then study thermalization. 
	\item[--] \textbf{Dependence on the choice of the observable used to measure equilibration:} In most of the literature, the analysis of equipartition was done for the energy of normal modes of the corresponding integrable problem, the harmonic chain. The analysis is in some sense ``global", since normal modes involve all the particles in the chain. The present work attempts to analyse the problem ``locally" - by checking equipartition theorem at different sites in the chain and tries to compute the time scale $\tau_{eq}$ the system needs to reach equipartition. Our method would be especially relevant	when we consider strong nonlinearity, for which the normal mode picture becomes	invalid. Several earlier works have discussed this in the context of harmonic chains \cite{mazur1960,titulaer1973a,titulaer1973b} and anharmonic chains \cite{Danieli2017}. We will then compare the dependence of $\tau_{eq}$ on the nonlinearity parameter $\epsilon$ of both the observables.
	\item[--] \textbf{Role of averaging in determing equilibration:} We also numerically investigate the role played by averaging over an initial ensemble in determining whether we observe thermalization in our system, and compare the results with the time averaging protocol. 
	\item[--] \textbf{Relation between thermalization and chaos in the system:} Finally, we investigate the question of dependence of the equilibration process on the width of the initial distribution. By doing so, we aim to find a relation between equilibration of local observables and sensitive dependence of the $\alpha-$FPUT system on initial conditions. We also compare this relation of the $\alpha-$FPUT system to the harmonic chain and the integrable Toda chain (the potential $V$ depends on $r$ as $V(r) = g(e^{br} -br)/b$. The parameter choice $b=2\alpha$ and $g=b^{-1}$ would then approximate to the $\alpha-$FPUT potential upto leading nonlinearity). 
	
\end{enumerate}

\subsection{Blast Wave in an Alternate Mass Hard Particle Gas}
The evolution of a blast wave emanating from an intense explosion has drawn much interest since the Second World War. The rapid release of a large amount of energy in a localized region produces a shock wave propagating into the non perturbed gas. The dynamics of the system can then be studied in terms of shock propagation in the system. Taylor \cite{Taylor19501,Taylor19502}, von Neumann \cite{VonNeumann1963} and Sedov \cite{Sedov1946, Sedov2014} (TvNS) understood that the compressible gas dynamics \cite{LandauBook,ZeldovichBook} provides an appropriate framework and described this problem using the evolution of conserved quantities of the gas system: the density field $\rho (r,t) $, the velocity field $v (r,t) $ and the energy field $e(r,t)$.  They also found an exact solution as long as the pressure in front of the shock wave can be neglected compared to the pressure behind the shock wave. In this situation, the hydrodynamic equations have a scaling solution and the hydrodynamic variables acquire a self-similar form (in time) that are functions only of the scaling variable. The hydrodynamic equations are thus converted from partial differential equations (PDE) to ordinary differential equations (ODE) depending only on this scaling variable, and they admit an exact solution.

It will be an interesting study to look for these scalings in the microscopic simulations of a many-particle system. Such studies have been attempted only recently, which can be partly attributed to the recent increase in computational performance. As mentioned in the introduction, finding an agreement between molecular dynamics simulations of a Hamiltonian system and hydrodynamics remains an open question. Studies done in two and three dimensions \cite{Barbier2016, Joy2021, Joy2021a, Joy2017} showed deviations from hydrodynamics. This was never done in one dimensional systems.


This work attempts to understand the blast wave dynamics in a one dimensional gas of hard point particles. In this system, the only interactions between the particles is through point collisions between the nearest neighbors, which are perfectly elastic and instantaneous. In between collisions the particles move ballistically with constant velocities. 
Consider two colliding particles with masses $(m_1,m_2)$ and initial velocities $(v_1,v_2)$. Their post-collisional velocities are given by:
\begin{equation}
\label{elasticcollision1}
\begin{split}
v_1^\prime = \frac{2m_2 v_2 + (m_1-m_2)v_1}{m_1+m_2},\ \
v_2^\prime = \frac{2m_1 v_1 + (m_2-m_1)v_2}{m_1+m_2}. \end{split}
\end{equation}
These follow from the conservation of momentum and energy. If $m_1=m_2$, then one can notice from the above equations that after the collision, $v_1^\prime = v_2, v_2^\prime = v_1 $, i.e. 
the particles merely exchange velocities. This means that the system effectively behaves as a non-interacting system.  This is an integrable system, whose evolution cannot be described by the usual hydrodynamics. We choose a system that only has a few conserved quantities (e.g. particle number, energy and momentum). This is the case with typical non-integrable systems, in which case it is possible to describe the evolution using the usual hydrodynamics. Thus, we take a non-integrable system, where all odd numbered particles are taken to have mass $m_1$ and all even numbered particles are taken to have mass $m_2$($\neq m_1$). This system only has three conserved quantities, and so we expect the usual hydrodynamics to be sufficient to describe it. We excite the system to a sudden blast wave $-$ energy concentrated in a small region somewhere in the middle of the system, and evolve the system only upto times the energy reaches the boundary. The problems that this work addresses, are listed below:

\begin{enumerate}
	
	\item[--] \textbf{The TvNS solution of the 1D Euler
		equations for an ideal gas:} We aim to derive the expression for the position of the shock front in terms of the initial total energy of the blast $E$ and the ambient density of the system $\rho_\infty$. We also derive the expressions for scaling solution of the conserved quantities. Together, this is the complete TvNS solution in one dimension, which is an exact scaling solution of the Euler equations with an ideal gas equation of state.
		\item[--] \textbf{Comparison of microscopic simulations of the 1D alternate mass hard particle gas (AHP) and the TvNS solution:} We first compare the results of molecular dynamics of the  alternate mass hard particle gas with the TvNS solution.  We also study the subtleties of the microscopic simulations like comparison of ensemble averaging and spatial coarse graining protocols, and also the comparison of the microcanonical and canonical ensemble averaging.


	\item[--] \textbf{Modelling using the NSF equations:} In order to understand some of the discrepancies between the microscopics and the TvNS solution, we then consider the evolution of the conserved fields of AHP system with the  Navier-Stokes Fourier (NSF) equations (Euler equations with added dissipative terms due to viscosity and heat conduction). 
	 We then move on to studying the various scaling solutions that are possible for the hydrodynamic equations and show TvNS scaling as one of the scaling solutions, and that it specifically results from a blast wave initial condition. 
	
	\item[--] \textbf{Study of deviations from the TvNS scaling and the cause of deviation:} 	We 
	attribute the discrepancies between microscopic simulations and the TvNS solution to the non-negligible contribution of the dissipation terms in the NSF equations. We then perform a detailed analysis in this regime, where these terms are not neglected and try to explain the deviation of the microscopics from the TvNS scaling.

\end{enumerate}

\section{Methods Used in this Study}
\subsection{Symplectic Integrator} Simulations of the $\alpha-$FPUT system were done by using a $6^{th}$ order symplectic integrator \cite{Yoshida1990}. This is more accurate than the Runge-Kutta methods. The reason for this is that the Runge-Kutta methods treat position and momentum of each degree of freedom as independent of each other at each instant of time. We note that an evolution in momentum at a certain time step will automatically affect its corresponding canonical position through $\dot{q} = p$/m. The Runge-Kutta methods do not take this interdependence into consideration and thus introduce additional error in the solution, which is avoided in symplectic integrators upto the desired level of accuracy. The time-step size was taken as $0.01$ and we checked the relative energy change in the system at the end of the computation to be of the  order of $10^{-11}$. Computations of the  Lyapunov exponent  were performed by solving the coupled system of $2N+2N$ nonlinear and linearized equations using a fourth order  Runge-Kutta integrator.  In this case the time-step size was taken as $0.001$ and  the relative energy change in the system at the end of the computation is around $10^{-9}$. We have  checked that the results  do not change significantly on	decreasing the time step  by a factor of two.

\subsection{Equipartition Theorem}
We estimate equilibration time by checking Equipartition theorem at different sites. According to the theorem, for a system in thermal equilibrium, the following is true:

\begin{equation}
\left\langle q_i\frac{\partial H}{\partial q_j} \right\rangle_{eq}=\left\langle p_i\frac{\partial H}{\partial p_j}\right\rangle_{eq}= c \delta_{ij},  \label{eqp1}
\end{equation}
for all $i,j=0,1,\ldots,N-1$ and where $\left\langle...\right\rangle_{eq}$ represents an  average over an equilibrium ensemble (the different types of averaging are described below) and  $c$ is a constant equal to $k_B(\partial S/\partial E)^{-1}$ for the microcanonical ensemble and equal to $k_BT$ for a canonical ensemble. 
Our method is in contrast to checking Equipartition among normal modes because normal modes of the harmonic chain only approximately describe the $\alpha-$FPUT system, while Eq.~\eqref{eqp1} is exact for the latter system and can be used for a wide range of parameters and for different systems. Choosing initial conditions localized in real space has the distinct advantage that all the members of the ensemble used in the ensemble averaging protocol lie on the same energy surface and hence, we will be able to link thermalization and chaos. This cannot be done by initially exciting the normal modes because all the members of the ensemble would then not lie on the same energy surface - we cannot fix the normal mode energies and the total energies at the same time.
\subsection{Averaging Methods}
One can use various averaging methods to check whether a system has reached thermal equilibrium. It is clear that checking for equilibration during the time evolution requires some kind of averaging, and we discuss three possible approaches:

\subsubsection{ Averaging over Initial Conditions}
\label{AvIC}
This approach involves performing an averaging over initial conditions \cite{Onorato2015}, with the expectation that such averages would represent the typical behaviour.  In our work, we perform an average over the initial conditions $({\bf q}_0,{\bf p}_0)$ which are now chosen from a narrow distribution centred around a specified point and that are still on the microcanonical surface of constant energy $E$, momentum $P$ and number of particles $N$. Denoting the initial distribution by $\rho_I({\bf q}_0,{\bf p}_0)$, we then obtain   the following average for any observable $A=A({\bf q},{\bf p})$:
\begin{equation}\label{eq:eavg}
\langle A \rangle(t)= \int d {\bf q}_0 d{\bf p}_0~ A({\bf q}(t),{\bf p}(t)) \rho_I({\bf q}_0,{\bf p}_0)~.  
\end{equation}
In our simulations we generate $R$ initial configurations from this distribution and evolve each of them with Hamiltonian dynamics. The ensemble average of a physical observable $A({\bf q},{\bf p})$ is estimated from Eq.~\eqref{eq:eavg} as: 

$$ \langle A \rangle (t) = \frac{\sum_{r=1}^R A_r(t)}{R}, $$ where the sum is over the $R$ members of the ensemble. 


\subsubsection{Temporal Coarse Graining} A second approach would be again to start with a fixed initial condition and performing an averaging over time \cite{Benettin2013,Danieli2017,Benettin2018}. In this case, one can look at microscopic variables, e.g the kinetic energy of individual particles, and ask whether the time averaged value corresponds to the expected equilibrium value. Starting the time evolution of the system from an arbitrary initial condition $({\bf q}_0,{\bf p}_0)$, with energy $E$, we can define a time averaged quantity for the observable $A=A({\bf q},{\bf p})$ as 
\begin{equation}\label{eq:tavg}
\bar{A}(t)= \frac{1}{t}\int_0^t ds A({\bf q}(s),{\bf p}(s))~.
\end{equation}
For a non-integrable system we would expect that for generic initial conditions, at long times we should get thermal equilibration or 
$\bar{A}(t\to \infty) = \langle A \rangle$. The time to reach the equilibrium value should give a measure of equilibration time scales.

\subsubsection{ Spatial Coarse Graining} We now discuss a protocol which is perhaps most relevant from the physical point of view if one remembers that real observations are made on single systems and on short time scales. It then makes sense to consider the system being in a single microscopic configuration, but we now  measure coarse grained physical observables, involving some spatial averaging \cite{Lebowitz1999,Goldstein2017}.  
In this approach one should eventually look at large systems and  the ideas related to typicality and law of large numbers are relevant for understanding equilibration. 
As an example, consider the problem of free expansion of a gas on removal of a partition inside a box. We take as our observable to be the number of particles in a tiny box, that is still large enough to contain thousands of particles, and located in the initially empty  half of the box. We expect that the number of particles will increase with time to eventually equilibrate around an average value with small fluctuations.  

For the $\alpha-$FPUT system, we use the ensemble averaging protocol in order to estimate the equilibration time. As mentioned before, doing so would give us a way to relate equilibration of local observables to chaos. We briefly compare the results of ensemble averaging with those of the time averaging protocol.
We also use the ensemble averaging protocol in order to study the blast propagation in the alternate mass hard particle gas. Here, we compare these results with those of spatial averaging. We will describe the spatial coarse graining protocol in chapter $5$.

\subsection{Estimation of Equilibration Time}
As a measure of the level of equipartition that is achieved, we define the  following  function, which has been referred to in the literature as entropy:

\begin{equation}\label{eq:ent}
S(t) = -\sum_{i=0}^{N-1} f_i(t) \ln f_i(t),
\end{equation}
where $ f_i(t) = \nu_i(t)/\sum_{r=0}^{N-1} \nu_r(t)$ for $i=0,1,\dots,N-1$, and $\nu_i$ could be either $\la p_i\frac{\partial H}{\partial p_i}\ra $ or $\la q_i\frac{\partial H}{\partial q_i}\ra $ or $\la E_i \ra$. The value of $S(t)$ is bounded between $0$, corresponding to the highly nonequilibrium situation with all the energy in a single degree of  freedom, and $\ln N$, corresponding to the equilibrated system with equipartition between all degrees with $\left\{f_i\right\}$ defining a uniform distribution over the set $\left\{0,1,\dots,N-1\right\}$.
Since it can theoretically take an infinite amount of time to reach $\ln N$, we estimate the equilibration time as the time  required for $S(t)$ to reach a  predetermined value of entropy which is close to the equilibration value.
In this work, we consider the following criteria to determine the equilibration timescale:
\begin{equation}\label{criterion}
\left|\frac{S(t)-S_{max}}{S_{max}}\right| \leq 0.01~. 
\end{equation}
The above threshold must be satisfied for two consecutive values of the time that is sampled. The minimum value of $t$ for which the above criteria is satisfied is termed as equilibration time (denoted by $\tau_{\rm eq}$). We use this to quantify the equilibration time in the $\alpha-$FPUT system.

\subsection{Navier-Stokes-Fourier (NSF) Equations and the MacCormack Method} The molecular dynamics simulations of the alternate mass hard particle gas were done by an event-driven algorithm, which is much faster than the time-driven algorithm.
The only known conserved quantities of this system are the total number of particles, the total momentum and the total energy of the system, and we expect a hydrodynamic description in terms of the corresponding conserved fields namely, the mass density field $\rho(x,t)$, the momentum density field $\rho(x,t)v(x,t)$ ($v$ is the velocity field) and the energy density field $E(x,t)$. 
We compare microscopic simulations with the results obtained by solving the one-dimensional Navier-Stokes-Fourier (NSF) equations,
\begin{subequations}
	\label{tNSi}
	\begin{align}
	&\partial_t \rho + \partial_x (\rho v) = 0, \label{tNS1i} \\
	&\partial_t (\rho v) + \partial_x (\rho v^2 +p) = \partial_x (\zeta\partial_x v), \label{tNS2i} \\
	&\partial_t (\rho e) + \partial_x (\rho ev   + p v)= \partial_x (\zeta v\partial_x v + \kappa\partial_x T ),
	\label{tNS3i}
	\end{align}
\end{subequations}
which are supplemented by the thermodynamic relations: 
$T(x,t)= 2 \mu\, \epsilon(x,t)$ and
$p(x,t)= 2 \rho(x,t) \epsilon(x,t)=  \mu^{-1}\rho(x,t)\, T(x,t).$ Here $\epsilon(x,t)= e(x,t)-\frac{v(x,t)^2}{2}$ is the internal energy per unit mass, $\mu = (m_1+m_2)/2$, and we have set the Boltzmann's constant $k_B=1$. The dissipative effects are manifested in the bulk viscosity $\zeta$ and the thermal conductivity $\kappa$ appearing in Eqs.~\eqref{tNSi}. Consistent with the Green-Kubo formalism and earlier studies, in our numerical study we have used the forms $\zeta =D_1T^{1/2}$ and $\kappa=D_2\rho^{1/3}T^{1/2}$, where $D_1$ and $D_2$ are constants taken to be unity. 

The Navier-Stokes-Fourier (NSF) equations were solved using the MacCormack method  ~\cite{maccormack1969, maccormack1982}, which is an explicit two-step, second order, finite difference method. 
	As an illustration of the method, consider the following first order hyperbolic equation \cite{wiki}:
	\begin{equation}\label{eq:hyp}
	\frac{\partial u}{\partial t} + a \frac{\partial u}{\partial x}= 0.
	\end{equation}
The initial profile $u(x,0)$ is known, and we aim to find $u(x,t)$ by using this method. The application of MacCormack method to the above equation proceeds in two steps; a predictor step which is followed by a corrector step. The predictor step is obtained by replacing the spatial and temporal derivatives in Eq.~\eqref{eq:hyp} using forward differences. The output of this step is the provisional value of $u$ at time level $n+1$ for each value of the space level $i$ denoted by $\overline{u_i^{n+1}}$, which is given by:
	\begin{equation}\label{eq:forward}
	\overline{u_i^{n+1}} = u_i^n - a\frac{\Delta t}{\Delta x}(u_{i+1}^{n}-u_i^n).
	\end{equation}
	In the corrector step, the predicted value $\overline{u_i^{n+1}}$ is corrected by using backward finite difference approximation for the spatial derivative. Also the time step used in the corrector step is $\Delta t/2$, in contrast to the $\Delta t$ used in the predictor step. This results in the desired output value $u_i^{n+1}$, which is given by:
	\begin{equation}\label{eq:backward}
	u_i^{n+1} = \frac{u_i^n+\overline{u_i^{n+1}}}{2} - a\frac{\Delta t}{2\Delta x}(\overline{u_{i}^{n+1}}-\overline{u_{i-1}^{n+1}}).
	\end{equation}
The order of differencing can be reversed for the time step (i.e., forward/backward followed by backward/forward). In our work, we chose the former order, and we have taken $\Delta t = 0.001$ and $\Delta x = 0.1$.

\chapter{The Fermi-Pasta-Ulam-Tsingou Story}
\label{chap:fput1}
Ever since the original numerical experiment, there have been numerous studies that tried to explain the recurrent behaviour in the Fermi-Pasta-Ulam-Tsingou system. It all started with Fermi, Pasta, Ulam and Tsingou wanting to check if nonlinearity is enough for a system to thermalize. But the failure to observe thermal behaviour actually led to significant developments in statistical physics, nonlinear dynamics and mathematical physics in the coming years.
There were some basic questions regarding the problem that needed answers. One of them was to understand if there existed a stochasticity threshold in the FPUT system - the critical value of the nonlinearity parameter below which the system is not expected to thermalize. Maybe the parameter taken in the original problem was lower than this threshold and hence thermalization was not observed. May be the system needed to be simulated for much longer times so that we could observe thermalization. Then there must be a theory which explains what is that time scale. There is also a possibility of the existence of breather solutions - time-periodic and space-localized stable solutions in the FPUT system that could delay or even prevent thermalization.

In this chapter we discuss some of the existing literature on this problem. FPUT problem has a vast literature, and here we only discuss some of the literature and touch upon only those aspects which are relevant to the problem that this work tries to address. We only refer to several of the review articles on the FPUT problem \cite{weissert1997,Berman2005,Gallavotti2008,Benettin2013,Ford1992}. Some of the recent studies that focus on the specific aspects of the FPUT problem that are especially relevant to this paper include the role of breathers \cite{Danieli2017, Marin1996,Flach2005,Flach2006,Christodoulidi2010}, wave-wave interaction theory \cite{Onorato2015,Lvov2018,Pistone2019}, and the role of breakup of invariant tori and the stochastic threshold \cite{Izrailev1966,Casetti1996,Deluca1995,Livi1985}.  Our emphasis is on \cite{Onorato2015}, which explains the equilibration problem in terms of resonant interactions between the nonlinear waves, and also tries to estimate the time scale the system needs in order to attain equipartition. 
\section{Continuum Limit and the KdV Equation} Zabusky and Kruskal \cite{Zabusky1965} showed that
the continuum limit of the FPUT model leads to the Korteweg–de Vries (KdV) equation, which is an integrable model and has soliton solutions. Two or more solitons do not interact irreversibly, and they just ``pass through" each other virtually unaffected in size or shape. Because the solitons are remarkably stable entities, preserving their identity through numerous interactions, these authors expected the FPUT system to exhibit thermalization (complete energy sharing among the corresponding linear normal modes) only after extremely long times, if ever. One might then expect the long wavelength initial condition used by FPUT to remain close to the continuum description for long times and so there should not be any thermalization. This is indeed consistent with what is observed in the original problem, which was however done for N = 32. 
We will come back to the KdV  equation later when we discuss breather solutions.


\section{Overlap of Resonances}
\label{sec:Overlap}
As mentioned in the beginning of this chapter, one important aspect that was not understood at the time the FPUT problem was first posed, was if there existed a stochasticity threshold - critical value of the nonlinearity parameter below which the system does not thermalize at all, and above which there is stochasticity in the system, which would eventually lead to thermalization. 
Izrailev and Chirikov \cite{Izrailev1966} proposed that there exists a stochasticity threshold value of the energy density ($e=E/N$) beyond which one expects thermalization. This idea was developed using the criterion of overlapping resonances. At the onset of stochasticity, the separation between the resonances ($\Delta_\omega \equiv \omega_{k+1}-\omega_k $) is of the order of resonance width of each resonance, arising due to the nonlinearity. Here $\omega_k$ is the frequency of the $k^{th}$ mode of the harmonic oscillator. It can be shown that for small values of $k\ (k<<N)$, $\Delta_\omega \approx 2\pi/N$, while for larger values of $k\ (N-k<<N)$, $\Delta_\omega\approx\pi^2/2N^2$, which is much less than that of low frequency modes. So, the separation between the neighbouring frequencies is smaller for larger values of k. If the system is initially excited in one particular normal mode such that its frequency shift arising due to nonlinearity is much less than $\Delta_\omega$, then one can neglect the influence of neighbouring resonances, since there is no overlap. For smaller values of k, the separation between the resonances is high and hence the stochasticity threshold needed for equipartition is also high. The initial condition taken by FPUT is below this stochastic border and hence there is a recurrent behaviour. On the other hand, if a higher mode is excited initially, then the separation between the resonances is relatively small. With the stochasticity threshold relatively small, one can easily observe irregular motion with strong energy sharing among a large number of high frequency modes.

The stochastic threshold has also been studied numerically \cite{Casetti1996,Deluca1995,Livi1985}. It is numerically found that for generic initial conditions \cite{Casetti1996}, the time evolution of the maximum Lyapunov exponent $\Lambda$ is indistinguishable for the $\alpha$ -FPUT and the Toda systems upto a characteristic time (called the trapping time $\tau_{\rm tr}$) that increases with decreasing energy (at fixed N). Then suddenly a bifurcation appears, which can be discussed in relation to the breakup of regular, solitonlike structures. After this bifurcation, $\Lambda$ of the $\alpha$ -FPUT model appears to approach a constant, while $\Lambda$ of the Toda system appears to approach 0, consistent with its integrability. This difference is attributed to
the untrapping of the FPUT system from its regular region in
phase space and escaping to the chaotic component of its
phase space. They infer that once a trajectory enters the
chaotic component of phase space, equipartition will eventually be attained. 
The authors computed the equilibration time $\tau_{\rm eq}, \tau_{\rm tr}$ and $\Lambda$ for system sizes $N=32,64,128$ at different energy densities and found the existence of a threshold $e_c(N)$ such that for $e < e_c$, $\tau_{\rm eq}$ and $\tau_{\rm tr}$ seemed to diverge while $\Lambda$ vanishes. The threshold decreases with system size as $e_c(N) \sim 1/N^2$. Above $e_c$, power law dependences of the form $\tau_{\rm eq} \approx 1/e^3, \tau_{\rm tr} \approx 1/e^{2.5}$ and $\Lambda \approx e^2$ were noted.  The FPUT parameters correspond to $e \ll e_c$.

\section{Role of Breather Solutions}

Breathers are time-periodic and space-localized solutions of nonlinear dynamical systems \cite{Marin1996}. By nature, they are non-thermal and one might expect them to
play a role in preventing thermalization. In a certain sense the idea is similar to the one relating the presence of solitons in the KdV system to the absence of equilibration in the FPUT system—the difference being that breathers are stable solutions of the discrete system while the KdV is a continuum approximation. The unexpected recurrences in the FPUT problem have been linked to the choice of initial conditions used by FPUT, which are set close to exact coherent time-periodic (or even quasiperiodic) trajectories, e.g., q-breathers, which show exponential localization of energy in normal mode space. Even if these trajectories have a measure zero in the phase space, they might have a finite measure impact simply by being linearly stable \cite{Flach2005}. Several other studies admit coherent time-periodic states localized in real space, which are known as discrete breathers or intrinsic localized modes. 

In \cite{Danieli2017}, Danieli and others have shown how one can check for the presence of breather solutions. They called trapping time, as the time the trajectory spends off the equilibrium manifold before piercing it again. By measuring the probability distribution functions of the trapping times in the FPUT chain, they infer that the trajectory is with high probability getting trapped in some parts of the phase space for long times. They also conjecture that these trapping events are due to visiting regions of the phase space which are substantially close to some regular orbits, which are supported by numerical experiments. 

\section{Ideas from Wave Turbulence}

The FPUT problem has recently been studied \cite{Onorato2015,Lvov2018,Pistone2019} using the approach of  wave turbulence \cite{Zakharov1992,Nazarenko2011}. Based on requirement of resonance between sets of normal modes, a detailed prediction has been made for the time-scale for equilibration and it is argued that this is finite for any non-zero strength of non-linearity, for finite sized systems. We now describe the work \cite{Onorato2015} in some detail. 
Let us first describe the $\alpha-$FPUT system in terms of the normal mode coordinates $a_k$:
\begin{equation}
a_k = \frac{1}{\sqrt{2\omega_k}}(P_k-i\omega_kQ_k).
\end{equation} 
If $\alpha \neq 0$, there will be coupling among the normal modes and let us write the equations of motion. Writing in terms of the dimensionless variables,

$$ a_k^\prime = \frac{(\mu/m)^{1/4}}{\sqrt{\sum_k \omega_k \mid a_k(t=0)\mid^2 }}a_k,\ t^\prime = \sqrt{\frac{\mu}{m}}t,\ \omega_k^\prime=\sqrt{\frac{m}{\mu}}\omega_k,$$
we can write the equations of motion after removing the primes for brevity:
\begin{equation}\label{eq:evol}
i\frac{\partial a_1}{\partial t} = \omega_1 a_1 + \epsilon \sum_{k2,k3} V_{1,2,3} (a_2a_3\delta_{1,2+3}+2a_2^\star a_3\delta_{1,3-2}+a_2^\star a_3^\star\delta_{1,-2-3})~,
\end{equation}
where the matrix $V_{1,2,3}$ weights the transfer of energy between wave numbers $k_1$, $k_2$ and $k_3$ and is given by:
\begin{equation}\label{eq:transfer}
V_{1,2,3} = -\frac{1}{2\sqrt{2}}\frac{\sqrt{\omega_1\omega_2\omega_3}}{sign(sin(\frac{\pi k_1}{N})sin(\frac{\pi k_2}{N})sin(\frac{\pi k_3}{N}))}~.
\end{equation}
The dimensionless parameter $\epsilon$ is given by:
\begin{equation}\label{eq:nonlinear}
\epsilon = \frac{\alpha}{m}(\frac{\mu}{m})^{1/4}\sqrt{\sum_k \omega_k \mid a_k(t=0)\mid^2 } ~,
\end{equation}
where $a_i\equiv a(k_1,t)$ and the delta function is 1 also if the argument differs mod N. The variables $k_2$ and $k_3$ run from $-N/2+1$ to $N/2$. It is easy to notice that this definition of $\epsilon$ reduces to the same as that we have used in the previous chapter. Eq.~(\ref{eq:evol}) describes the evolution of the normal mode variables in the $\alpha$-FPUT system and as expected, it has coupling between the different modes. We are now interested in the weak nonlinearity limit $\epsilon << 1$. The nonlinear behaviour of the $\alpha$-FPUT system is due to three-wave interactions, with the strength of coupling determined by $\epsilon$.  First we try to find out if these three waves interactions are resonant. 
The way to check for resonances is to first do a canonical transformation and then try to remove the three wave interactions. 
If a canonical transformation exists (does not diverge) then we say that the three wave interactions are not resonant. Otherwise, we conclude that there are resonances and these interactions cannot be removed by a canonical transformation. For the $\alpha$-FPUT system, this canonical transformation, in terms of $b$ variables, up to first order is given by:
\begin{equation}
a_1=b_1+\epsilon\sum_{k2,k3}(A_{1,2,3}^{(1)}b_2b_3\delta_{1,2+3}+A_{1,2,3}^{(2)}b_2^\star b_3\delta_{1,3-2}+A_{1,2,3}^{(3)}b_2^\star b_3^\star\delta_{1,-2-3})+O(\epsilon^2)~.
\end{equation}
Here,
$$ A_{1,2,3}^{(1)} = V_{1,2,3}/(\omega_3+\omega_2-\omega_1),$$
$$ A_{1,2,3}^{(2)} =2V_{1,2,3}/(\omega_3-\omega_2-\omega_1),$$
$$ A_{1,2,3}^{(3)} =V_{1,2,3}/(-\omega_3-\omega_2-\omega_1).$$
It is easy to verify using the trigonometric identities that for the frequencies $\{\omega_i\} $ of the harmonic chain 
the denominators in the above transformations are never zero. Therefore, a canonical transformation exists, the three wave interactions are not resonant and hence we observe the quasiperiodic behaviour at short time scales.
Substituting this transformation in the equations of motion, we get:
\begin{equation}\label{eq:evol2}
i\frac{\partial b_1}{\partial t} = \omega_1 b_1 + \epsilon^2 \sum_{k2,k3,k4} T_{1,2,3,4} b_2^\star b_3b_4\delta_{1+2,3+4} +O(\epsilon^3)~.
\end{equation}
Now proceeding in a similar way, we try to remove these four wave interactions using a suitable canonical transformation. This will be possible if these interactions are not resonant.
The above equation also has terms including the Kronecker deltas $\delta_{1,3+4+2}$,$\delta_{1,-3-4-2}$ and $\delta_{1,4-3-2}$. However, those terms are not resonant and can be removed by higher order terms in the transformation, and are of no use in this analysis. 
It is easy to verify that the resonances of Eq.~\eqref{eq:evol2} are described by the equations:
\begin{equation}\label{eq:res2}
k_1 + k_2 - k_3 -k_4 = 0\ mod\ N,\ \rm{and}\  \omega_1+\omega_2-\omega_3-\omega_4=0~.
\end{equation}
Now there are two kinds of solutions for this equation:
(i) Trivial solutions: These are obtained when
$ k_1 = k_3$ and $ k_2 = k_4$, or $ k_1 = k_4$ and $ k_2 = k_3$. These trivial solutions are responsible for a nonlinear frequency shift and do not contribute to the energy transfer between the modes.
(ii) Nontrivial solutions: These are obtained when $ (k_1,k_2,k_3,k_4) = (k_1,k_2,-k_1,-k_2)$ or $k_1 + k_2 = mN/2$, $m=0,\pm1,\pm2,...$. These solutions allow for a transfer of energy between the the modes. However, these solutions are isolated from one another. For example, two sets of solutions $(k_1,k_2,k_3,k_4) = (1,15,-1,-15)$ and $(k_1,k_2,k_3,k_4) = (3,13,-3,-13)$ have no wave number in common. 
Hence energy transfer can only take place between the modes that satisfy these conditions and energy transfer does not take place across the spectrum. Thus we say that these resonances are isolated and this would not lead to equipartition of energy. Nevertheless, these resonances do exist and cannot be removed by a higher order canonical transformation. Now, we move on to higher order resonances. Doing another canonical transformation would lead to five wave interactions, which can be shown to be non resonant. So, these can be removed by an appropriate choice of a canonical transformation, again just like a three wave interaction. So, we go to the next order interactions, which are the six wave interactions. The evolution of the system is described in terms of the following equation:



\begin{equation}\label{eq:evol3}
i\frac{\partial b_1}{\partial t} = \omega_1 b_1 + \epsilon^2 \sum T_{1,2,3,4} b_2^\star b_3b_4\delta_{1+2,3+4} + \epsilon^4 \sum W_{1,2,3}^{4,5,6} b_2^\star b_3^\star b_4 b_5 b_6\delta_{1+2+3,4+5+6} +O(\epsilon^5)~.
\end{equation}
In the above equation, the four wave interactions cannot be removed because they are resonant and a canonical transformation suitable for removing those modes would diverge. Resonances for this equation can be found by finding the solutions of:
\begin{equation}\label{eq:res3}
k_1 + k_2 + k_3 -k_4 - k_5 -k_6 = 0\ mod\ N,\ \rm{and}\  \omega_1+\omega_2+\omega_3-\omega_4-\omega_5-\omega_6=0~.
\end{equation}
There are three kinds of solutions for this equation now:
(i) Trivial solutions: As before, they would only lead to a nonlinear frequency shift. 
(ii) Nontrivial symmetric resonances: These are given by $ (k_1,k_2,k_3,k_4,k_5,k_6) = (k_1,k_2,k_3,-k_1,-k_2,-k_3)$ or $k_1 + k_2 +k_3= mN/2$, $m=0,\pm1,\pm2,...$. For example, $(k_1,k_2,k_3,k_4,k_5,k_6) = (1,2,13,-1,-2,-13)$ and  $ (k_1,k_2,k_3,k_4,k_5,k_6) = (1,8,7,-1,-8,-7)$ have the wave number $k=1$ in common.
(iii) Nontrivial quasisymmetric resonances: These are given by 
$(k_1,k_2,k_3,k_4,k_5,k_6)=(k_1,k_2,k_3,-k_1,-k_2,k_3)$ or $k_1 + k_2= mN/2$, $m=0,\pm1,\pm2,...$. For example, \newline$ (k_1,k_2,k_3,k_4,k_5,k_6) = (1,15,13,-1,-15,13)$ is connected with $ (k_1,k_2,k_3,k_4,k_5,k_6) = (13,3,7,-13,-3,7)$ through $k=13$ mode. One can also find a common wave number between nontrivial symmetric resonances e.g., $ (k_1,k_2,k_3,k_4,k_5,k_6) = (2,4,10,-2,-4,-10)$ and nontrivial quasisymmetric resonances e.g., $ (k_1,k_2,k_3,k_4,k_5,k_6) = (7,9,10,-7,-9,10)$. Thus, these resonances are interconnected and they lead to the equipartition of energy. By using BBGKY hierarchy (as given in \cite{Onorato2015}), one can estimate that the equilibration time ($\tau_{\rm eq}$) of the $\alpha-$FPUT system depends on the nonlinearity parameter as:
\begin{equation}\label{tauepsilon}
\tau_{\rm eq} \propto \frac{1}{\epsilon^{8}}~.
\end{equation}
Had the four wave resonances been interconnected, as is the case in the thermodynamic limit, the system would thermalize sooner. In \cite{Onorato2015} it is argued that in the thermodynamic limit the dependence would instead have been: $\tau_{\rm eq} \sim 1/\epsilon^4$. A more recent study \cite{Lvov2018} of the $\beta$-FPUT chain suggests that wave-wave resonances lead to thermalization at small $\epsilon$, where $\tau_{\rm eq} \sim 1/\epsilon^4$, while at larger $\epsilon$ the overlap of resonances (Sec.~\ref{sec:Overlap}) leads to $\tau_{\rm eq} \sim 1/\epsilon$.

\chapter{Thermalization of Local Observables in the $\alpha-$FPUT Chain}
\label{chap:fput2}
In this chapter, we study the problem of equilibration of the $\alpha-$FPUT system using a different approach. As mentioned in the problem statement in chapter \ref{chap:intro}, our work  attempts to analyse the problem ``locally"- by checking equipartition theorem at different sites in the chain instead of checking the normal mode energy which is  ``global", since normal modes involve all the particles in the chain. Also, most studies look at initial conditions specified in the normal mode space - they excite the first or the first few normal modes and then study the evolution of the FPUT chain. However, we study the evolution by using initial conditions that are localized in the real space - we excite a few particles initially and then study thermalization. We study various aspects of thermalization like the dependence of the equilibration time on the choice of initial conditions, choice of observables, choice of averaging and the role of chaos in equilibration, and also compare our results with the predictions of the wave turbulence theory \cite{Onorato2015}. 

\section{Initial Conditions and the Observables that are Considered in this Study}
Let us first describe the initial conditions that we have used in our study. Here we consider the case where the initial distribution lies on the constant $(E,P,N)$ surface in $2N$ dimensional phase space and has a small spread, with the size of the spread characterized by  a dimensionless number $\gamma$. The initial condition is also chosen to correspond to the initial energy of the chain being localized in a small region called space localized excitations (SLE). We set $q_i=0$, for $i=1,2,\ldots,N$ and $p_i=0$ for $i=5,6\ldots,N-1,N$. The total energy $E$ is then distributed amongst the four remaining particles in the following way:
\begin{eqnarray}
E_1&=&E_2=(1-\gamma)E/4 +\nu \gamma E/2,~ \label{eq:ic1}\\
E_3&=&E_4=(1-\gamma)E/4 +(1-\nu) \gamma E/2,~\label{eq:ic2} 
\end{eqnarray}
where $0<\gamma <1 $ is a number which specifies the ``width" of the distribution and  $\nu$ is a uniformly distributed random number in the interval $(0,1)$ which basically gives us some randomness in the initial conditions. $\gamma=0$ corresponds to a fixed initial state while $\gamma=1$ corresponds to the broadest distribution. Here we consider initial conditions that have zero momentum. The first and the second particles are given velocities in opposite directions, as are the third and the fourth. We shall
refer to these initial conditions as space localized excitations (SLE) as opposed to normal mode localized excitations (NMLE) commonly used in most studies, where the first or the first few normal modes are originally excited. In our simulations we generate $R$ initial configurations from this distribution and evolve each with Hamiltonian dynamics.  We then consider the ensemble average of a physical observable $A({\bf q},{\bf p})$ estimated as 
$$ \langle A \rangle (t) = \frac{\sum_{r=1}^R A_r(t)}{R}, $$ where the sum is over the $R$ members of the ensemble. Instead of normal modes, the observables that we are interested in are given by the equiparition theorem.  In the rest of this chapter we use the following notation: 
\begin{align}
\left\langle T_i \right\rangle = \frac{1}{2} \left\langle p_i\frac{\partial H}{\partial p_i}\right\rangle \, \quad \textrm{and} \quad \left\langle V_i \right\rangle =  \left\langle q_i\frac{\partial H}{\partial q_i} \right\rangle \,. \label{eq:avg-t-v1}
\end{align}
$\left\langle T_i\right\rangle$ is nothing but the average kinetic energy of the $i^{th}$ particle. Note however that $\left\langle V_i\right\rangle$ is not the average potential energy in general.

\section{Evolution of Local Observables from Space Localized Initial Conditions}

\begin{figure}[ht]
	\centering
	\hspace{-10mm}
	\includegraphics[width=0.54\textwidth]{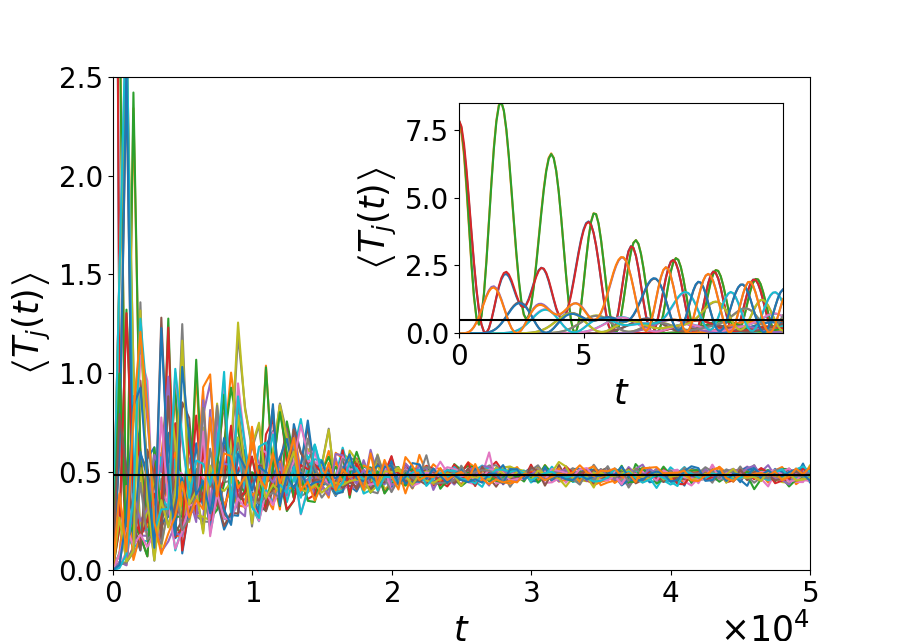}
	\put (-140,163) {$(a)$}
	\hspace{-6mm}
	\includegraphics[width=0.54\textwidth]{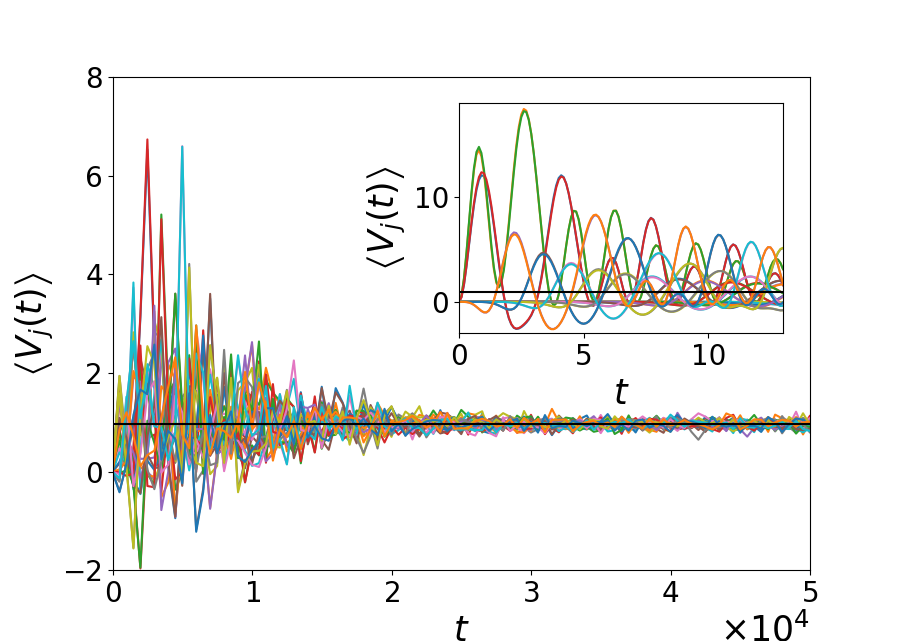}
	\put (-140,163) {$(b)$}
	\hspace{-10mm}
	\caption{$\alpha-$FPUT chain: Plot shows time evolution of $\left\langle T_j\right\rangle$ (left panel) and $\left\langle V_j\right\rangle$ (right panel) starting from space localized initial conditions for $j = 0,1,2,...N-1$.  Parameter values for this plot are $N=32$, $E=31$, $\alpha= 0.0848$ ($\epsilon=0.0834$), $\gamma=0.9$ and  $R=1000$. The solid black line corresponds to the equipartition value $\langle T_j \rangle= 0.484$ and $\langle V_j \rangle= 0.969$. The inset shows the time evolution at the earliest times.}
	\label{kvgp9}
\end{figure}
\begin{figure}[!]
	\centering
	\hspace{-10mm}
	\includegraphics[width=0.54\textwidth]{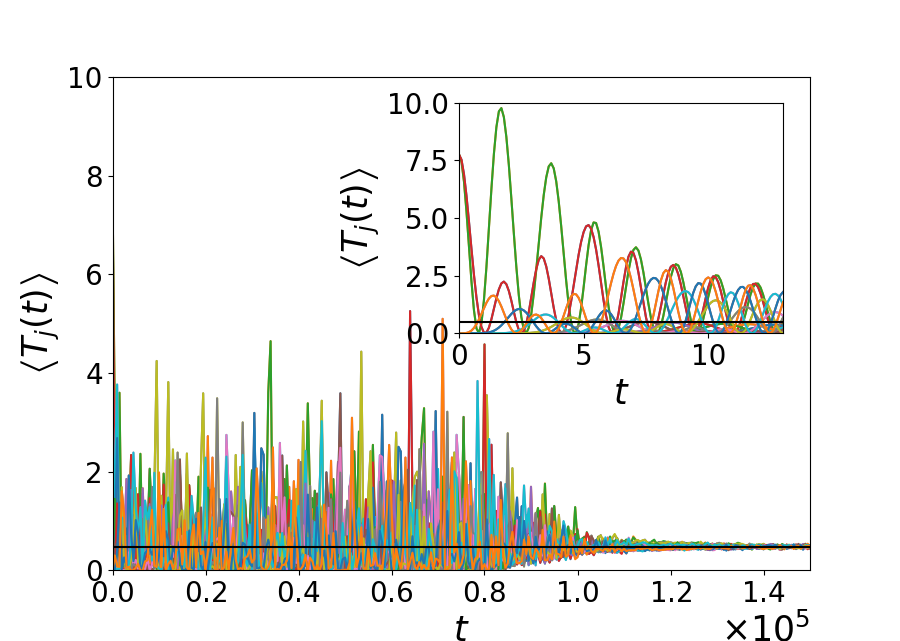}
	\put (-140,163) {$(a)$}
	\hspace{-6mm}
	\includegraphics[width=0.54\textwidth]{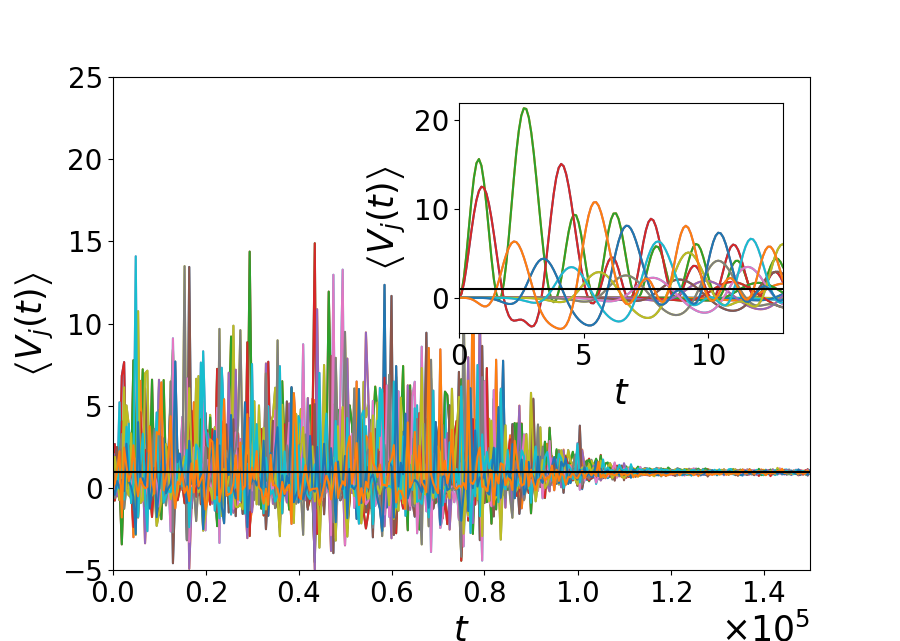}
	\put (-140,163) {$(b)$}
	\hspace{-10mm}
	\caption{$\alpha-$FPUT chain: Same as Fig.~\ref{kvgp9} but with $\gamma=10^{-8}$. Note that the range of the time $t$ and the averages $\langle T_j \rangle$ and $\langle V_j \rangle$ are different from those in Fig.~\ref{kvgp9}.}
	\label{kvgpm8}
\end{figure}

\begin{figure}[ht]
	\centering
	\hspace{-10mm}
	\includegraphics[width=0.54\textwidth]{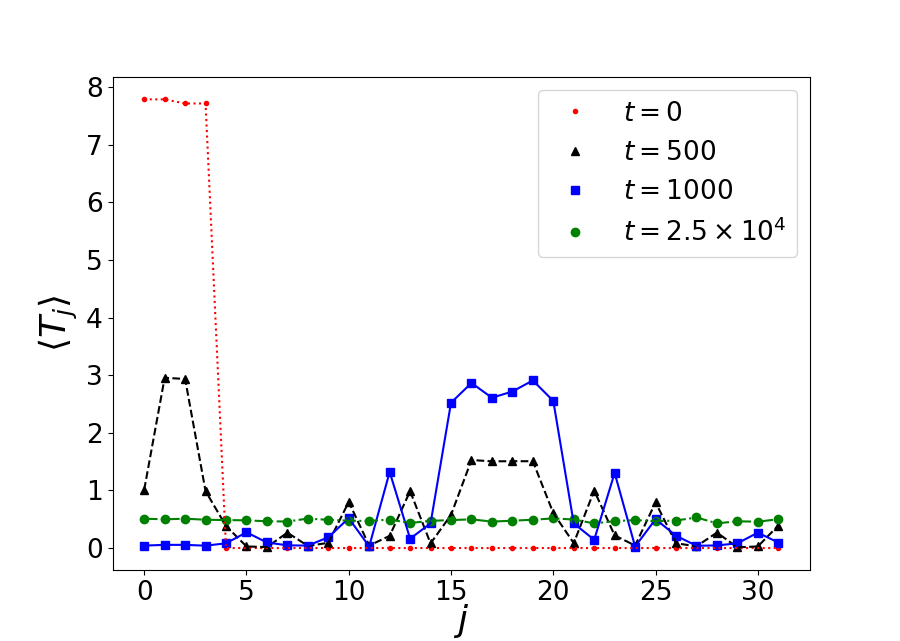}
	\put (-140,163) {$(a)$}
	\hspace{-6mm}
	\includegraphics[width=0.54\textwidth]{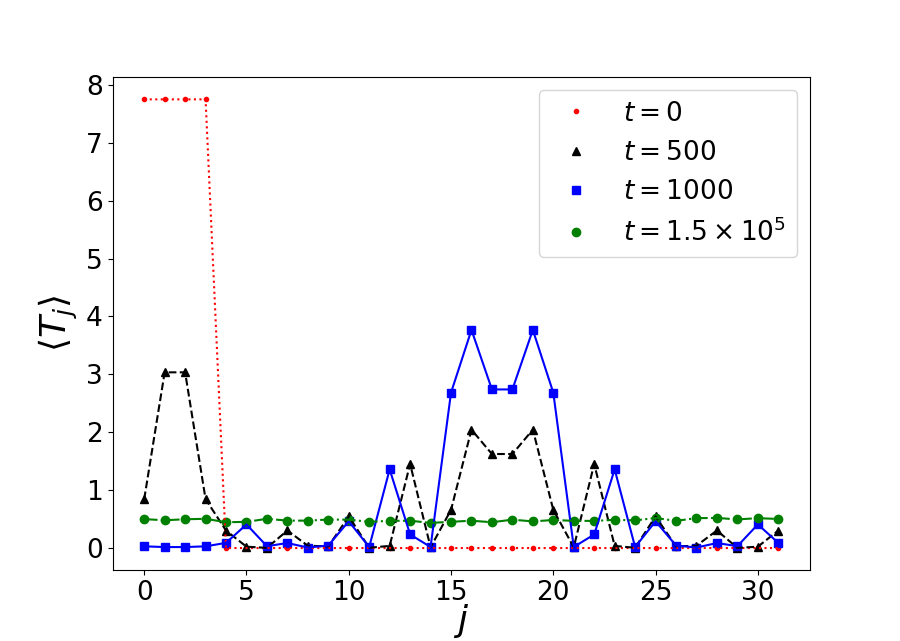}
	\put (-140,163) {$(b)$}
	\hspace{-10mm}
	\caption{$\alpha-$FPUT chain: Spatial profile of $\left\langle T_j\right\rangle$ at different times for   $\gamma=0.9$ (left panel) and $ \gamma=10^{-8}$  (right panel) and other parameters $N = 32$, $E = 31$, $\alpha=0.0848$ and $R=1000$. We see that the initially localized energy quickly spreads through the chain while equipartition is achieved at much longer time scales.}
	\label{evolution}
\end{figure}
In Fig.~\ref{kvgp9} we show the time evolution of $\left\langle T_j\right\rangle$ and $\left\langle V_j\right\rangle$ for $\gamma =0.9$.  We see that there is a long transient period and then we see equipartition at times  $\sim 2\times10^4$. At the earliest times, the inset in Fig.~\ref{kvgp9} shows near-recurrent behaviour. The results for $\gamma =10^{-8}$ are plotted in Fig.~\ref{kvgpm8}, where we now see that equipartition is achieved at somewhat longer times, around $t \sim 10^5$. We will return to the question of dependence of the equilibration time on the width of the initial distribution $\gamma$ later, when we try to relate thermalization and chaos. In Fig.~\ref{evolution} we plot the averaged kinetic energy profile at different times. Here it can be seen that the energy spreads quickly through the entire system, while equipartition is achieved at much longer time scales. 
\begin{figure}[!]
	\centering
	\hspace{-10mm}
	\includegraphics[width=0.56\textwidth]{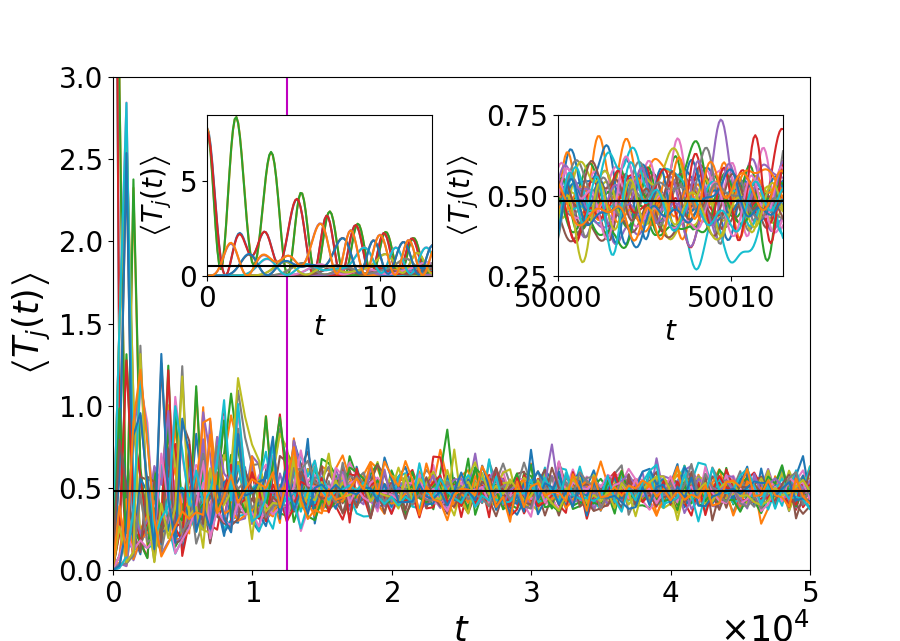}
	\put (-140,173) {$(a)$}
	\hspace{-10.7mm}
	\includegraphics[width=0.443\textwidth]{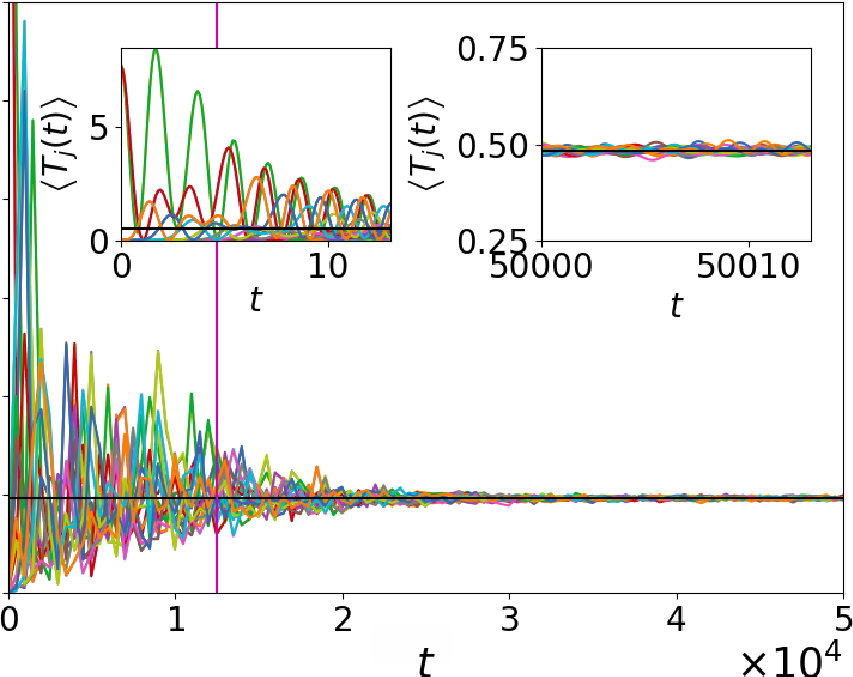}
	\put (-130,173) {$(b)$}
	\hspace{-10mm}
	\caption{$\alpha-$FPUT chain: Here we examine how the fluctuations seen in $\left\langle T_j\right\rangle$ depend on the number of realizations $R$. We plot $\left\langle T_j\right\rangle$  for $R=10^2$ (left panel) and $R=10^4$ (right panel). Other parameters were taken as  $N = 32$, $E = 31$, $\alpha= 0.0848$,  $\gamma=0.9$. The insets show zoom-ins at short and long times. A vertical line is drawn at $t=1.2\times10^4$, which is the equilibration time. Up to this time, both the plots look nearly the same.}
	\label{Rdependence}
\end{figure}

A closer examination of the plots in Figs.~\ref{kvgp9}, \ref{kvgpm8} shows that even at late times, the averaged quantities continue to fluctuate around their equilibrium values. We show 
in Fig.~\ref{Rdependence} that these fluctuations in fact decrease with increase in the number of realizations $R$ used to compute averages. We also see that  $\left\langle T_j\right\rangle$ at pre-thermalization times does not depend significantly on $R$ and the plots are near identical for $R=10^2$ and $R=10^4$ (up to the vertical line in Fig.~\ref{Rdependence}).  

\section{Normal Mode Localized Initial Conditions}
\begin{figure}[htp]
	\centering
	\hspace{-10mm}
	\includegraphics[width=0.54\textwidth]{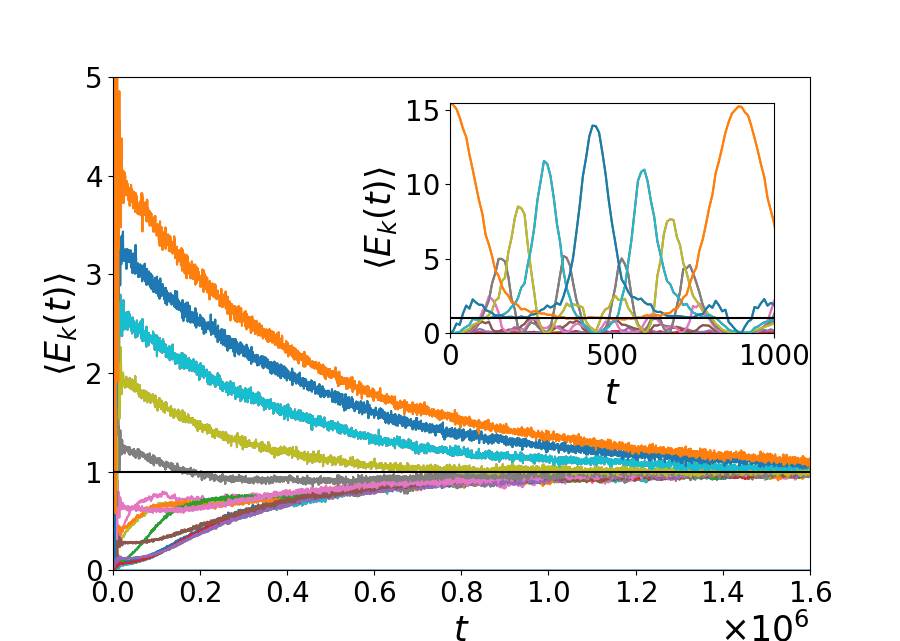}
	\put (-140,163) {$(a)$}
	\hspace{-5.8mm}
	\includegraphics[width=0.54\textwidth]{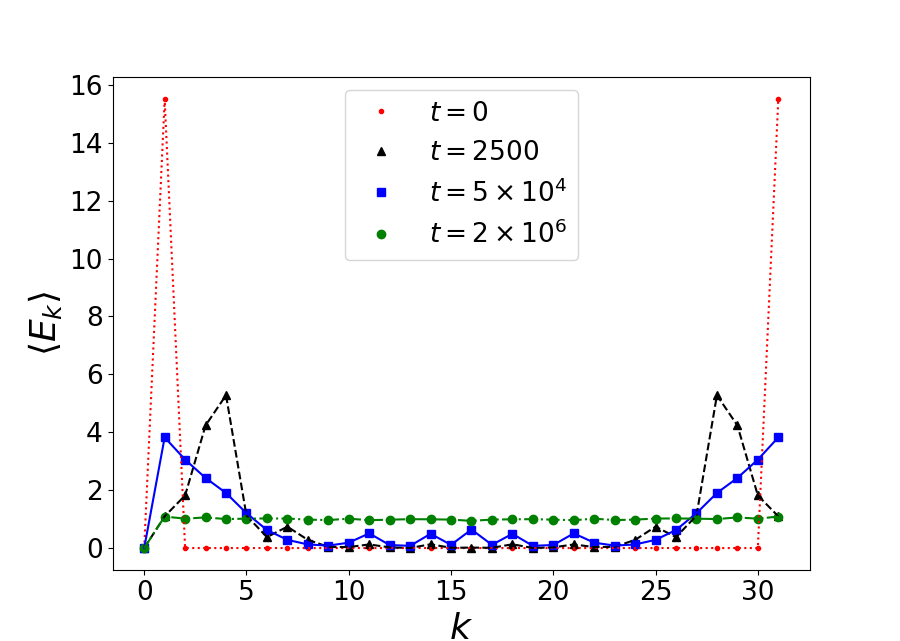}
	\put (-140,163) {$(b)$}
	\hspace{-10mm}
	\caption{$\alpha-$FPUT chain: Left panel shows time evolution of $\left\langle E_k\right\rangle$  for $N=32, E=31, \alpha=0.0848$ and  $R=1000$. Only modes $k=1,31$ are initially excited. The inset shows quasiperiodic behaviour at short time scale. Right panel shows the mode energy profile at different  times.}
	\label{NMevol}
\end{figure}

We now discuss and compare our results with those obtained in studies on equipartition of normal mode energies using initial conditions which are localized in the Fourier space \cite{Onorato2015}. 
The authors in \cite{Onorato2015}  considered  initial conditions  where the  energy was  distributed between the modes $k=1$ and $k=31$ with frequencies  $\Omega_{1}=\Omega_{31}$.  Averages were done over $1000$ initial conditions by choosing  random phases for 
the modes.  The time evolution of the normal mode energies was monitored to check for equipartition. Here we reproduce their numerical results and compare with the results in the previous section. We consider again  $N=32$ particles with total energy $E=31$. In Fig.~\ref{NMevol} we show the time evolution of the energy of all the modes in the system. Comparing with Figs.~\ref{kvgp9}, \ref{kvgpm8} it is clear that equilibration now occurs at a  time scale ($\sim  10^6$) that is about an order of magnitude  longer.   

\section{Dependence of Equilibration on the Averaging Procedure}
\begin{figure}[ht]
	\centering
	\hspace{-10mm}
	\includegraphics[width=0.54\textwidth]{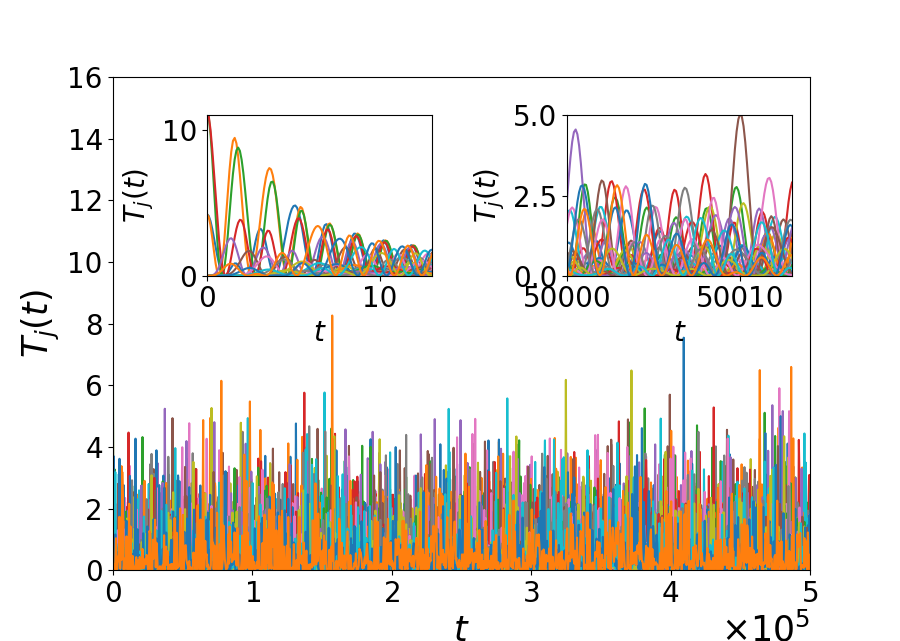}
	\put (-140,163) {$(a)$}
	\hspace{-6mm}
	\includegraphics[width=0.54\textwidth]{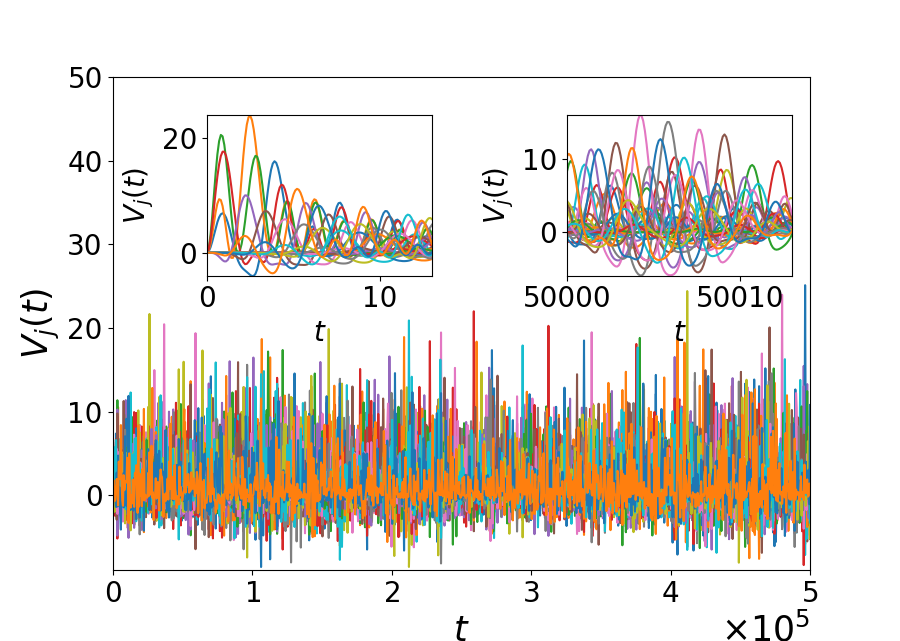}
	\put (-140,163) {$(b)$}
	\hspace{-10mm}
	\caption{$\alpha-$FPUT chain: Plot shows time evolution of $ T_j$ (left panel) and $ V_j$ (right panel) starting from a single initial condition.  Parameter values for this plot are $N=32$, $E=31$, $\alpha= 0.0848$ ($\epsilon=0.0834$). The insets show zoom-ins of the time evolution at early times and at late times, and we see oscillatory behaviour in both cases. Thus in this case, no equipartition is achieved.}
	\label{singleR}
\end{figure}

\begin{figure}[!]
	\centering
	\includegraphics[width=0.56\textwidth]{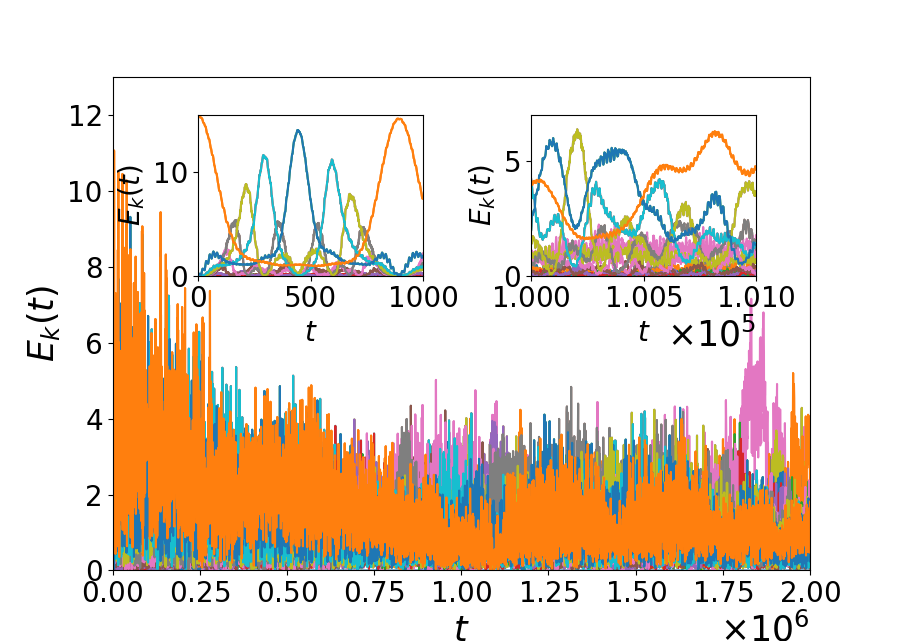}
	\caption{$\alpha-$FPUT chain: Plot shows time evolution of $E_k $  starting from a single initial condition with energy in two normal modes.  Parameter values for this plot are $N=32$, $E=31$, $\alpha= 0.0848$. The insets show zoom-ins  of  the time evolution at early times and at late times and we see oscillatory behaviour in both cases. Thus in this case, no equipartition is achieved.}
	\label{NMsingle}
\end{figure}

To demonstrate that the averaging procedure is crucial to the equilibration process in this set-up, we show the evolution of  $ T_j$ and $V_j$ for a single initial condition. In this case, we see in Fig.~\ref{singleR} that no equilibration is achieved and the oscillatory behaviour persists up to $t =5 \times 10^5$. In Fig.~\ref{NMsingle} we see again that one does not see any signs of equilibration in the evolution of $E_k$ for a single realization. 

\begin{figure}[!]
	\centering
	\hspace{-10mm}
	\includegraphics[width=0.54\textwidth]{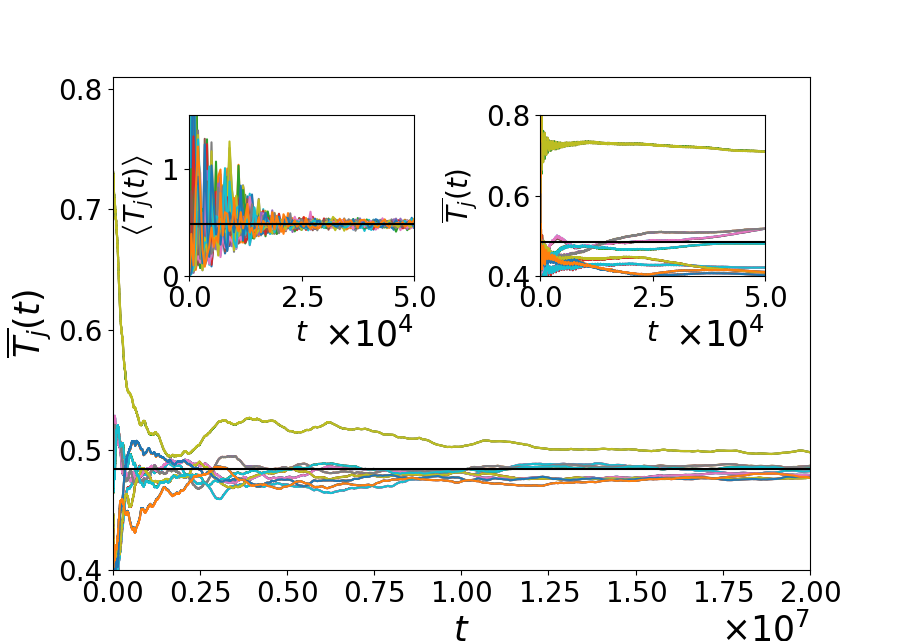}
	\put (-140,163) {$(a)$}
	\hspace{-6.5mm}
	\includegraphics[width=0.54\textwidth]{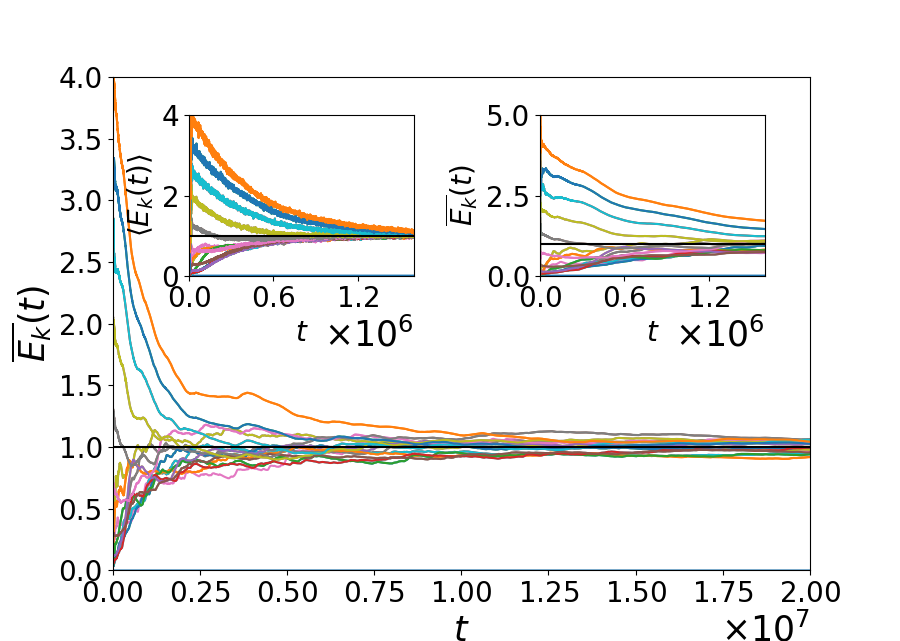}
	\put (-140,163) {$(b)$}
	\hspace{-10mm}
	\caption{$\alpha-$FPUT chain: Plot shows time evolution of running time averages of $ T_j$ for SLE (left panel) and $ E_k $ for NMLE (right panel) starting from a single initial condition.  Parameter values for this plot are $N=32$, $E=31$, $\alpha= 0.0848$ ($\epsilon=0.0834$).  The insets show comparison between time averages and ensemble averages, and illustrates that the latter procedure leads to faster thermalization.}
	\label{Timeaverage}
\end{figure}

We can also discuss thermalization using a different protocol where one starts with a single initial condition and then considers a time average of any given observable given by Eq.~\eqref{eq:tavg}. In Fig.~\ref{Timeaverage} we show results obtained by using this protocol for both the space-local and normal mode observables. The insets in the figures show that thermalization time scales are completely different from those obtained by the ensemble averaging protocol. This shows the important role played by the choice of averaging protocol in determining how fast the system equilibrates. 

\section{Quantification of the Equilibration Time}

\begin{figure}[ht]
	\centering
	\hspace{-10mm}
	\includegraphics[width=0.54\textwidth]{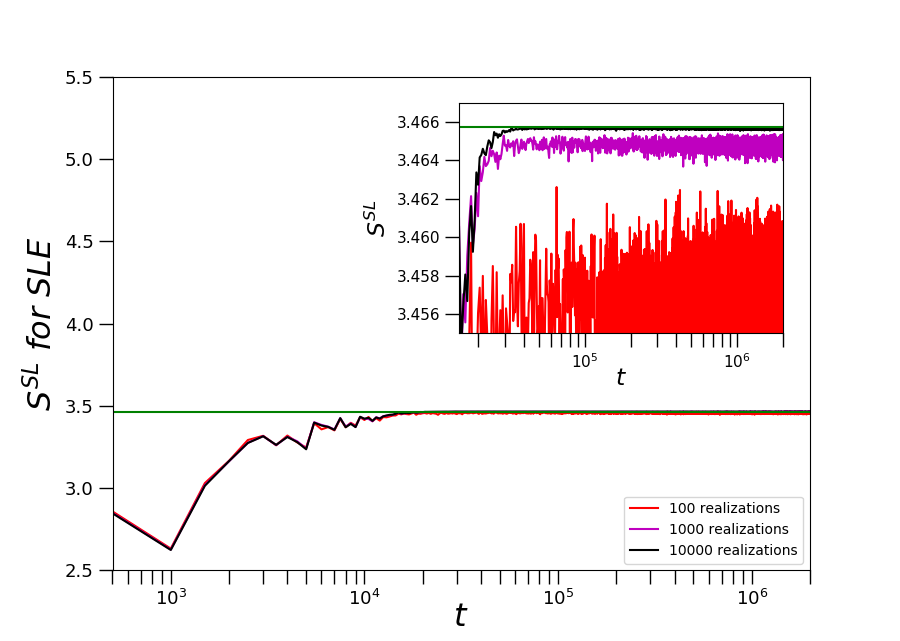}
	\put (-140,163) {$(a)$}
	\hspace{-6mm}
	\includegraphics[width=0.54\textwidth]{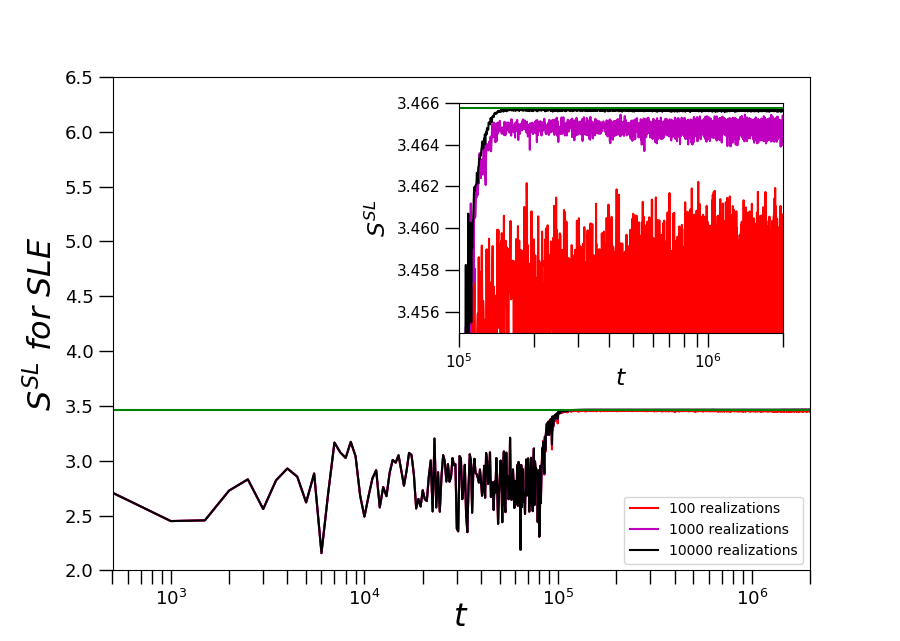}
	\put (-140,163) {$(b)$}
	\hspace{-10mm}
	\caption{$\alpha-$FPUT chain: Plot of $S^{SL}(t)$ corresponding to $\left\langle T_j\right\rangle$ for $\gamma=0.9$ (left panel) and $\gamma=10^{-8}$ (right panel) with other parameters $N=32, E=31, \alpha=0.0848$. The insets show the convergence of the equilibration process as the number of realizations is increased.  }
	\label{EntropyEv}
\end{figure}

To get a more systematic and quantitative estimate of the equilibration time we now look at the entropy function defined in Eq.~\eqref{eq:ent}. This in some sense, performs an average over all the degrees of freedom and attains its maximum value $\ln N$ when all degrees have equilibrated. 
In Fig.~\ref{EntropyEv}, we plot the evolution of entropy for the two parameter values $\gamma=0.9$ and $\gamma=10^{-8}$. The insets show zoom-ins near the equilibrium value $\ln N$, showing the approach to  equilibration and its dependence on the number of realizations. For higher number of realizations the fluctuations in the entropy are lower and also the mean is closer to the equilibrium value. 

We  use the criterion of Eq.~\eqref{criterion} to estimate the equilibration time from the entropy $S^{SL}$ corresponding to $\langle T_j\rangle$ and find $\tau_{\rm eq}\approx100300$ and $12500$ for $\gamma=10^{-8}$ and $0.9$ respectively.
In general we find that as the ``width" of the distribution $\gamma$ is increased, thermalization happens faster. We will discuss this again later. 

\section{Dependence of Equilibration on the Choice of Initial Conditions and Observables}

\begin{figure}[ht]
	\centering
	\includegraphics[width=0.6\textwidth]{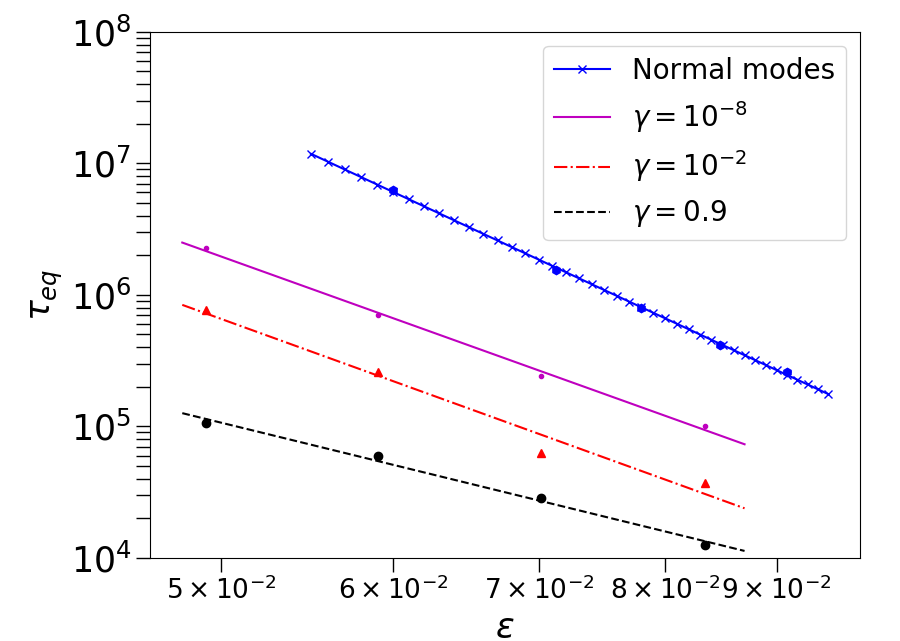}
	\caption{$\alpha-$FPUT chain: Graph showing $\tau_{\rm eq}$ of $\left\langle T_j\right\rangle$  for $N=32,E = 31$ as a function of $\epsilon$. The slopes of the fitting lines give  $a=5.9, 6.0, 4.0$  for $\gamma=10^{-8}, 0.01$ and $0.9$ respectively. We also show the equilibration times obtained from the normal mode entropy function $S^{NML}$ for normal mode localized initial conditions, which leads to an exponent $a\approx 7.7$.}
	\label{relaxationtimeepsilon}
\end{figure}

Using our method, we compute the equilibration time for different values of the dimensionless parameter $\epsilon$ for $N=32$. These results are plotted in Fig.~\ref{relaxationtimeepsilon} where we find a power-law dependence of equilibration time, $\tau_{\rm eq}$, on $\epsilon$  of the form
\begin{equation}\label{tauform}
\tau_{\rm eq} \propto \frac{1}{\epsilon^{a}}~.
\end{equation}
The value $a$ is found to depend on $\gamma$ and lies between 4 and 6. This is significantly different from the form 1/$\epsilon^8$ obtained in \cite{Onorato2015}, by considering equilibration of normal modes. In Fig.~\ref{relaxationtimeepsilon} we also indicate the relaxation time results for the normal modes which give  $a \approx 7.7$. 

\begin{figure}[!]
	\centering
	\includegraphics[width=1.1\textwidth]{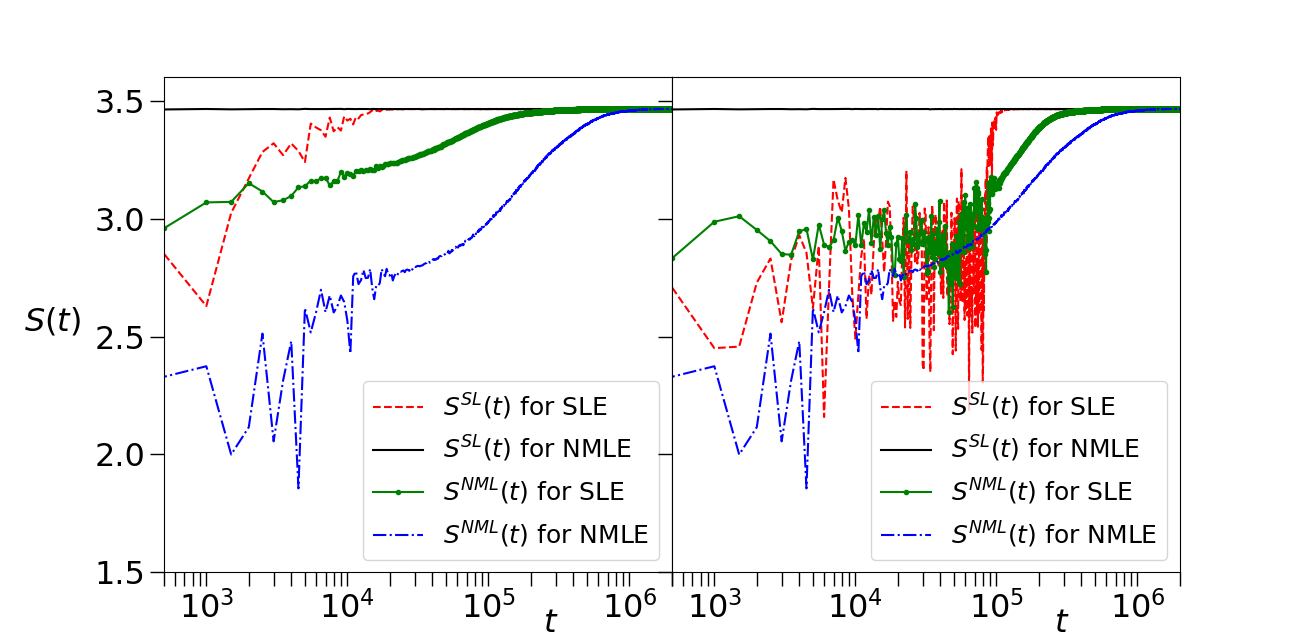}
	\put (-350,230) {$(a)$}
	\put (-150,230) {$(b)$}
	\caption{$\alpha-$FPUT chain: Graphs showing the evolution of the entropy corresponding to space-local observables and normal modes, for $\gamma=0.9$ (left panel) and  $\gamma=10^{-8}$ (right panel), with other parameters given by  $N=32, E=31, \alpha=0.0848$.  Results for both space localized initial conditions (SLE) and normal mode localized initial conditions (NMLE) are shown.}
	\label{combo8}
\end{figure}

It is to be expected that the equilibration time scale should depend  not only on the initial ensemble in which the system is prepared, but also on the observable for which equipartition is being tested. We investigate this question further by computing the entropy functions $S^{SL}$ and $S^{NML}$ for both types of initial conditions, namely space localized (SLE) and normal mode localized (NMLE). 
These results have been plotted in Fig.~\ref{combo8}. We see clearly that the relaxation of normal mode coordinates is slower than that of the space localized observables, irrespective of initial distribution. This is consistent with the higher value of the exponent $a$.


\section{Comparison with Other Integrable Models}
\begin{figure}[ht]
	\centering
	\hspace{-10mm}
	\includegraphics[width=0.54\textwidth]{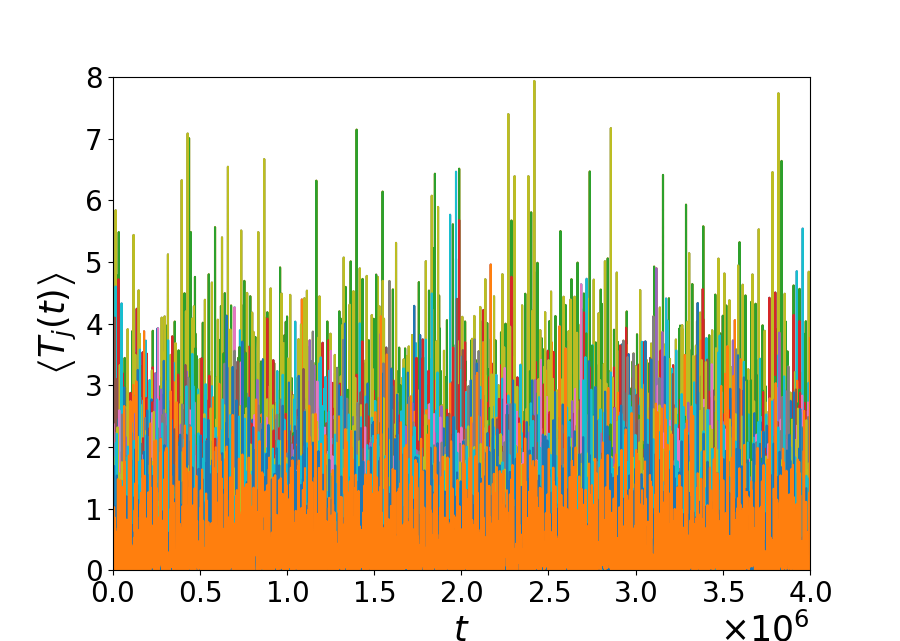}
	\put (-140,163) {$(a)$}
	\hspace{-5.8mm}
	\includegraphics[width=0.54\textwidth]{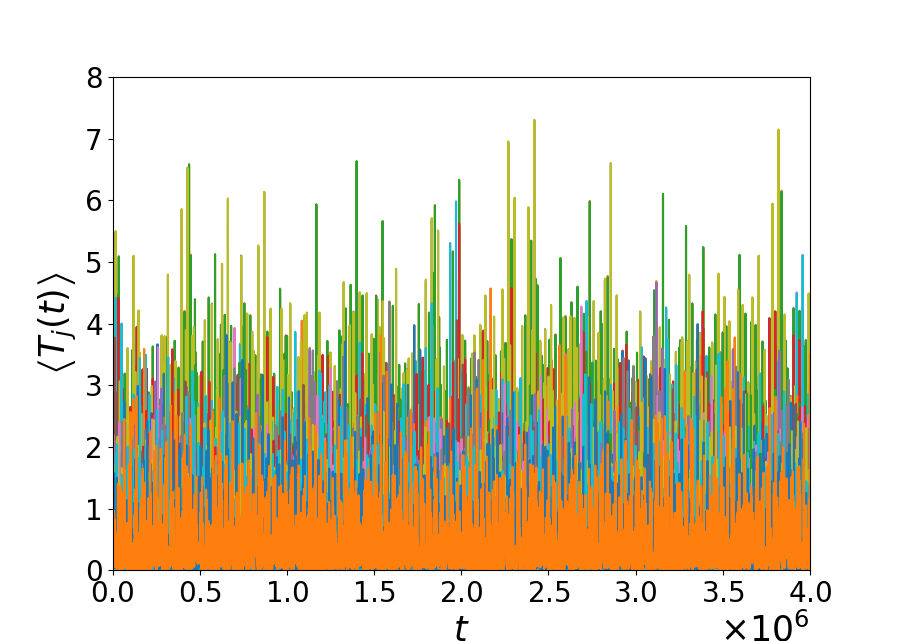}
	\put (-140,163) {$(b)$}
	\hspace{-10mm}
	\caption{Harmonic chain: Plot of $\left\langle T_j\right\rangle$ as a function of time, for $N = 32, E = 31, R=1000$ and $\gamma=10^{-8}$ (left panel) and $\gamma=0.9$ (right panel). In both cases there is no sign of thermalization.}
	\label{KEharmonic}
\end{figure}

\begin{figure}[ht]
	\centering
	\hspace{-10mm}
	\includegraphics[width=0.54\textwidth]{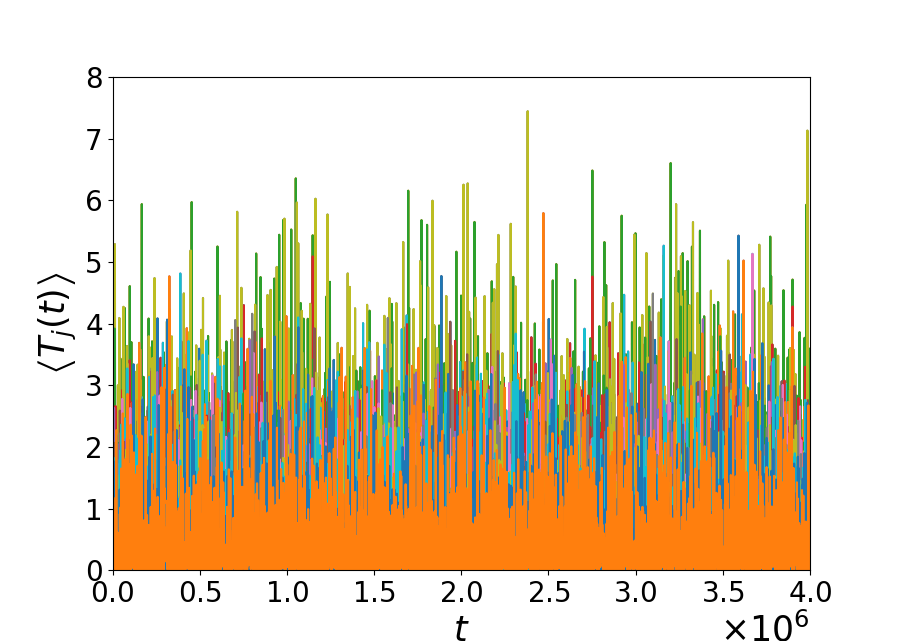}
	\put (-140,163) {$(a)$}
	\hspace{-5.8mm}
	\includegraphics[width=0.54\textwidth]{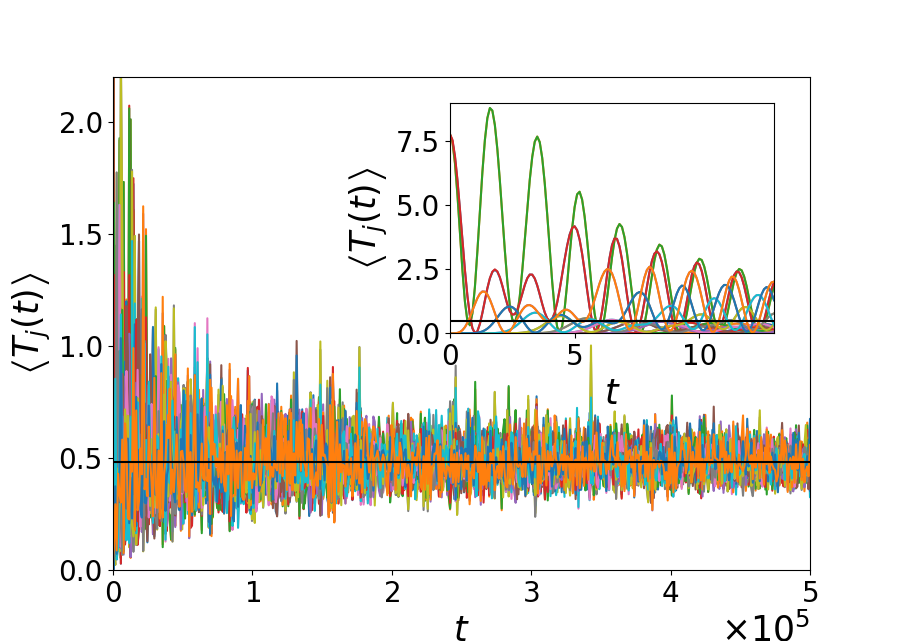}
	\put (-140,163) {$(b)$}
	\hspace{-10mm}
	\caption{Toda chain: Plot of $\left\langle T_j\right\rangle$ as a function of time, for $N = 32, E = 31, \alpha=0.0848, b=2\alpha, g=b^{-1}, R=1000$ and $\gamma=10^{-8}$ (left panel), and $\gamma=0.9$ (right panel). Now we observe thermalization in the right panel.}
	\label{KEtoda}
\end{figure}

\begin{figure}[!]
	\centering
	\hspace{-10mm}
	\includegraphics[width=0.53\textwidth]{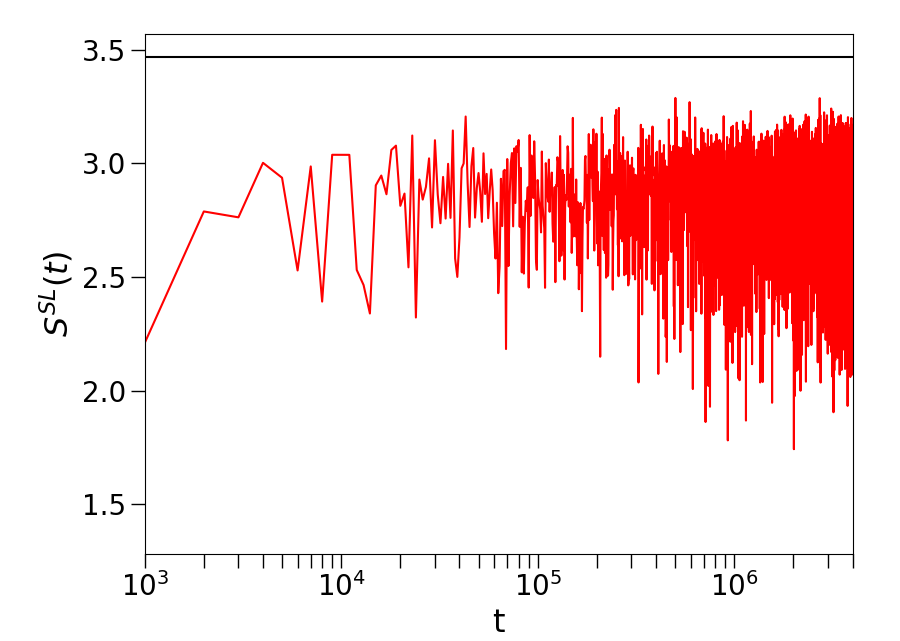}
	\put (-130,173) {$(a)$}
	\hspace{-3.8mm}	
	\includegraphics[width=0.53\textwidth]{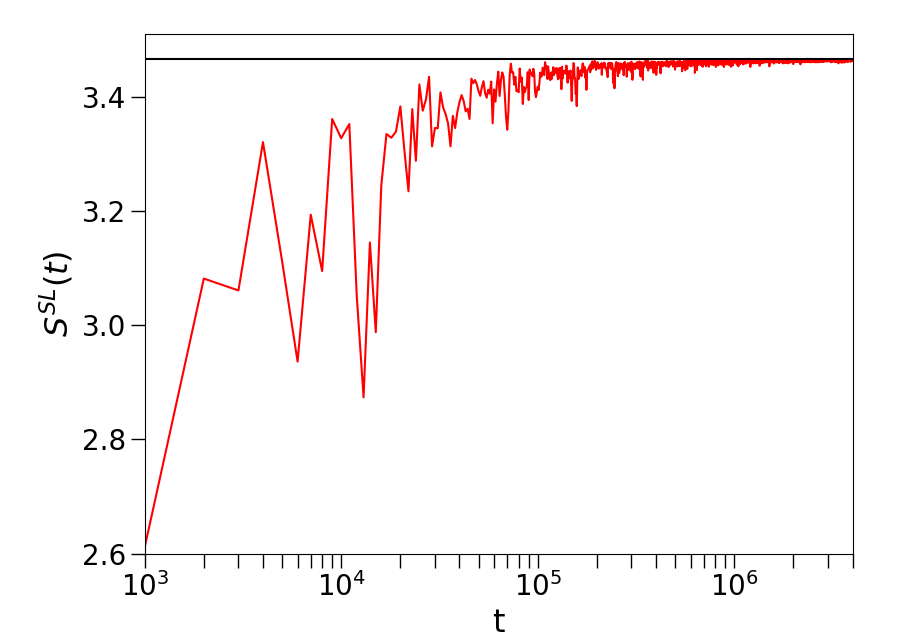}
	\put (-130,173) {$(b)$}	
	\hspace{-10mm}
	\caption{Toda chain: Plot of the entropy $S^{SL}(t)$ for $\left\langle T_j\right\rangle$ of the Toda chain with same parameters as Fig.~\ref{KEtoda} and $\gamma=10^{-8}$ (left panel), and $\gamma=0.9$ (right panel). The second case shows clear equilibration.}
	\label{Enttoda}
\end{figure}

To investigate the role of integrability, we now repeat the above computations in two integrable models that are related to the $\alpha-$FPUT system in the limit of weak nonlinearity. We consider the harmonic chain which is described by the Hamiltonian in Eq.~\eqref{eq:hamharm} and the Toda chain, described by the Hamiltonian
\begin{equation}\label{eq:todaham}
H(\boldsymbol{p},\boldsymbol{q})=\sum_{i=0}^{N-1}\left[\frac{p_i^2}{2}+ \frac{g}{b} e^{b (q_{i+1}-q_{i})} \right]~. 
\end{equation}
The Toda system is known to be integrable \cite{Toda1967,Toda1975} and has been much studied as the integrable limit of the FPUT chain \cite{Benettin2013,Casetti1996,Goldfriend2019,Fu2019}. 
The parameter choice $b=2\alpha$ and $g=b^{-1}$ would then approximate the $\alpha-$FPUT potential to leading nonlinearity.  Starting with the   same space-localized initial conditions as in the previous sections, we now check equipartition of kinetic energy $T_i$. In Fig.~\ref{KEharmonic} we see that no equilibration is achieved for the harmonic chain. On the other hand, somewhat surprisingly, we see in Fig.~\ref{KEtoda} that the Toda chain does equilibrate, provided we start with a wider initial distribution ($\gamma=0.9$). In Fig.~\ref{Enttoda} we plot the entropy function $S^{SL}$ and using the criterion in Eq.~\eqref{criterion}, estimate the equilibration time and find $\tau_{\rm eq} \approx 66000$ for $\gamma=0.9$. 

\section{Dependence of Equilibration on the Width of the Initial Distribution}
\begin{figure}[!]
	\centering
	\hspace{-10mm}
	\includegraphics[width=0.53\textwidth]{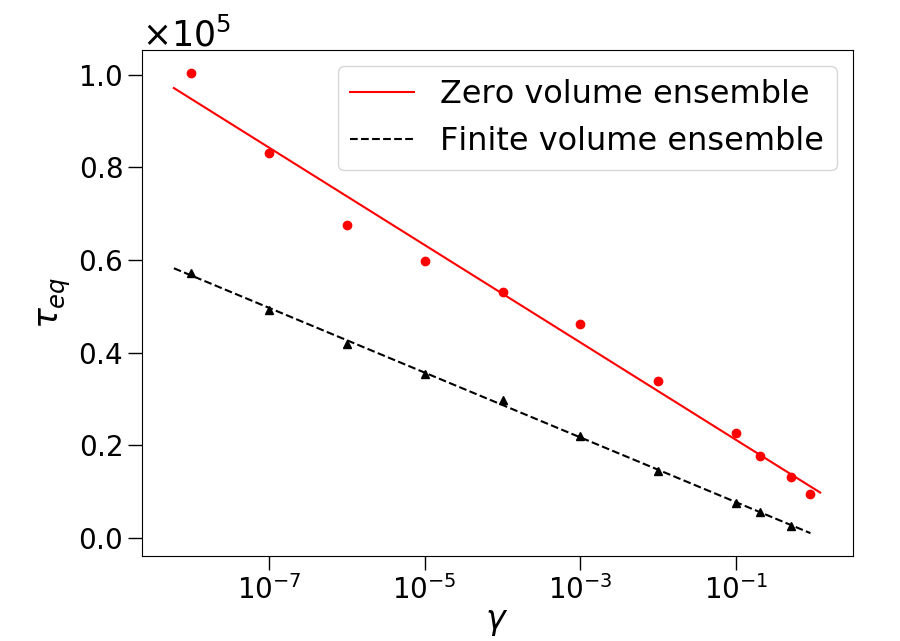}
	\put (-130,168) {$(a)$}
	\hspace{-3.8mm}
	\includegraphics[width=0.52\textwidth]{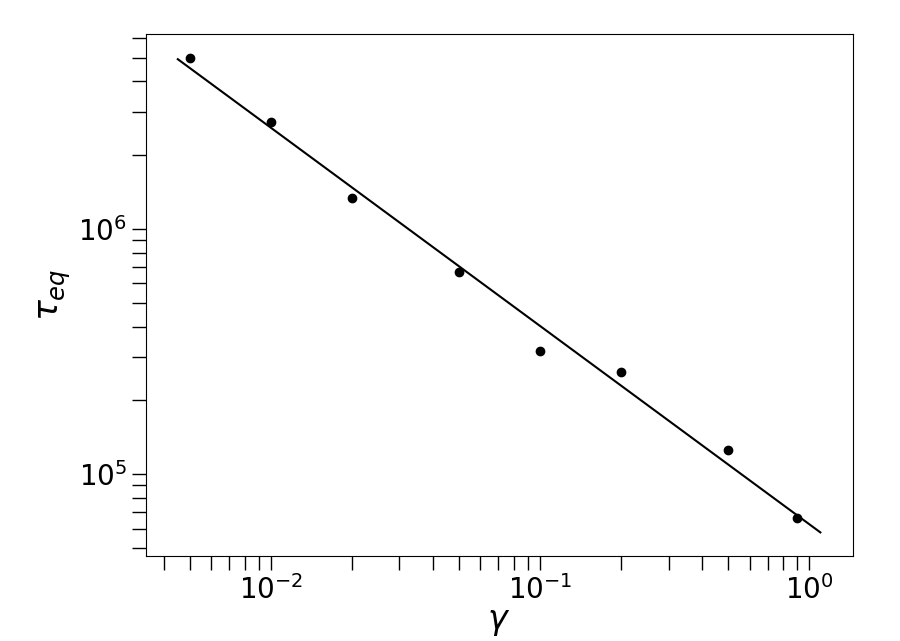}
	\put (-120,168) {$(b)$}
	\hspace{-10mm}
	\caption{Graphs showing the dependence of equilibration time of $\left\langle T_j\right\rangle$ on $\gamma$ for the $\alpha$-FPUT chain (left panel) and the Toda chain for $\alpha=0.0848$ (right panel). Other parameters are $N = 32$, $E = 31$,  and $R=1000$.  For the $\alpha$-FPUT chain, plotted on a linear-log scale, there are two different ensembles. The solid line represents an ensemble of initial conditions occupying zero volume in the phase space. It is described by Eqs.~\eqref{eq:ic1}-\eqref{eq:ic2}. The slope of this line is $-4.5\times 10^3$. The dashed line represents an ensemble of initial conditions occupying a finite volume in the phase space. The slope of this line is $-3.0\times 10^3$. The slope for the Toda  chain, plotted on a log-log scale, is $-0.81$.}
	\label{Fchaos}
\end{figure}

We now discuss the dependence of the equilibration process in the $\alpha-$FPUT system on the width of the initial distribution $\gamma$ used to compute the ensemble averages. For the $\alpha-$FPUT system, we now take $N = 32, E = 31, R = 1000$ and $\alpha = 0.0848$. We compute the equilibration time $\tau_{\rm eq}$ for different values of $\gamma$.  We find that broader the initial distribution (higher the $\gamma$), faster the equilibration. More precisely, we observe that $\tau_{\rm eq} \propto log(\gamma)$. In Fig.~\ref{Fchaos}a, the solid line describes an ensemble of initial conditions described by Eqs.~\eqref{eq:ic1}-\eqref{eq:ic2}, referred to as zero volume ensemble, since the latter occupies zero volume in the phase space. The magnitude of the slope of this line on a semilog plot is found to be $4.5\times 10^{3}$. Its inverse is $2.2\times 10^{-4}$, close to the Lyapunov exponent of the system $\Lambda\approx4.1\times 10^{-4}$.
We have also studied an ensemble of initial conditions that has randomness in  all the degrees of freedom (while maintaining momentum conservation), thereby occupying a finite volume in the phase space,  quantified by the number $\gamma$. 
The results for this are shown by the dashed line in  Fig.~\ref{Fchaos}a where again we see the logarithmic dependence on $\gamma$ 
and in fact find a  closer agreement between the magnitude of the slope (the inverse of which is $3.3\times 10^{-4}$) and the Lyapunov exponent.
For the Toda system however, we observe a different behaviour. In Fig.~\ref{Fchaos}b we show the dependence of $\tau_{\rm eq}$ on  $\gamma$ for the Toda chain, again with $\alpha=0.0848$. The slope on a log-log plot is close to $-1$ suggesting $\tau_{\rm eq} \propto 1/\gamma$. In the next section, we show how this dependence links equilibration of local observables to the  growth of perturbations of initial conditions, and hence to chaos. This idea then also explains the absence of equilibration in the harmonic chain and slow equilibration in the Toda chain.

\begin{figure}[ht]
	\centering
	\includegraphics[width=\textwidth]{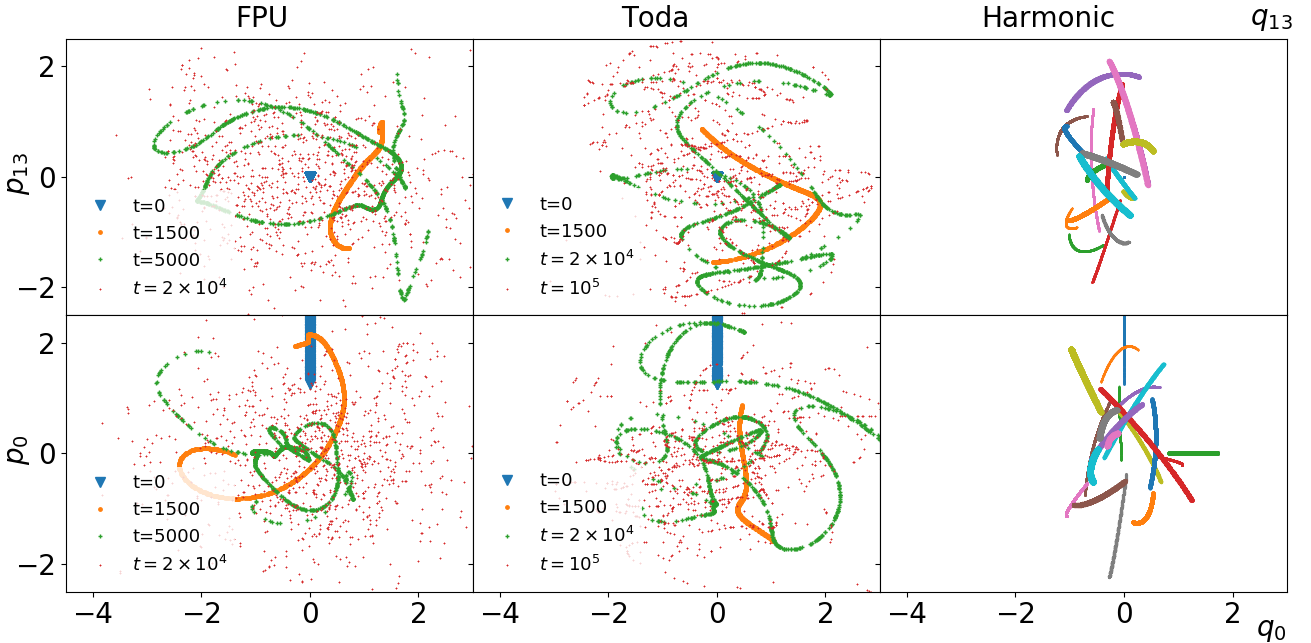}
	\caption{The time-evolution of the ensemble of initial conditions ($\gamma=0.9$) obtained using Eqs.~\eqref{eq:ic1}-\eqref{eq:ic2} is shown for the FPUT chain with $\alpha = 0.0848$ (left column), the corresponding Toda chain (middle column), and the harmonic chain (right column), in the $(p_{13},q_{13})$ and $(p_0,q_0)$ planes (top and bottom rows, respectively). The different lines for harmonic chain are at increased times, with increasing marker size.}
	\label{schematic}
\end{figure}

\section{Relation to Chaos}
We first quantify the growth of perturbations in the system. Let us consider an infinitesimal perturbation $\{ \delta q_i(0),\delta p_i(0) \}$ of an initial condition $\{ q_i(0), p_i(0)\}$. We compute the quantity
\begin{align}
Z(t)&=\frac{\sum_{i=1}^N \left[ \delta q_i^2(t) +\delta p_i^2(t) \right]}{\sum_{i=1}^N \left[ \delta q_i^2(0) +\delta p_i^2(0) \right]}\,.
\end{align}
We then compute the ensemble averaged time-dependent Lyapunov exponent $\lambda(t)$ defined as
\begin{align}
\lambda(t) =\frac{1}{2t} \langle \ln Z(t) \rangle~,
\end{align}
where  $\langle...\rangle$ denotes an average over initial conditions  $\{ q_i(0), p_i(0)\}$ chosen from the distribution $\rho_0({\bf q},{\bf p})$. As described earlier, the numerical integration of $\{ q_i(t), p_i(t), \delta q_i(t),\delta p_i(t) \}$ is done by solving $2N + 2N$ nonlinear and linearized equations. The largest Lyapunov exponent $\Lambda$ is then given by $\Lambda = \lim_{t \to \infty} \lambda(t)$. Since the harmonic chain and the Toda chain are integrable, $\Lambda=0$ for both of them. So, let us first understand why the Toda chain thermalizes for a broad initial distribution and the harmonic chain never thermalizes. Since they are both integrable, they can be written in terms of action-angle variables. The $N$ action variables ($\textbf{I}$) are constants of motions, while the $N$ angle variables ($\boldsymbol{\theta}$) evolve linearly with time as:
\begin{equation}\label{eq:angle}
	\theta_i(t) = \theta_i(0) + \omega_i(\textbf{I}) t~.
\end{equation}
The separation between any two initial conditions ($\boldsymbol{I_1},\boldsymbol{\theta_1}$) and ($\boldsymbol{I_2},\boldsymbol{\theta_2}$) evolves as:
\begin{equation}\label{eq:anglediff}
\theta_{1i}(t)-\theta_{2i}(t) = \theta_{1i}(0) - \theta_{2i}(0) + (\omega_{1i}(\boldsymbol{I_1}) - \omega_{2i}(\boldsymbol{I_2})) t~.
\end{equation}
for $i=0,1,2,...N-1$. In case of the harmonic chain, the $\omega$'s are not dependent on the action variables and are constants for different initial conditions. Thus, the separation between the two initial conditions is bounded. For the Toda chain however, the $\omega$'s are dependent on the action variables and are thus different for different initial conditions. Here the separation between the two initial conditions grows linearly with time. Hence we expect the Toda system to thermalize for a broad enough initial distribution.

The basic picture that illustrates the difference between the three models is shown in Fig.~\ref{schematic}, where we show the time-evolution of the ensemble of initial conditions obtained using Eqs.~\eqref{eq:ic1}-\eqref{eq:ic2}. The FPUT chain shows a fast growth in phase space because of its positive Lyapunov exponent, while it takes much longer for the Toda chain because of its integrability (and linear temporal growth of perturbations). There is no spread in the harmonic chain. 

\begin{figure}[!]
	\centering
	\hspace{-10mm}
	\includegraphics[width=0.54\textwidth]{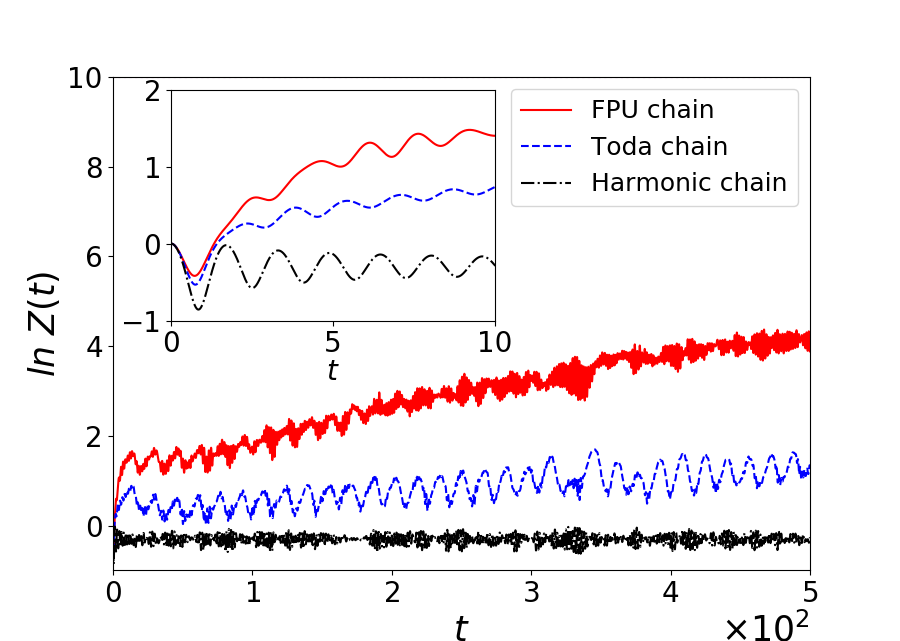}
	\put (-140,163) {$(a)$}
	\hspace{-6mm}
	\includegraphics[width=0.54\textwidth]{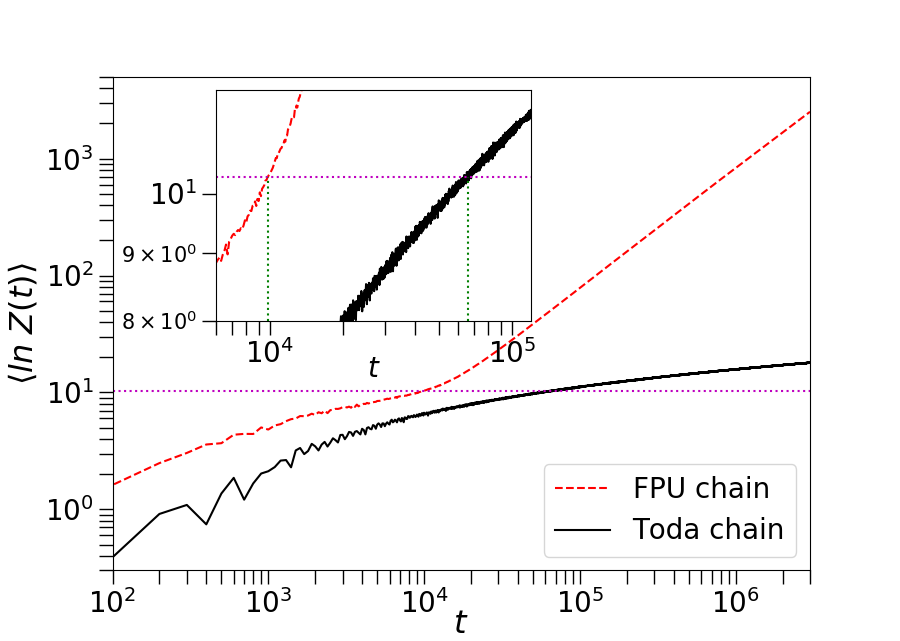}
	\put (-140,163) {$(b)$}
	\hspace{-10mm}
	\caption{Left panel shows $Z(t)$ of $\alpha-$FPUT chain, Toda chain and the harmonic chain at short times. It can be seen that the growth of perturbations is bounded for the harmonic chain and linear for the Toda chain. Right panel shows $Z(t)$ of $\alpha-$FPUT chain and the Toda chain for long times. The growth of perturbations in the $\alpha-$FPUT chain though linear at short times, becomes exponential at long times. It can be seen that the growth of perturbations is linear for the FPU chain and logarithmic for the Toda chain in the logarithmic scale. A horizontal line drawn at $\langle \log Z(t) \rangle=10.3$ intersects both the curves. Its significance is explained in the text.}
	\label{lyap-comp}
\end{figure}

In Fig.~\ref{lyap-comp} we plot $\langle \log Z(t) \rangle$ for the FPUT chain as well for the corresponding harmonic chain and Toda chain (with $b=2 \alpha$, $g=b^{-1}$). We confirm the expected exponential growth of $Z(t)$ for the FPUT chain at large times, it's linear growth for the Toda chain and the lack of growth in case of the harmonic chain. 
In the inset of the right panel, we also show the line $\langle \log Z(t) \rangle = 10.3$, which corresponds to the equilibration time of the  $\alpha-$FPUT chain ($\approx12000$, which we found earlier in this chapter). 
It turns out that the point of intersection of this horizontal line with the Toda chain is very close to its equilibration time ($\approx66000$). This gives us a means to relate the thermalization properties of both the systems to the growth of their perturbations.

Let us define $D(t)=\left[\sum_{i=1}^N \left( \Delta q_i^2(t) +\Delta p_i^2(t) \right) \right]^{1/2}$ to be the distance at time $t$, of points that are initially separated by a small but {\emph{finite}} separation $D(0)$, due to a finite perturbation of initial conditions. 
In particular, we will interpret $D(t)$ as the ``spread'' of an ensemble of trajectories, for example, using the ensemble of initial conditions as described in Eqs.~\eqref{eq:ic1}-\eqref{eq:ic2}.
At early times we expect that the growth can be described by
\begin{align}
D(t) \sim
\begin{cases}
D(0) e^{\Lambda t} & \textrm{for FPUT,} \\
D(0) t  & \textrm{for Toda.}
\end{cases}
\end{align}
We say early times because $\{\Delta{\bf q}(t), \Delta{\bf p}(t)\}$ cannot grow  forever since $\{{\bf q}(t),{\bf p}(t)\}$ are constrained to be on the constant energy surface and are themselves bounded. 
The initial width $D(0)$ should be proportional to $\gamma$, which characterizes the width of our  initial phase-space distribution. Thus we have, taking $\log$ on both sides:

\begin{align}
\log(D(t)) \sim
\begin{cases}
 \Lambda t + \log(\gamma)& \textrm{for FPUT,} \\
 \log(t) + \log(\gamma)  & \textrm{for Toda.}
\end{cases}
\end{align}
Thus $D(t)$ is bounded - we expect it to attain the maximum value after thermalization has been reached.
Therefore, if we plot $\tau_{\rm eq}$ as a function of $\gamma$ for the $\alpha-$FPUT system then we should expect a straight line on a log-linear plot, which is what is observed in Fig.~\ref{Fchaos}a. As reported earlier, the magnitude of the slope of the line is close to the inverse of the Lyapunov exponent. Thus, we have made a strong point for our claim regarding the relation between the equilibration of local observables and the sensitive dependence of the system on initial conditions. Similarly, if we plot the $\tau_{\rm eq}$ as a function of $\gamma$ for the Toda system then we should expect a straight line on a log-log plot of slope $-1$, which is again, what is observed in Fig.~\ref{Fchaos}b. 
Thus, we have explained the results of Fig.~\ref{Fchaos}.


The system is translationally invariant. So, the results are identical if we initially distribute the energy to a different set of four successive particles, while maintaining the order of the initial  distribution. If the energies are distributed in a different permutation  then we find that while  the precise equilibration times are different, the dependence of $\tau_{\rm eq}$ on $\epsilon$ and $\gamma$ are still the same.

For the $\alpha-$FPUT system, as mentioned in the previous section, we found that $\tau_{\rm eq} \propto \log(\gamma)$. The inverse of the proportionality constant $3.3\times 10^{-4}$, is close to the Lyapunov exponent $\Lambda\approx4.1\times 10^{-4}$ of the system. We believe that the properties of the initial conditions, such as its symmetries and its vicinity from breather solutions can affect the exponential growth of perturbations of the initial conditions, which would lead to a retardation of thermalization. This could explain why we did not get a closer agreement. Far away from breathers we can expect there is no such effect. This needs to be investigated further and is beyond the scope of this work. Nevertheless, this method gives us a way to link chaos and thermalization in the $\alpha-$FPUT system, and we expect this method to also work for other chaotic systems. 

\section{Summary}
We studied the time scale of thermalization of local variables in the $\alpha$-FPUT chain and its two  limiting integrable versions, namely the harmonic chain and the Toda chain. Considering systems with $N=32$ particles and total energy $E=31$, we estimated the thermalization time $\tau_{\rm eq}$  by measuring $\langle z_i \partial H/\partial z_i \rangle$ ($z_i$ indicating phase space coordinates) and finding the time to attain equipartition. The averaging is done over initial conditions chosen from a distribution whose width is characterized by the parameter $0 \leq \gamma \leq 1$, with $\gamma=1$ corresponding to the broadest distribution and $\gamma=0$ corresponding to a fixed initial condition. The initial distribution is taken to be one where energy is localized initially in  real space instead of normal mode space.   
The system is described by a single dimensionless parameter $\epsilon=\alpha(E/N)^{1/2} $ characterizing the effective nonlinearity.  Some of our main findings are as follows:
\begin{itemize}

\item[--]For the $\alpha$-FPUT chain we find  $\tau_{\rm eq} \propto {1}/{\epsilon^{a}}$ with $a$ between 4 and 6, and $\gamma$ dependent, in contrast to normal mode equilibration times  \cite{Onorato2015}, where one finds $a\approx8$.

\item[--] We find that local variables equilibrate at much shorter time scales than normal modes for both normal mode localized excitations (NMLE) and phase space localized excitations (SLE).

\item[--] For the $\alpha$-FPUT chain we observe the time scales for equilibration to be completely different when we perform ensemble averaging and time averaging protocols.

\item[--]The thermalization time depends on the initial ensemble and we find $\tau_{\rm eq} \propto \ln(\gamma)$, with the proportionality constant being close to the inverse of the maximal Lyapunov exponent of the system, thus quantifying the relation between thermalization and chaos for the $\alpha$-FPUT system.

\item[--] Surprisingly, we find that the Toda chain equilibrated on very long time scales if the width of the initial distribution is broad enough. In fact we obtain $\tau_{\rm eq} \sim 1/\gamma$. On the other hand, the harmonic chain never equilibrates.

\item[--] We provide a simple geometric understanding of  these results -- the equilibration time is simply related to the time it takes for an ensemble of initial conditions in the $2N$ dimensional phase space to spread over the microcanonical energy surface. For the FPUT chain, for energies such that the system is chaotic with a positive Lyapunov exponent, a fast exponential (in time) spreading occurs. 
For the Toda chain the growth is linear and so thermalization takes more time.  We provide  numerical evidence to support  this picture. 
\end{itemize}

\chapter{The Taylor$-$von Neumann$-$Sedov Blast Wave Solution}
\label{chap:tvns1}
Up until now, we have studied what causes a system to thermalize and we explored the roles of initial conditions, averaging, choice of observables and chaos in thermalization. Now we move on to a different problem. 
For a one dimensional system, instead of describing the evolution by solving $2N$ equations, can we somehow use significantly less number of equations to describe the evolution? As usual, we are only looking at a coarse-grained description of the system when we are trying to describe the evolution. We therefore enter the realm of hydrodynamics, which describes the evolution of conserved fields, which are the slow degrees of freedom. We say slow because these fields evolve slower than the evolution of individual particles. 
So, we are in the non-equilibrium regime where, we study the evolution of conserved quantities by using the Euler equations. 

In this chapter, we study the blast problem. 
In Sec.~\ref{sec:TvNSd} we introduce the blast problem and the Euler equations in $d$ dimensions. Now solving Euler equations, which are PDEs is not always be easy. For the blast problem, where the energy is initially given only to a small region, the Euler equations however admit solutions that are self-similar (in time) up to the times the perturbation reaches the boundary. In this case the Euler equations can be converted into ODEs (the scaling variable would then be the only independent variable) and we can get an exact solution. This was first noticed in three dimensions by Taylor \cite{Taylor19501,Taylor19502}, von Neumann \cite{VonNeumann1963} and Sedov \cite{Sedov1946, Sedov2014} (TvNS), who studied the self-similar solutions in the context of atomic explosions. For our purposes, we derive the exact scaling solution in one dimension in Sec.~\ref{sec:TvNS1}, which will be valid upto the times the shock reaches the boundary of the system. 
We then cite some of the literature which tried to find an agreement between hydrodynamics and molecular dynamics simulations of a macroscopic Hamiltonian system in Sec.~\ref{sec:mdtvns}. Some technical details like derivation of the Rankine-Hugoniot conditions and exploring what are the various scaling regimes of the Euler equations are presented in the appendices.

\section{Euler Equations and its Self-similar Solutions}
\label{sec:TvNSd}
Consider a system of interacting particles in which the only known conserved quantities  are the total number of particles, the total momentum and the total energy of the system. We begin by discussing the general case of propagation of a blast wave in a $d$ dimensional system for which, because of  the radial symmetry of the problem there are three Euler equations for the hydrodynamic fields which depend only on the radial coordinate $r$. We therefore expect a hydrodynamic description in terms of the corresponding conserved fields namely, the mass density field $\rho(r,t)$, the momentum density field $\rho(r,t)v(r,t)$ ($v$ is the velocity field) and the energy density field $E(r,t) = \rho(r,t)e(r,t) $.
 
The blast is caused by the instantaneous release of energy $E$ in a very small region that creates a spherical shock wave propagating through the quiescent gas. The blast is infinitely strong if the pressure behind the shock wave can be neglected. In normal conditions, this is valid up to a certain time; for the gas at zero temperature, it remains valid forever. Below we consider the infinitely strong blast. Dimensional analysis \cite{BarenblattBook} alone gives the position of the shock wave $R=R(t)$ in terms of time $t$, the released energy $E$, and the background density $\rho_\infty$:
 \begin{equation}
 \label{R-d}
 R(t)= \left(\frac{Et^2}{A \rho_\infty}\right)^\frac{1}{d+2}.
 \end{equation}
 The dimensionless amplitude $A$ must be determined somehow, but the dependence of the position of the shock front $R(t)$ on the basic parameters automatically follows from dimensional considerations. Interestingly, whether the shock propagation in the system is diffusive, sub-diffusive or super-diffusive depends on the dimension $d$ the system is set up in, as can be seen from the above equation.

 The density is uniform and equals to $\rho_\infty$ everywhere in front of the shock wave, i.e., for $r>R(t)$. The velocity of the shock wave is
 \begin{equation}
 \label{U-d}
 U = \frac{dR}{dt} = \delta\,\frac{R}{t}, \qquad \delta\equiv \frac{2}{d+2}\,.
 \end{equation}
Behind the shock wave $0\leq r<R(t)$,  the radial velocity $v(r,t)$, density $\rho(r,t)$ and pressure $p(r,t)$ satisfy: 
 \begin{subequations}
 	\label{Euler-eqs}
 	\begin{align}
 	\label{cont-eq}
 	&\partial_t \rho + \partial_r (\rho v) +\frac{d-1}{r}\,\rho v  = 0,\\
 	\label{E-eq}
 	&(\partial_t + v \partial_r) v +\frac{1}{\rho}\, \partial_r p = 0, \\
 	\label{entropy-eq}
 	&(\partial_t + v \partial_r)\ln\frac{p}{\rho^\gamma} =0,
 	\end{align}
 \end{subequations}
 where $\gamma$ is the adiabatic index. These equations are the Euler equations for conserved fields and we used the entropy form for the ideal gas. Instead of energy we have used pressure, which is related to energy by the equation of state. 
 The Rankine-Hugoniot conditions  \cite{LandauBook}, describing the jump between the states on both sides of the shock wave are given by: 
 \begin{equation}
 \label{RH}
 \frac{\rho(R)}{\rho_\infty}=  \frac{\gamma+1}{\gamma-1}\,, ~~ 
 \frac{v(R)}{U}=\frac{2}{\gamma+1}\,,~~
 \frac{p(R)}{\rho_\infty U^2} = \frac{2}{\gamma+1}
 \end{equation}
 in the case of the infinitely strong blast. 
 The pressure $p_\infty$ in front of the shock wave can be neglected as long as $p(R)\gg p_\infty$. Using Eqs.~\eqref{R-d}--\eqref{U-d} and \eqref{RH} one finds that this is valid in the time range  
 \begin{equation}
 t\ll t_*, \quad  t_* \sim \left(\frac{E}{p_\infty}\right)^\frac{1}{d}\sqrt{\frac{\rho_\infty}{p_\infty}}.
 \end{equation}
 In the case when the particles surrounding the blast are initially at rest, $p_\infty=0$, so $t_* = \infty$ and the blast forever remains infinitely strong. Now we convert the PDEs given by Eqs.~\eqref{Euler-eqs} to ODEs.
 
 Instead of pressure, we can use the temperature field given by (for the ideal gas) $T= \mu p/\rho$. Dimensional analysis ensures that the hydrodynamic variables acquire a self-similar form 
 \begin{equation}
 \label{scaling}
 \rho=\rho_\infty G(\xi),\quad v=\delta\,\frac{r}{t}\,V(\xi),  \quad T=\frac{\mu \delta^2}{\gamma}\,\frac{r^2}{t^2}\,Z(\xi).
 \end{equation}
The factors $\delta,\gamma$ are inserted for convenience; e.g., from Eq.~\eqref{U-d} we see that the velocity of the shock wave is $\delta {R}/{t}$ and this suggests the use of the factor $\delta$ for $v$. The fields now depend on the single dimensionless  variable
\begin{equation}
\label{xi-def}
\xi = \frac{r}{R}.
\end{equation}
One seeks to find the scaling functions $G(\xi), V(\xi)$ and $Z(\xi)$ behind the shock wave, $0\leq \xi\leq 1$. The Rankine-Hugoniot conditions Eq.~\eqref{RH} (also derived in Appendix~\ref{sec:RHappendix}) become:

 \begin{subequations}
 	\label{RH-gamma}
 	\begin{align}
 	\label{RH-G}
 	& G(1)= \frac{\gamma+1}{\gamma-1},\\
 	\label{RH-V}
 	& V(1)=\frac{2}{\gamma+1},\\
 	\label{RH-Z}
 	& Z(1)=\frac{2\gamma(\gamma-1)}{(\gamma+1)^2}.
 	\end{align}
 \end{subequations}
 The conservation of energy allows to express $Z$ through the scaled velocity $V$:
 \begin{equation}
 \label{ZV}
 Z = \frac{\gamma(\gamma-1)(1-V)V^2}{2(\gamma V -1)}.
 \end{equation}
 This integral of motion is usually established \cite{LandauBook} for $d=3$, but the same derivation works in arbitrary dimension and yields the universal result Eq.~\eqref{ZV}. 
 
 Plugging the ansatz Eqs.~\eqref{scaling}--\eqref{xi-def} into Eq.~\eqref{cont-eq} we obtain
 \begin{equation}
 \label{VG-eq}
 \frac{dV}{d\ell}+(V-1)\,\frac{d\ln G}{d\ell}    = - dV,  
 \end{equation}
 where $\ell=\ln\xi$. Similarly we transform Eq.~\eqref{entropy-eq} into
 \begin{equation}
 \label{ZG-eq}
 \frac{d\ln Z}{d\ell} - (\gamma-1)\,\frac{d\ln G}{d\ell} = \frac{d+2-2V}{V-1}.
 \end{equation} 
 Equations \eqref{VG-eq}--\eqref{ZG-eq} are solvable for arbitrary $d$ and $\gamma>1$. One must solve these equations even if one merely wants to determine the amplitude $A=A(d,\gamma)$ in Eq.~\eqref{R-d}. The energy conservation gives
 \begin{eqnarray}
 \label{energy}
 E  &=& \int_0^R dr\,\,\Omega_d\, r^{d-1}\,\rho\left[\frac{v^2}{2}+\frac{c^2}{\gamma(\gamma-1)}\right] \nonumber \\
 &=& \rho_\infty\,\Omega_d\, \delta^2\,\frac{R^{d+2}}{t^2} \int_0^1 d\xi\,\xi^{d+1}\,\frac{(\gamma-1)V^3}{2(\gamma V -1)}\,G,
 \end{eqnarray}
 where $c=\sqrt{\gamma p/\rho}$ is the sound speed, $\Omega_d$ is the surface area of the unit sphere and we have used Eq.~\eqref{ZV}. Combining Eq.~\eqref{energy} with Eq.~\eqref{R-d} we obtain
 \begin{equation}
 \label{A-int}
 A =\Omega_d\, \delta^2\,\frac{\gamma-1}{2} \int_0^1 d\xi\,\xi^{d+1}\,\frac{V^3}{\gamma V -1}\,G.
 \end{equation}
 \section{The TvNS Blast Wave Solution in One Dimension}
 \label{sec:TvNS1}
In the classical literature, the blast problem is studied in three dimensions; the two-dimensional solutions are mentioned in \cite{Sedov2014}. Here we present the derivation of the solution in $d=1$. The problem is solvable for an arbitrary adiabatic index $\gamma>1$, but we use $\gamma=3$ following the general prediction, $\gamma=1+2/d$, of kinetic theory for monoatomic gases \cite{Resibois}. The Rankine-Hugoniot conditions Eqs.~\eqref{RH-gamma} become
 \begin{equation}
 \label{BC}
 G(1) = 2, \quad V(1) = \tfrac{1}{2}, \quad  Z(1) = \tfrac{3}{4}.
 \end{equation}
 We insert Eq.~\eqref{ZV} into  Eq.~\eqref{ZG-eq} and find
 \begin{equation}
 \label{ZG-eq-d}
 \left[\frac{2}{V}+\frac{1}{V-1}-\frac{3}{3V -1}\right]\frac{dV}{d\ell} 
 ={2}\,\frac{d\ln G}{d\ell} + \frac{3-2V}{V-1}.
 \end{equation}
 Using Eqs.~\eqref{VG-eq} and \eqref{ZG-eq-d}, we express the derivatives of $V$ and $\ln G$ through $V$:
 \begin{subequations}
 	\begin{align}
 	\label{V-diff}
 	&  2\,\frac{dV}{d\ell}  = - V\,\frac{3-13 V + 12 V^2}{1-4V+6 V^2},       \\
 	\label{G-diff}
 	& 2\,\frac{d\ln G}{d\ell} = V\,\frac{5V - 1}{1-5V+10 V^2 - 6 V^3}.
 	\end{align}
 \end{subequations}
 Dividing Eq.~\eqref{G-diff} by Eq.~\eqref{V-diff} yields
 \begin{equation}
 \frac{d\ln G}{d V} = \frac{5V-1}{(3V-1)(3-4V)(1-V)}
 \end{equation}
 which is integrated to give
 \begin{subequations} 
 	\label{GVZ-xi:1}
 	\begin{equation}
 	\label{GV:1}
 	G = 2^\frac{16}{5}\, (1-V)^2\, (3V-1)^\frac{1}{5}\, (3-4V)^{-\frac{11}{5}}.
 	\end{equation}
 	The amplitude is fixed by the boundary conditions Eq.~\eqref{BC}. Similarly integrating Eq.~\eqref{V-diff} one gets
 	\begin{equation}
 	\label{xi-V:1}
 	\xi^5 = 2^{-\frac{4}{3}}\,(3V-1)^2\, V^{-\frac{10}{3}}\,(3-4V)^{-\frac{11}{3}}
 	\end{equation}
 	that implicitly determines $V=V(\xi)$. Finally $Z$ is given by Eq.~\eqref{ZV} which when $\gamma=3$ becomes 
 	\begin{equation}
 	\label{ZV:1}
 	Z = \frac{3(1-V)V^2}{3V - 1}.
 	\end{equation}
 \end{subequations}
 Equations \eqref{GVZ-xi:1} constitute the exact solution. Now all that is remaining is to find the value of the dimensionless parameter $A$ in Eq.~\eqref{R-d}. When $d=1$, we have $\Omega_1=2, ~\delta=2/3$ and $\gamma=3$. Therefore Eq.~\eqref{A-int} reduces to 
 \begin{equation}
 \label{A:int-1}
 A = \left(\frac{2}{3}\right)^2\int_0^1 d\xi\,\xi^2\,\frac{2V^3}{3V-1}\,G.
 \end{equation}
 Using Eqs.~\eqref{GV:1}--\eqref{xi-V:1} one can reduce the integral in Eq.~\eqref{A:int-1} to a rather complicated integral over $V$ which is computed to give
 \begin{equation}
 \label{A-1}
 A=\frac{152}{1071}.
 \end{equation}
 Our basic analytical prediction about the position of the shock wave becomes 
 \begin{equation}
 \label{R-1}
 R(t)= \left(\frac{1071}{152}\, \frac{Et^2}{\rho_\infty}\right)^\frac{1}{3}.
 \end{equation}
Equations \eqref{GVZ-xi:1} and \eqref{R-1} provide the complete TvNS solution in one dimension. In the next chapter we compare this solution with results from direct simulations of our microscopic model. 

\section{Earlier Attempts to Observe TvNS Scaling in Microscopic Models}
\label{sec:mdtvns}
The  scaling law for $R(t)$ (Eq.~\eqref{R-d}) is obeyed in nuclear explosions \cite{Taylor19501,Taylor19502}. This growth law and its generalization arise in numerous applications, e.g. in astrophysical blast waves \cite{McKee1988,McKee1995,Waxman10,Kushnir14,Wheeler2017,Sari21}, laser-driven blast waves in gas jets \cite{Ditmire2001}, plasma \cite{Smith2005}, granular materials \cite{Kellay2009,Kellay2013} and relativistic blast problems \cite{blandford1976,best2000,sari2006,tian2018,Coughlin19}. It is interesting to compare the scaling law  with the microscopic simulations of the underlying many-particle system. Such studies have been attempted only recently, which can be partly
attributed to the recent increase in computational performance. An earlier analysis \cite{Antal2008} of the blast in a homogeneous elastic hard-sphere gas in one and two dimensions veriﬁed the growth law Eq.~\eqref{R-d}. The authors also studied the dependence of the number of moving particles at time $t$ and the number of collisions up to time $t$. These growth laws satisfied the predictions of the  hydrodynamic theory for the shock wave emanating from an explosion. More detailed studies of the scaling functions, defined in Eqs.~(\ref{R-d}), (\ref{scaling}) through molecular dynamic simulations have recently been performed for the hard-sphere gas in various dimensions \cite{Jabeen2010,Barbier2015,Barbier2015a} both with elastic and inelastic collisional dynamics \cite{Barbier2016,Joy2021,Joy2021a,Joy2017}. For elastic systems in two dimensions, a reasonable agreement between the TvNS solution and microscopic simulations has been reported \cite{Barbier2016}  when the ambient density is not too large. More extensive simulations of the hard-sphere gas in three dimensions~\cite{Joy2021} and two dimensions~\cite{Joy2021a} however found significant departure of the simulation results from the TvNS solution both near the center of the explosion and at the shock front. The agreement between molecular dynamics simulations of a Hamiltonian system and hydrodynamics still remains an open question. While the scaling form seems robust, the scaling functions $G, V$ and $Z$ appear to be different from the TvNS prediction. Recently, \cite{Joy2021} tried to find a good match between the microscopic simulations and the hydrodynamic predictions by replacing the ideal gas equation of state with the equation of state that included up to tenth order virial coefﬁcients. The hydrodynamics would then predict a modified scaling solution with the same scaling exponents. It was then observed that while there is an improvement, they still failed to get a good match. This deviation was attributed to the violation of the assumption of local equilibrium and the neglect of heat conduction, which is not accounted for by the Euler equations. If the later is indeed the reason, then we have to use a different model to describe the hydrodynamic behaviour. One way is to use some additional dissipation terms in the Euler equations in order to capture this effect, which is the theme of our work, to be described in the next chapter.

\begin{subappendices}

\section{Rankine-Hugoniot Conditions}
\label{sec:RHappendix}
In this section we derive the Rankine-Hugoniot conditions. The Rankine–Hugoniot conditions, also referred to as Rankine–Hugoniot jump conditions or Rankine–Hugoniot relations, describe the relationship between the states on both sides of a shock wave. They are needed as boundary conditions of the scaling solutions so that we can construct the scaling solutions of the Euler equations. We start with the Euler equations in $d$ dimensions. Because of the
radial symmetry of the problem, these three equations for the hydrodynamic ﬁelds depend only on the radial coordinate $r$. Thus we have the three one dimensional equations:
	\begin{subequations}
		\label{Euler-eqs1}
		\begin{align}
		\label{cont-eq1}
		&\partial_t \rho + \partial_r (\rho v) +\frac{d-1}{r}\,\rho v  = 0,\\
		\label{E-eq1}
		&(\partial_t + v \partial_r) v +\frac{1}{\rho}\, \partial_r p = 0, \\
		\label{entropy-eq1}
		&(\partial_t + v \partial_r)\ln\frac{p}{\rho^\gamma} =0,
		\end{align}
	\end{subequations}
	where $r$ is the radial coordinate. Now consider Eq.~\eqref{cont-eq1}. Let the solution exhibit a shock at $r = R(t)$. We now integrate this equation over a small region, $r \in (r_1,r_2) $. Using the Leibnitz rule, the first term becomes:
	$$ \int_{r_1}^{r_2} \partial_t \rho\ dr= \partial_t \int_{r_1}^{r_2} \rho\ dr+ \rho(r_1,t)\dot{r}_1 - \rho(r_2,t)\dot{r}_2~.$$
	Then Eq.~\eqref{cont-eq1} can be written as:
	\begin{equation}\label{eqn:fluxLeib}
	\partial_t \int_{r_1}^{r_2} \rho\ dr+ \rho(r_1,t)\dot{r}_1 - \rho(r_2,t)\dot{r}_2 + \rho(r_2,t)v(r_2,t) - \rho(r_1,t)v(r_1,t) + \int_{r_1}^{r_2} dr\ \frac{d-1}{r}\rho v = 0.
	\end{equation}
	We now go to the limit where $r_1$ is just before the shock front and $r_2$ is just after the shock front. We take $r_1 = R(t) - \epsilon$ and $r_2 = R(t) + \epsilon$, with $\epsilon \rightarrow 0$. Then $\dot{r}_1 = \dot{r}_2 = \dot{R}(t) $, the velocity of the shock front. Also, the first term and the last term in Eq.~\eqref{eqn:fluxLeib} vanish because $r_1$ and $r_2$ are getting close to each other in this limit. Considering that the shock has not reached $r=r_2$ at time $t$ yet, we have $v(r_2,t)=0$. This leads to the following equation:
	\begin{equation}	\label{eq:RHrho}
	\frac{\rho(R)v(R)}{\rho(R)-\rho_\infty} = \dot{R}.
	\end{equation}
	Similarly, we get the following equations by integrating Eq.~\eqref{E-eq1} and Eq.~\eqref{entropy-eq1}:
	\begin{align}
	&\frac{ \rho(R) v^2(R) + P(R)}{\rho(R) v(R)}=\dot{R}, \label{eq:RHv} \\
	&\frac{ \rho(R) v(R) e(R) + P(R) v(R)}{\rho(R) e(R)}=\dot{R}\label{eq:RHe}.
	\end{align}
	Now we use the forms of the scaling solutions as given in Eq.~\eqref{scaling}:
	$$\rho=\rho_\infty G(\xi),\quad v=\delta\,\frac{r}{t}\,V(\xi),  \quad T=\frac{\mu \delta^2}{\gamma}\,\frac{r^2}{t^2}\,Z(\xi).$$
	The position of the shock front corresponds to $\xi = 1$. Also, the velocity of the shock front is related to its position by $\dot{R} = \delta R/t$, as given by Eqs.~\eqref{R-d}, \eqref{U-d}. Substituting these forms at $\xi = 1$ in Eqs.~\eqref{eq:RHrho}, \eqref{eq:RHv}, \eqref{eq:RHe} we get the Rankine-Hugoniot conditions. They are given in terms of $G(1), V(1)$ and $Z(1)$ by: 	
	\begin{subequations}
		\label{RHderiv}
		\begin{align}
		\frac{GV}{G-1} &= 1, \\
		\frac{V^2 + Z/\gamma}{V} &= 1, \\
		\frac{V^3/2 + ZV/(\gamma - 1)}{V^2/2 + Z/(\gamma(\gamma - 1))} &= 1.
		\end{align}
	\end{subequations}
	Solving these three equations would finally lead us to the Rankine-Hugoniot conditions:	
	\begin{subequations}
		\begin{align}
		G(1) = \frac{\gamma + 1}{\gamma - 1}, \\
		V(1) = \frac{2}{\gamma + 1}, \\
		Z(1) = \frac{2\gamma(\gamma-1)}{(\gamma + 1)^2}.
		\end{align}
	\end{subequations} 
	These boundary conditions can then be used to fix the integration constants and we thus derive the TvNS solution.
	
\section{Other Scalings of the Euler Equations} 
\label{sec:scalings}
In this chapter we have derived the complete scaling solution of the Euler equations in one dimension. In this section, we emphasize the role played by the blast wave initial conditions and re-derive the form of the TvNS scaling in one dimension using a different method. We also explore if there are other scaling solutions that are possible in the Euler equations. We start with the Euler equations in one dimension:
	\begin{subequations}
		\label{Eul}
		\begin{align}
		&\partial_t \rho + \partial_x (\rho v) = 0, \label{Eula} \\
		&\partial_t (\rho v) + \partial_x (\rho v^2 +p) = 0, \label{Eulb} \\
		&\partial_t (\rho e) + \partial_x (\rho ev   + p v)= 0,
		\label{Eulc}
		\end{align}
	\end{subequations}
	where $p(x,t)$ is the pressure and we need the equation of state that relates pressure to the conserved fields. We consider one dimensional systems that can be described by ideal gas thermodynamics: 
	\begin{equation}
	\begin{split}
	\label{therm2}
	T(x,t)&= 2 \mu\, \epsilon(x,t),   \\
	p(x,t)&= 2 \rho(x,t) \epsilon(x,t)=  \mu^{-1} \rho(x,t)\, T(x,t),  
	\end{split}
	\end{equation}
	where $\epsilon= e-v^2/2$ is the internal energy per unit mass and $\mu=(m_1+m_2)/2$. Using these the Euler equations now become:
	\begin{subequations}
		\label{Eul2}
		\begin{align}
		&\partial_t \rho + \partial_x (\rho v) = 0, \label{Eul2a} \\
		&\partial_t (\rho v) + \partial_x (2\rho e) = 0, \label{Eul2b} \\
		&\partial_t (\rho e) + \partial_x (3\rho ev -\rho v^3)= 0.
		\label{Eul2c}
		\end{align}
	\end{subequations}
Let us assume the solution to the above equations be of the form:
	\begin{subequations}
		~\label{eq:scaling}
		\begin{align}
		\rho(x,t) &= \rho_\infty+ t^b\tilde{\rho}(xt^a),~ \label{eq:rhoscale}\\
		v(x,t) &= t^c\tilde{v}(xt^a),~\label{eq:uscale}\\ 
		e(x,t) &= t^d\tilde{e}(xt^a).~\label{eq:escale} 
		\end{align}
	\end{subequations}
Here $\rho_\infty$ is the background density into which the blast is propagating. $\tilde{\rho}, \tilde{v}$ and $\tilde{e}$ are the scaling functions of the conserved fields (which are related to $G,\ V$ and $Z$ dervied in this chapter). We aim to find the values of $a,b,c$ and $d$ here. 
	Now substituting these scaling forms into Eqs.~\eqref{Eul2a}-\eqref{Eul2c}, and defining the scaling variable $z=x t^a$, we arrive at the following equations:

	\begin{align}
	&b\tilde{\rho} + a z \tilde{\rho}^\prime + t^{a+c+1}(\tilde{\rho}\tilde{v})^\prime+\rho_\infty t^{a-b+c+1}\tilde{v}^\prime= 0,~ \label{rhoscaletilde}\\
	&(b+c)\tilde{\rho}\tilde{v} + a z (\tilde{\rho}\tilde{v})^\prime +\rho_\infty ct^{-b}\tilde{v}+
	\rho_\infty a z t^{-b}\tilde{v}^\prime \nn \\ &~~~~~~~~+ 2 t^{a-c+d+1}( \tilde{\rho}\tilde{e})^\prime   
	+2\rho_\infty t^{a+d-c-b+1}\tilde{e}^\prime=0,~\label{uscaletilde}  \\
	&(b+d)\tilde{\rho}\tilde{e} + a z ( \tilde{\rho}\tilde{e})^\prime +\rho_\infty dt^{-b}\tilde{e}+ \rho_\infty a z t^{-b}\tilde{e}^\prime  + 3t^{a+c+1}(\tilde{\rho}\tilde{e}\tilde{v})^\prime\nn \\&~~~~~~~~+ 3\rho_\infty t^{a+c-b+1}(\tilde{e}\tilde{v})^\prime  -t^{a+3c-d+1} (\tilde{\rho}\tilde{v}^3)^\prime -\rho_\infty t^{a+3c-b-d+1}(\tilde{v}^3)^\prime= 0.~\label{escaletilde} 
	\end{align}
	Here prime denotes derivative wrt $z$. In order to get the scaling solution all we need is that there should not be $x$ and $t$ anywhere in the equations except in the form $z=x t^a$. Requiring now that there exists a scaling solution, we set all powers of the time variable to zero, which then gives us the conditions $b=0$, $a+c+1=0$ and $d=2c$. We still need one more condition to determine the exponents completely and, as  first noticed by Taylor~\cite{Taylor19501,Taylor19502}, the energy conservation condition turns out to be sufficient. The constancy of energy means that the integral
	\begin{equation}
	\int_{-\infty}^{\infty} dx \rho(x,t)e(x,t)  = E_0,
	\end{equation}
	should be a constant, independent of time. Plugging in our scaling forms for $\rho$ and $e$, and making a change of variables, the above equation gives 
	\begin{equation}
	t^{d-a}\int_{-\infty}^{\infty} dz \tilde{\rho}(z) \tilde{e}(z)  = E_0.
	\end{equation}
	Demanding time-independence then requires $d=a$ and together with the earlier equations we finally get:
	\begin{equation}
	a=-2/3,~b=0,c=-1/3,~ d=-2/3.~ \label{c4}
	\end{equation}
	This is the TvNS scaling in one dimension. It is easy to see that with this choice of the exponents, Eqs.~\eqref{eq:scaling} reduce to Eqs.~\eqref{scaling}. For instance, $\widetilde{\rho}=\rho_\infty G-\rho_\infty$ and $\widetilde{v}=t^{1/3} v(x,t) =(2 \alpha/3) \xi V$, where  $\alpha = [E/(A \rho_\infty)]^{1/3}$.

	Note that we got the TvNS scaling for a blast wave initial condition - energy is initially excited in a small region and there is no energy in the ambient medium. If however, there is energy in the surrounding medium initially then we get a  different scaling. For example, the following scaling solution leads to a ballistic scaling and we get $a = -1, b= 0, c = 0$ and $d = 0$:
	\begin{subequations}
		~\label{eq:ballisticscaling}
		\begin{align}
		\rho(x,t) &= \rho_\infty+ t^b\tilde{\rho}(xt^a),~ \label{eq:rhoscaleb}\\
		v(x,t) &= t^c\tilde{v}(xt^a),~\label{eq:uscaleb}\\ 
		e(x,t) &= e_0 + t^d\tilde{e}(xt^a).~\label{eq:escaleb} 
		\end{align}
	\end{subequations}
	The above scaling will be observed if we are interested in studying the propagation of a step discontinuity in the system (Riemann problem), which results in a ballistic scaling.

\end{subappendices}

\chapter{Blast Wave in the One Dimensional Cold Gas}
\label{chap:tvns2}
In the previous chapter we derived the TvNS solution in one dimension, which is the scaling solution of the Euler equations for a blast wave initial condition. 
	In this chapter we try to show an agreement between the TvNS scaling and microscopic simulations. We consider the well-studied alternate mass hard particle (AHP) gas system \cite{Antal2008, Dhar2001, Grassberger2002, Casati2003, Cipriani2005, Chen2014, Hurtado2016, lepri2020}. We describe the relation between hydrodynamic fields and the microscopic variables of this system in Sec.~\ref{sec:model}. We then go on to brief some of the earlier works which studied various aspects of the AHP gas in Sec.~\ref{sec:lit}. We study the blast problem in the AHP gas system. We excite the system to a ``blast wave" initial condition, which will be described in Sec.~\ref{sec:ic}. This will result in the propagation of a shock wave in the system. We first compare the results of microscopic simulations with the TvNS solution in Sec.~\ref{sec:micro}. We find a very good agreement between them everywhere except in a small region near the core of the blast. We attribute this deviation near the core to the non-negligible contribution of heat conduction, which is not captured by the Euler equations. So, we model our system by adding viscosity and the heat conductivity terms to the Euler equations, which then become the Navier-Stokes-Fourier (NSF) equations \cite{Zeytounian2012}, which are described in Sec.~\ref{sec:hydro}. Now we observe a very good agreement between the scaling solutions of the microscopic simulations and the NSF equations. We thus propose that the NSF equations capture the hydrodynamics of the one dimensional AHP gas more accurately than the Euler equations. Then in Sec.~\ref{sec:core} we analyse the role of the conductivity term in more detail and study its role in the evolution of the conserved quantities in the system near the core of the blast.

	\section{The Alternate Mass Hard Particle Gas}
	\label{sec:model}
	As mentioned in chapter \ref{chap:intro}, the reason for considering alternate masses is that we need some sort of integrability breaking. Having equal masses would essentially mean a system of non-interacting particles, which is integrable and leads to a macroscopic number of conservation quantities. Such systems are described by generalized hydrodynamics. We are however, interested in describing the hydrodynamics of a system that only has a few conserved quantities. We choose odd-numbered particles to have mass $m_1$, so even-numbered particles have mass $m_2$ ($\neq m_1$). In this case, the usual hydrodynamics is sufficient. The only known conserved quantities of this system are the total number of particles, the total momentum and the total energy, and we expect a hydrodynamic description in terms of the corresponding conserved fields. 
	The hydrodynamic fields $(\rho,v,E)$, are related to the microscopic variables:
	\begin{align}
	\begin{split}
	\label{rho-def1}
	&\rho(x,t)=\sum_j  \left\langle m_j \delta(q_j(t) - x)\right\rangle,\\
	&\rho(x,t) v(x,t)=\sum_j  \left\langle m_j v_j \delta(q_j(t) - x)\right\rangle, \\
	&E(x,t)=\rho(x,t)e(x,t)= \tfrac{1}{2} \sum_j  \left\langle m_j v_j^2 \delta(q_j(t) - x)\right\rangle,
	\end{split}
	\end{align}
	where $\la ...\ra$ indicates an average over an initial distribution of micro-states that correspond to the same initial macro-state.
	
	We study the blast problem in this system. We consider a gas with density $\rho_\infty$ and particles initially at rest (zero pressure and temperature). The pressure in front of the shock wave remains zero throughout the evolution if it is zero initially. 
	We suddenly inject, at time $t=0$, energy $E$ into a localized region. The details of injection are not important, the explosion quickly becomes radially symmetric and keeps \cite{Ryu1987,Sanz2016} its spherical shape, so the hydrodynamic functions depend on the  distance $r$ from the center of the explosion and time $t$. Since the temperature is zero in the unperturbed region, local equilibrium and hydrodynamic behaviors may emerge only sufficiently far behind the shock, namely in the region where particles have already undergone a few collisions.  Hence the blast problem with particles initially at rest is an excellent laboratory to study deviations from hydrodynamics. Also, at any time only a finite number of particles move thereby allowing us to simulate an infinite-particle system.

	\section{Earlier Studies done on the AHP gas}
	\label{sec:lit}
	Here we discuss some works which studied various properties of the alternate mass hard particle (AHP) gas. The AHP gas has also been extensively studied in the context of the breakdown of Fourier's law of heat conduction in one dimension \cite{Dhar2001,Grassberger2002,Casati2003,Cipriani2005,Chen2014,Hurtado2016,lepri2020,Garrido2001,Zhao2018}. Remarkably, the thermal conductivity $\kappa$ diverges with system size $L$ as $\kappa \sim L^{1/3}$. Some theoretical understanding of this has been achieved using nonlinear fluctuating hydrodynamics \cite{Narayan2002,Van2012,Mendl2013,Spohn2014}, which makes very detailed predictions about the form of equilibrium dynamical correlation functions. These predictions have been verified for two microscopic models, the AHP gas~\cite{Mendl2013,Mendl2014} and the FPUT chain \cite{Das2014}. As far as the evolution of macroscopic profiles is considered, the AHP gas was used recently \cite{Mendl2017} to study the so-called Riemann problem \cite{RiemannBook}, where one considers an infinite system with domain wall initial conditions. A comparison was made between the simulations and the predictions from the Euler equations. Note that the blast-wave initial conditions lead to a different class of self-similar solutions than those that one gets for the Riemann problem [ballistic scaling with $R(t) \sim t$]. We have already seen this in Appendix~\eqref{sec:scalings}, where we derived the ballistic scaling for the hydrodynamic equations for a different set of initial conditions. The one-dimensional hard-particle gas has also been used to probe the breakdown of the hydrodynamic description \cite{Kadanoff1995,Hurtado2006}. It was observed \cite{Hurtado2006} that the field profiles in front of a rapidly moving piston differ from hydrodynamic predictions. The splash problem in the 1D AHP gas has also been studied recently \cite{Chakraborti2022}.
	
\section{``Blast Wave" Initial Conditions for Microscopic Simulations}

\label{sec:ic}
We consider an initial macro-state where the gas has a  finite uniform density $\rho_\infty$, zero flow velocity $v$, and is at zero temperature everywhere except in a region of width $\sigma$ centered at $x=0$. This is the region of the blast and has a specified finite temperature profile. In numerical simulations, two different profiles for the initial energy have been used:
\begin{align}
&{\rm (i)~Gaussian~profile}~~E(x,0) = \frac{E }{\sqrt{2\pi\sigma^2}}e^{-x^2/{2\sigma^2}}, \\
&{\rm (ii)~Box~profile}~~E(x,0)=
\begin{cases}\frac{E }{2 \sigma},~-\sigma < x < \sigma,\\
0,~~|x| > \sigma.    
\end{cases}
~\label{eq:ezero}
\end{align} 
In both cases, the total energy of the blast is $E$. We now describe how one can  realize these macro-states in the microscopic simulations of the AHP gas. Different choices of the initial ensemble of micro-states can lead to the same average macroscopic profile. Here we discuss two possible ensemble choices.

\noindent  {\bf (A) Ensemble where  energy and momentum are fixed on average}: In this case the two initial macro-states can be realized as follows:

\noindent (i)  {\em Gaussian temperature profile:}  We distribute $N$ particles labelled $i=-N/2+1,-N/2+2,\ldots, N/2$ uniformly in the interval $(-L/2,L/2)$. The number density is $n_0=N/L=\rho_\infty/\mu$, where $\rho_\infty$ is the ambient mass density of the gas.  For the $N_c$ particles at the center, with $i=-N_c/2+1, \ldots, N_c/2$, the velocities are chosen from the Maxwell distribution with temperature $T=2 E /N_c$: $\text{Prob}(v_i)= \sqrt{m_i/(2 \pi T)}  e^{-m_i v_i^2/(2 T)}$. The velocities of other $N-N_c$ particles are set to zero. We note that $E(x,0)= \la \sum  m_i v_i(0)^2 \delta(x-q_i(0))/2 \ra $.
For large $N$,  each particle's position is a Gaussian with mean $\bar{q}_i=i/n_0$ and variance $\sigma^2=L/(4 n_0)$. Defining the length scale $s=N_c/n_0$,  one can express
\begin{align}
E(x,0) &= \sum_{i=-N_c/2+1}^{N_c/2} \f{T}{ 2 \sqrt{2 \pi \sigma^2}}  e^{-(x-\bar{q}_i)^2/ (2 \sigma^2)} \nn \\
&\approx \f{n_0 T}{2} \int_{-s/2}^{s/2} dy   \f{e^{-(x-y)^2/ (2 \sigma^2)}}{  \sqrt{2 \pi \sigma^2}} \nn \\
&=\f{n_0 T}{4} \left[ \erf{\left(\f{s-2x}{2 \sqrt{2} \sigma}\right)} +  \erf{\left(\f{s+2x}{2 \sqrt{2} \sigma}\right)}  \right] \label{Eerf}  
\end{align}
via the error function $\erf(z)=\frac{2}{\sqrt{\pi}}\int_0^z du\,e^{-u^2}$. 
When $s\ll \sigma$, Eq.~\eqref{Eerf} simplifies to  
\begin{align}
E(x,0)\approx E\,  \f{e^{-x^2/ (2 \sigma^2)}}{  \sqrt{2 \pi \sigma^2}}\,. \label{Egauss}
\end{align}

\noindent (ii) {\em Box temperature profile:} In this case we distribute $(N-N_c)/2$ particles uniformly in $(-L/2,-s/2)$, and  $(N-N_c)/2$ particles in $(s/2, L/2)$, with $s=N_c/n_0$. The remaining $N_c$ particles are distributed uniformly in $(-s/2,s/2)$.  The mean number density is again $n_0=N/L$. The velocities of the $N_c$ particles in the center are chosen from the Maxwell distribution with temperature $T=2 E /N_c$, viz. $\text{Prob}(v_i)= \sqrt{m_i/(2 \pi T)}  e^{-m_i v_i^2/(2 T)}$;  the initial velocities of all other particles are set to zero.

\noindent{\bf (B) Ensemble where initial energy is fixed exactly to $E $ and momentum exactly to $0$}: In this case we distribute particles in space using the same protocol as before. We now  choose the velocities of only $N_c/2$ particles ($-N_c/2+1,-N_c/2+2,\ldots,0$) from the Maxwell distribution, $\text{Prob}(v_i)= \sqrt{m_i/(2 \pi T)}  e^{-m_i v_i^2/(2 T)}$, where $T=2 E /N_c$. For every realization, we rescale these velocities by a constant factor such that the total energy of the $N_c/2$ particles is exactly $E /2$. The remaining particles ($i=1,2,\ldots,N_c/2$) are assigned mirror-image momenta: $p_i= p_{-i+1}$. Thus we have an ensemble of initial micro-states which all have total energy exactly equal to $E $ and total momentum exactly zero. 

\noindent{\bf Simulation details}: 
In all our simulations we took $m_1=1,\ m_2=2$  (so that $\mu=(m_1+m_2)/2=1.5$), $\rho_\infty=1.5$, $E =32$ and $N_c=32$. In largest simulations, we took $N=24000$, $L=24000$ and averaged over  an ensemble of ${\cal{R}}=10^5$ initial conditions. In all cases, simulations were carried out up to times such that the shocks have not reached the boundaries, so the results correspond to an infinite system.  The molecular dynamics simulations for this system can be done efficiently using  an event-driven algorithm that updates the system between successive collisions. 

\section{Microscopic Simulations and Comparison with the TvNS Solution}
\label{sec:micro}

We now present the results of the microscopic simulations. We mainly discuss the results of the Gaussian initial temperature profile and with the averaging over an initial ensemble of micro-states where the total energy and total momentum are fixed exactly (ensemble (B) in the previous section).  We also show a comparison with the empirical  profiles (obtained from spatial coarse-graining of single realization) and outline results obtained for the box profile and for the ensemble choice where energy and momentum are fixed on average.

\begin{figure}
	\begin{center}
		\leavevmode
		\includegraphics[width=8.cm,angle=0]{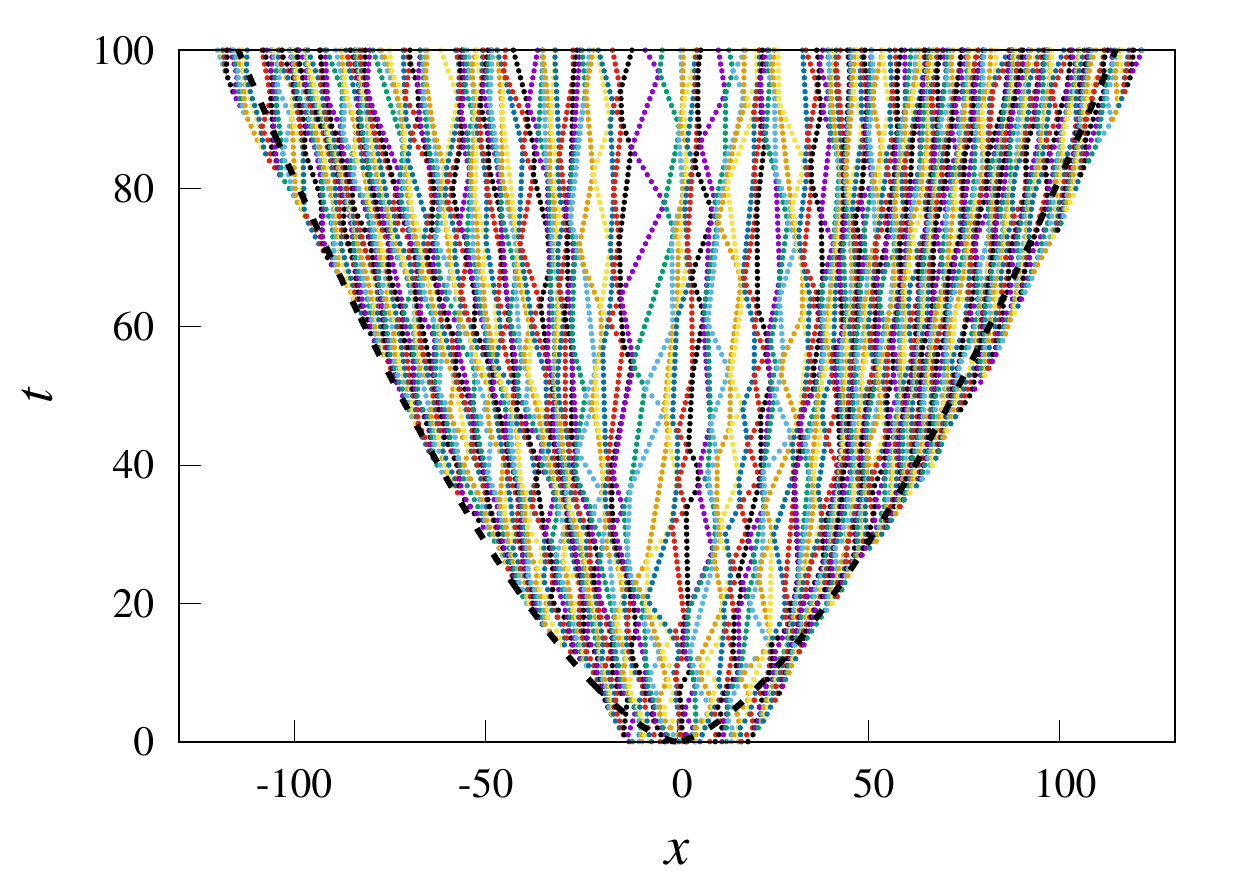}\hfill
		\caption{The trajectories of about $240$ moving particles in a typical blast, starting from initial conditions corresponding to a Gaussian initial 
			temperature profile. The expected shock positions, $\pm R(t)$, are shown as dashed lines. The stationary particles are not shown.
			In this realization, the left-most and right-most moving particles are quite a bit ahead of where they are expected to be according to 
			the deterministic hydrodynamics.}
		\label{trajectory}
		
	\end{center}
\end{figure}

In Fig.~\ref{trajectory} we illustrate at the microscopic level  a typical blast in one dimension, by showing the trajectories of the moving particles. Around $240$ particles are shown in this simulation; other parameters are $N_c=32$, $\rho(x,0)=\rho_\infty = 1.5$ and the initial energy profile was taken to be a Gaussian (see Sec.~\ref{sec:ic}). The expected shock positions $\pm R$, with $R(t)$ given by Eq.~\eqref{R-d}, are shown by the black dashed lines.

\begin{figure*}
	\begin{center}
		\leavevmode
		\includegraphics[width=5.25cm,angle=0]{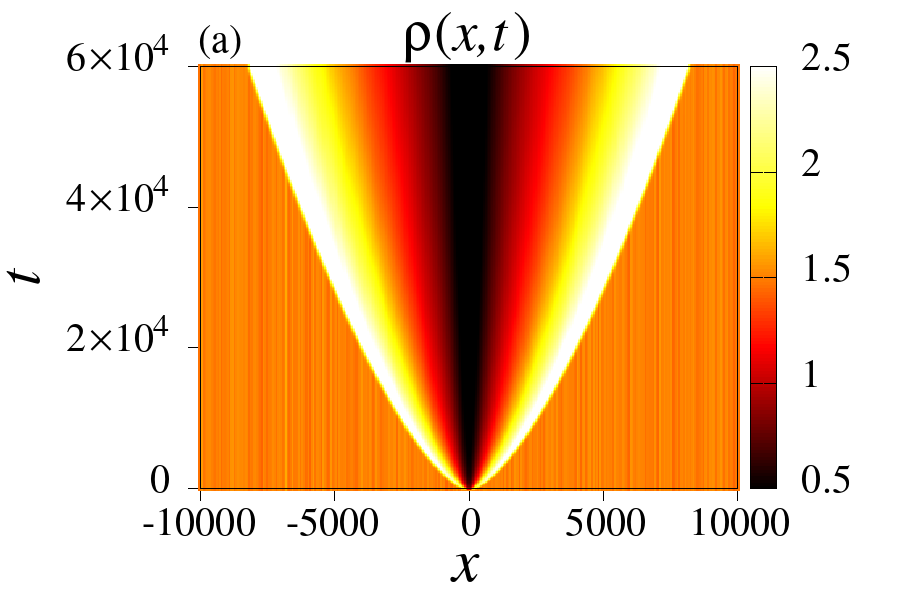}
		\includegraphics[width=5.25cm,angle=0]{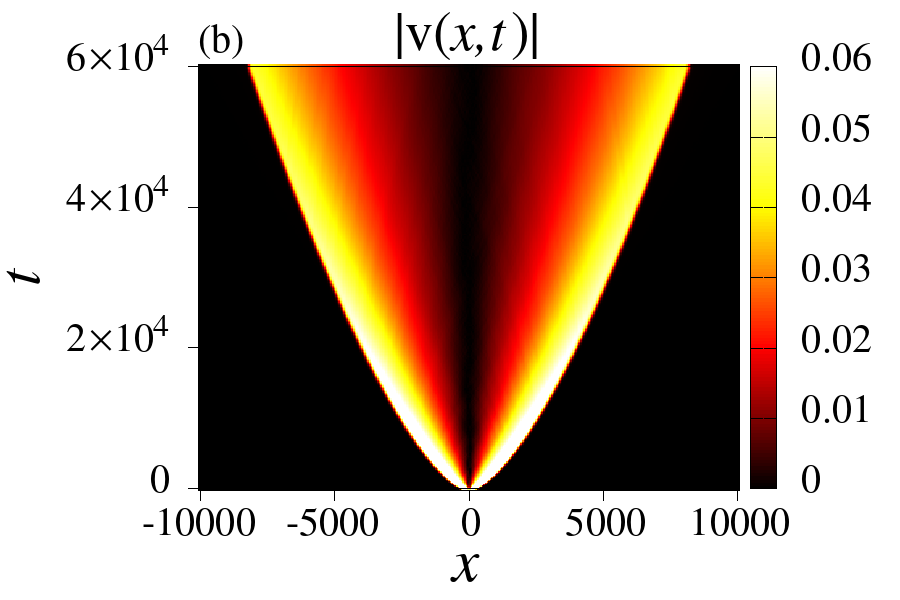}
		\includegraphics[width=5.25cm,angle=0]{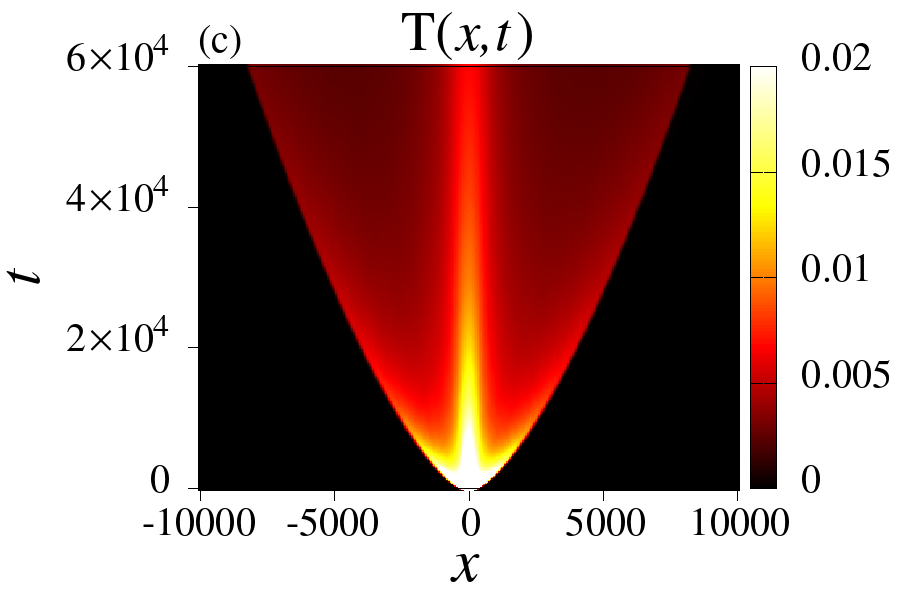}
		\caption{Heat maps of the spatio-temporal evolution of (a) density, (b) velocity, and (c) temperature starting from initial conditions corresponding to a Gaussian initial temperature profile and an ensemble average over $10^4$ initial conditions.}
		\label{heatmap}
	\end{center}
\end{figure*}

In Fig.~\ref{heatmap} we show the spatio-temporal evolution of $\rho(x,t)$, $v(x,t)$, $T(x,t)$ in a gas initially at rest perturbed by a localized Gaussian temperature profile. We see a clear sub-ballistic evolution of the fields.  The mass density and velocity are peaked around the shock front, while the temperature has an additional peak at the center. 

\begin{figure*}
	\begin{center}
		\leavevmode
		\includegraphics[width=5.5cm,angle=0]{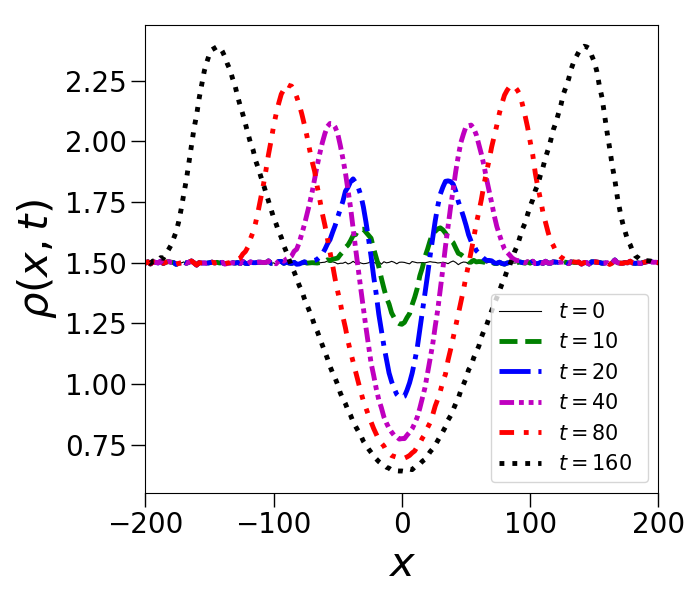}
		\put (-75,130) {$\textbf{(a)}$}
		\includegraphics[width=5.5cm,angle=0]{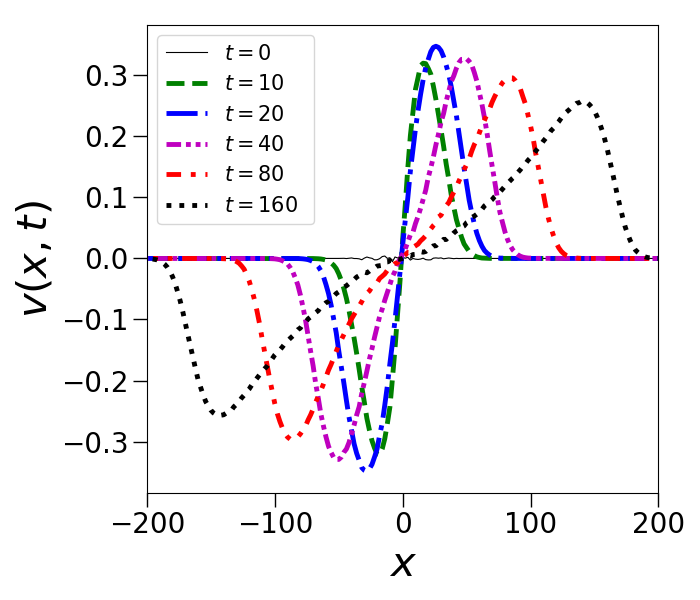}
		\put (-75,130) {$\textbf{(b)}$}
		\includegraphics[width=5.5cm,angle=0]{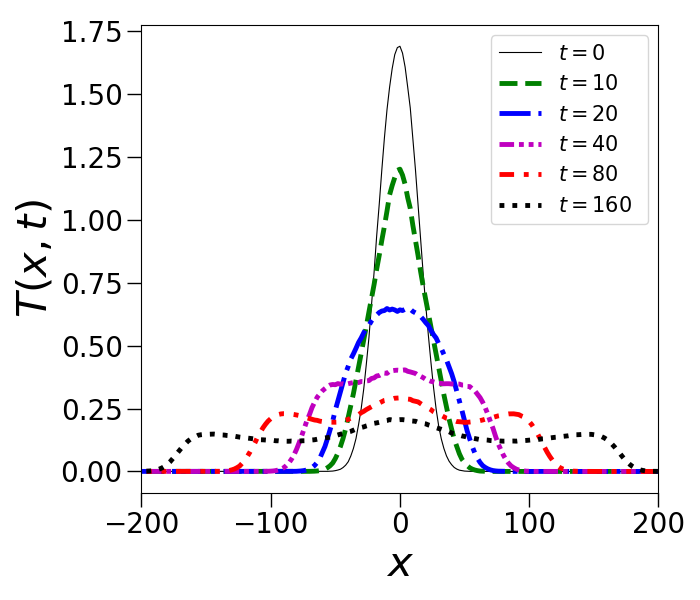}
		\put (-75,130) {$\textbf{(c)}$}
		\caption{Early time evolution of (a) density, (b) velocity and (c) temperature obtained by ensemble averaging starting from initial conditions corresponding to a Gaussian initial temperature profile and an average over $10^5$ initial conditions.}
		\label{earlytime}
	\end{center}
\end{figure*}

\begin{figure*}[!]
	\begin{center}
		\leavevmode
		\includegraphics[width=5.5cm,angle=0]{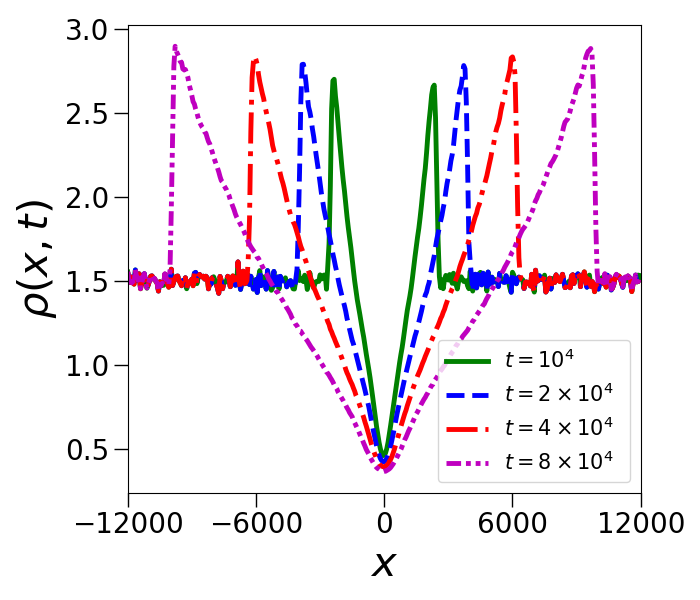}
		\put (-75,130) {$\textbf{(a)}$}
		\includegraphics[width=5.5cm,angle=0]{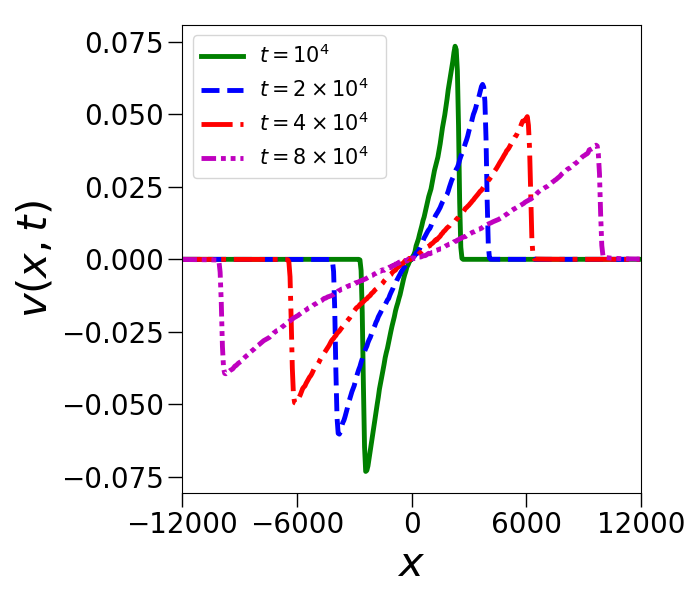}
		\put (-75,130) {$\textbf{(b)}$}
		\includegraphics[width=5.5cm,angle=0]{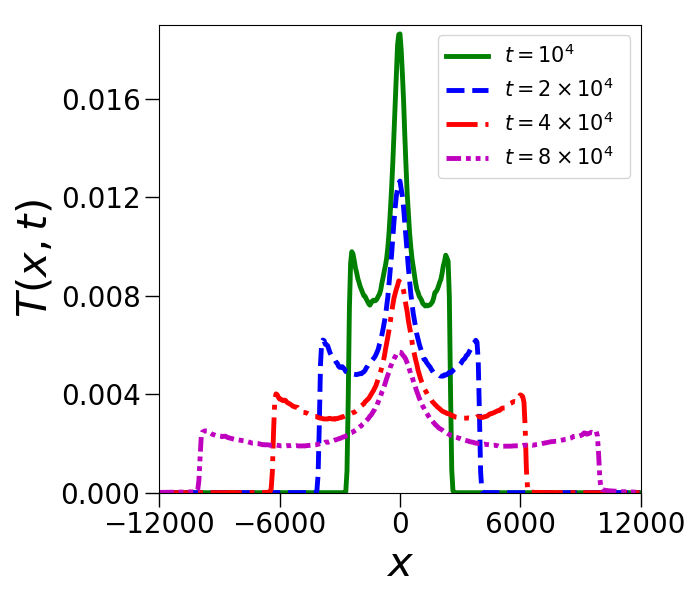}
		\put (-75,130) {$\textbf{(c)}$}
		
		\includegraphics[width=5.5cm,angle=0]{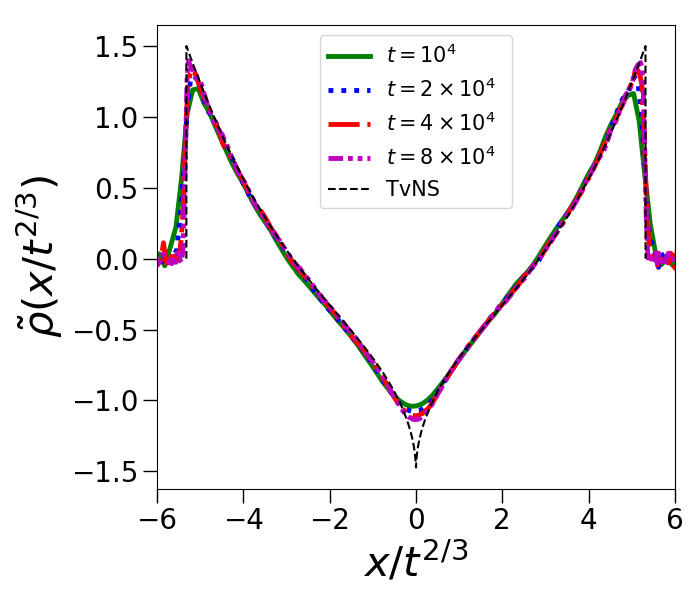}
		\put (-75,130) {$\textbf{(d)}$}
		\includegraphics[width=5.5cm,angle=0]{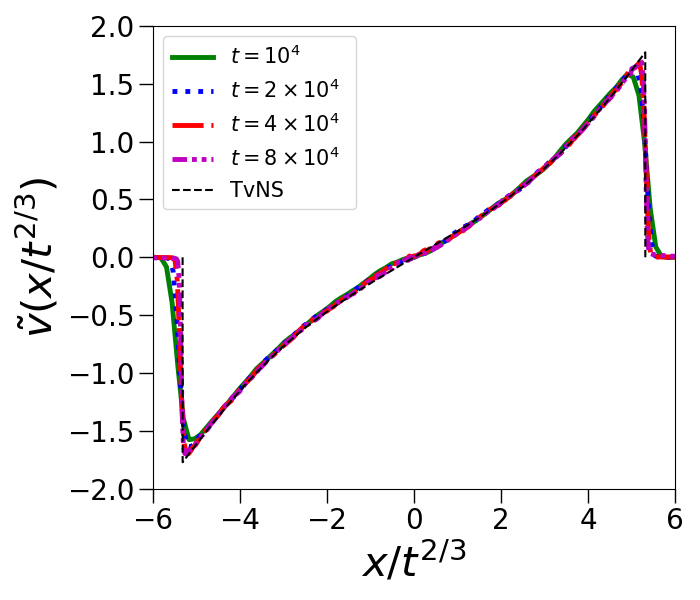}
		\put (-75,130) {$\textbf{(e)}$}
		\includegraphics[width=5.5cm,angle=0]{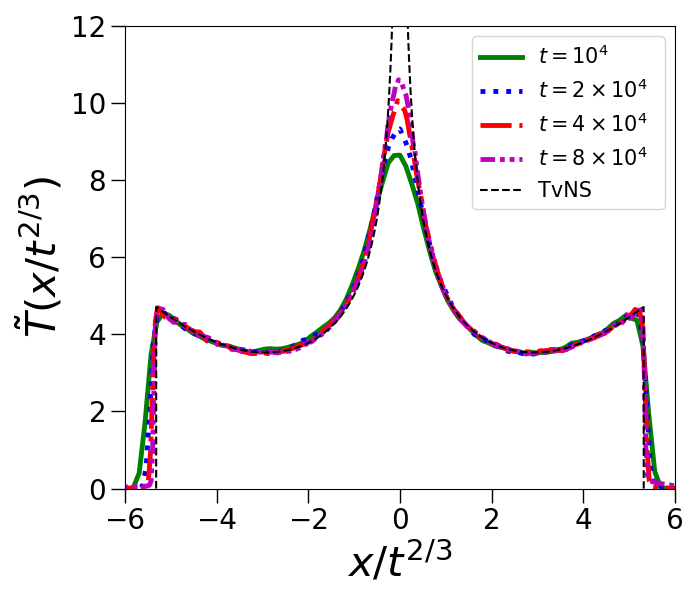}
		\put (-75,130) {$\textbf{(f)}$}
		\caption{ (a,b,c) Late time evolution of density, velocity, and temperature starting from the initial conditions corresponding to a Gaussian initial temperature profile and an average over $10^4$ initial conditions. (d,e,f) The $x/t^{2/3}$ scaling gives a good collapse of the data at the longest times everywhere except near the origin.  The exact scaling solution of the hydrodynamic Euler equations is shown by the black dashed line.}
		\label{longtime}
	\end{center}
\end{figure*}

In Fig.~\ref{earlytime} we plot the early time evolution of the hydrodynamic fields $\rho(x,t)$, $v(x,t)$,  and $T(x,t)$ as  functions of position $x$ at different times $t$. In Fig.~\ref{longtime} we plot the  evolution of the blast wave at long times. The lower panel in Fig.~\ref{longtime} shows $\widetilde{\rho}=\rho_\infty G-\rho_\infty,~\widetilde{v}=t^{1/3} v(x,t) =(2 \alpha/3) \xi V$ and $\widetilde{T}=t^{2/3} T(x,t)=(4 \mu \alpha^2/27) \xi^2 Z$, where  $\alpha = [E/(A \rho_\infty)]^{1/3}$, as  functions of the scaling variable $x/t^{2/3}$. We find a very good scaling collapse of the data confirming the expected TvNS scaling. For comparison, we plot the exact predictions Eq.~\eqref{GVZ-xi:1}. We find an almost perfect agreement between the simulation results and the TvNS solution except for deviations near the center of the blast. As we discuss later in Sec. \ref{sec:core}, the TvNS solution predicts a divergence of the temperature field with $\widetilde{T}(\xi) \sim \xi^{-1/2}$ and this implies that the effect of the heat conduction term in the hydrodynamic equation for energy conservation cannot be neglected. Thus one has to deal with the full Navier-Stokes-Fourier equations instead of the Euler equations. 

\begin{figure*}
	\centering
	\includegraphics[width=5.5cm,angle=0]{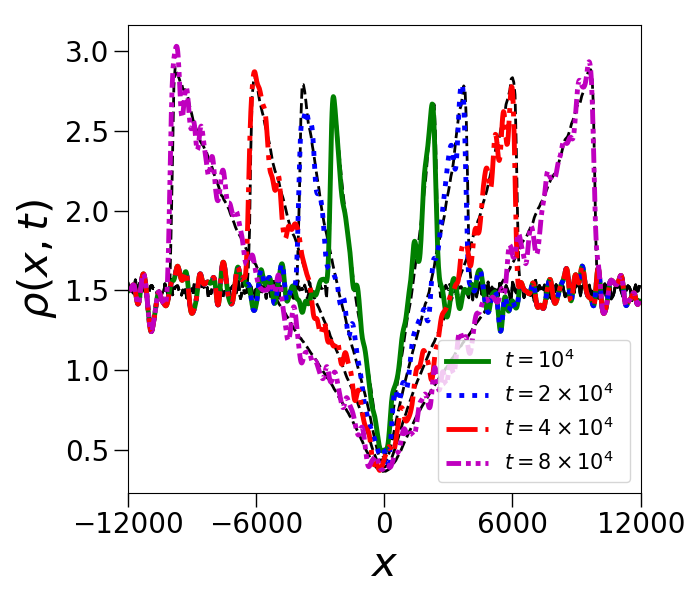}
	\put (-75,130) {$\textbf{(a)}$}
	\includegraphics[width=5.5cm,angle=0]{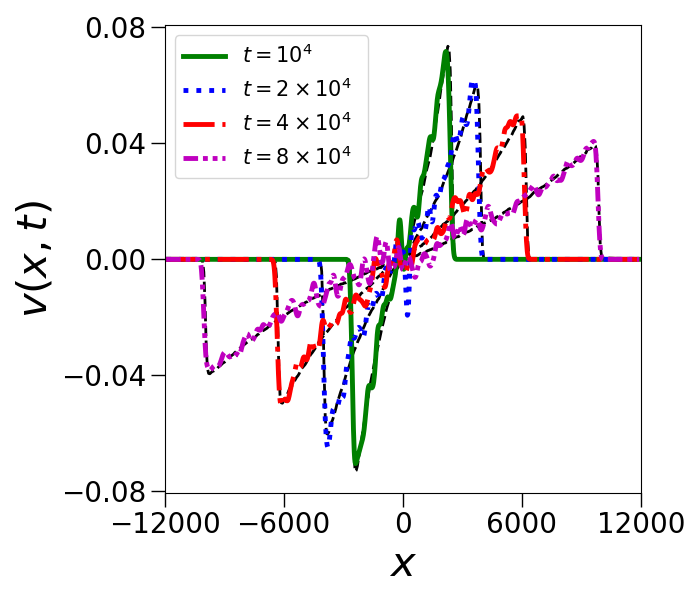}
	\put (-75,130) {$\textbf{(b)}$}
	\includegraphics[width=5.5cm,angle=0]{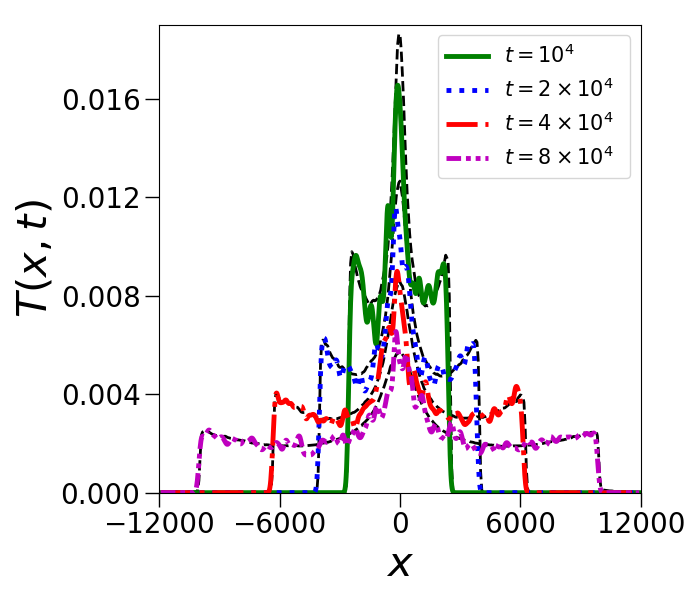}
	\put (-75,130) {$\textbf{(c)}$}
	\caption{Evolution of the empirical density, velocity, and temperature obtained from a single realization, and after spatial coarse-graining according to Eqs.~(\ref{br}-\ref{bE}) and using sliding bins.  The black dashed lines indicate the ensemble-averaged fields, and we see that apart from the fluctuations seen in the empirical fields, there is a good agreement between the two. The  coarse-graining length scale,  $\Delta=50$, and same initial conditions as in Fig.~\ref{longtime} are used. }
	\label{noscalespace}
\end{figure*}

\section{Various Aspects of Microscopic Simulations}
\label{sec:md}
\subsection{Comparison with the Empirical Fields}
Ideally, we would like to see a comparison of the hydrodynamic theory with simulations for the empirical fields that correspond to the physical situation where we observe a {\it{single}} realization, and there is only a spatial averaging of the microscopic degrees. We divide space into boxes of size $\Delta$ that are small compared to the length of the region with moving particles but still typically contain many particles. The empirical  fields are defined as:
\begin{align}
&\bar{\rho}(x,t)=\f{1}{\Delta}\sum m_j \delta[q_j(t),x], \label{br}\\
& \bar{\rho}(x,t) \bar{v}(x,t)=\f{1}{\Delta}\sum  m_j v_j \delta[q_j(t), x], \label{bp} \\
&\bar{E}(x,t)=\bar{\rho}(x,t)\bar{e}(x,t)=\f{1}{2\Delta}\sum  m_j v_j^2 \delta[q_j(t),x], \label{bE}
\end{align}
where $\delta[q_j,x]$ is an indicator function taking value $1$ if $ x-\Delta/2 \leq q_j \leq x+\Delta/2$ and zero otherwise. The system is initially excited in the same manner as the previous subsection with the same initial energy $E $ and zero momentum, but we now take only \emph{one realization} of the microscopic process.  The comparison  of the empirical density profiles with the ensemble-averaged profiles is shown in Fig.~\ref{noscalespace}, and we see a close agreement between the two. Here we used a bin size $\Delta=50$ and, in order to get more data points, we used sliding bins. The fluctuations seen in the empirical profile are expected to vanish in the thermodynamic limit. 

\subsection{Dependence on Initial Conditions}
\begin{figure*}[!]
	\begin{center}
		\leavevmode
		\includegraphics[width=5.5cm,angle=0]{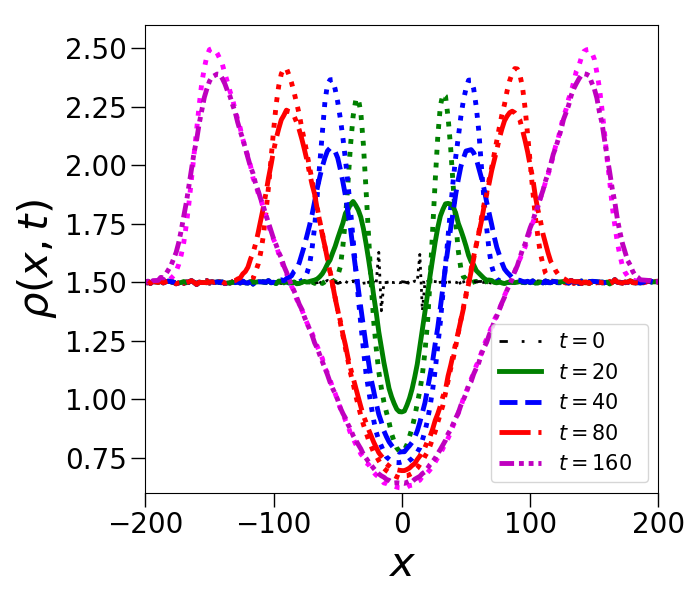}
		\put (-75,130) {$\textbf{(a)}$}
		\includegraphics[width=5.5cm,angle=0]{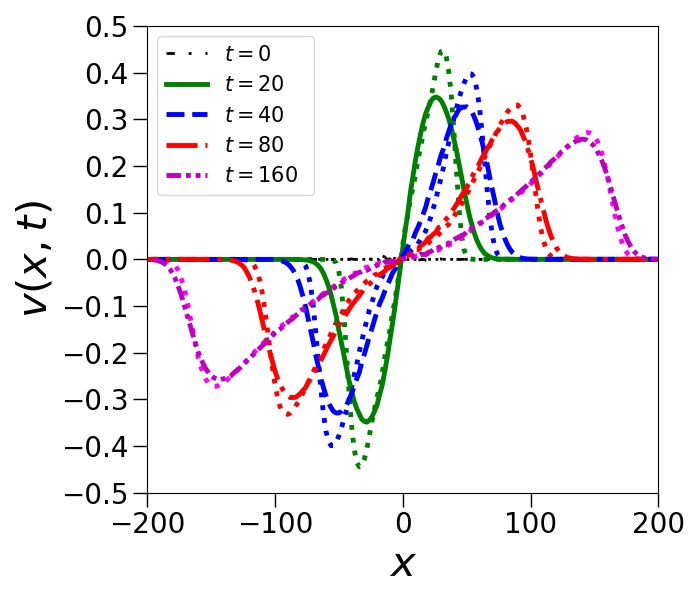}
		\put (-75,130) {$\textbf{(b)}$}
		\includegraphics[width=5.5cm,angle=0]{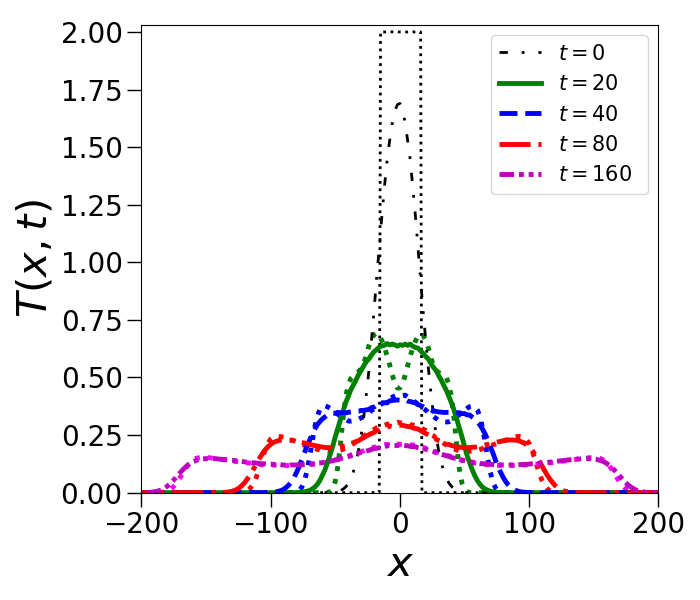}
		\put (-75,130) {$\textbf{(c)}$}
		\caption{The comparison of the evolution of three field profiles starting from two different initial conditions corresponding to either a Gaussian temperature profile (solid lines) or a box-profile (dotted lines). The total energy was fixed at $E =32$,  the background density was taken as $\rho_\infty=1.5$, and the average was taken over $10^5$ realizations.}
		\label{ICcomp}
	\end{center}
\end{figure*}

The long-time scaling solution should be independent of the details of the initial conditions. It may depend on two parameters, the total blast energy $E$ and the background density $\rho_\infty$. We verify this in simulations by starting with the box initial condition Eq.~\eqref{eq:ezero}, with the same values of $E =32$ and $\rho_\infty=1.5$. From the results plotted in Fig.~\ref{ICcomp}, it is clear that the profiles of all the three fields from the two different initial conditions quickly converge and become almost indistinguishable. 

\subsection{Dependence on the Initial Ensemble} 
\begin{figure*}
	\begin{center}
		\leavevmode
		\includegraphics[width=5.5cm,angle=0]{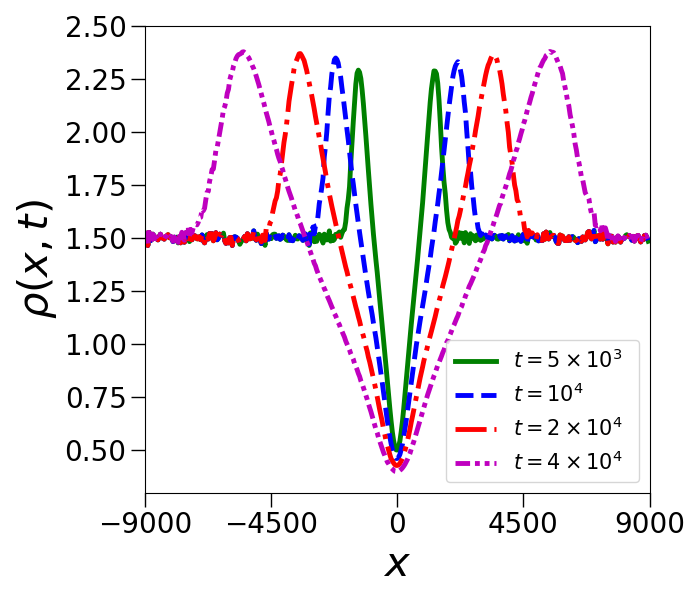}
		\put (-75,130) {$\textbf{(a)}$}
		\includegraphics[width=5.5cm,angle=0]{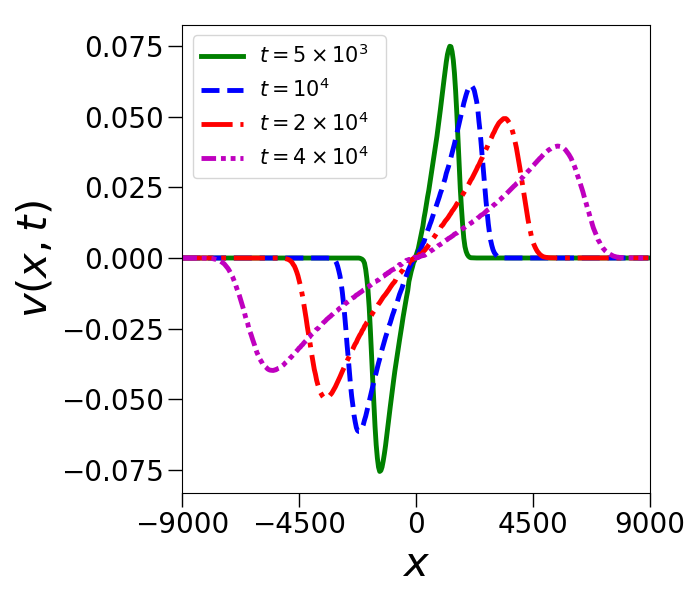}
		\put (-75,130) {$\textbf{(b)}$}
		\includegraphics[width=5.5cm,angle=0]{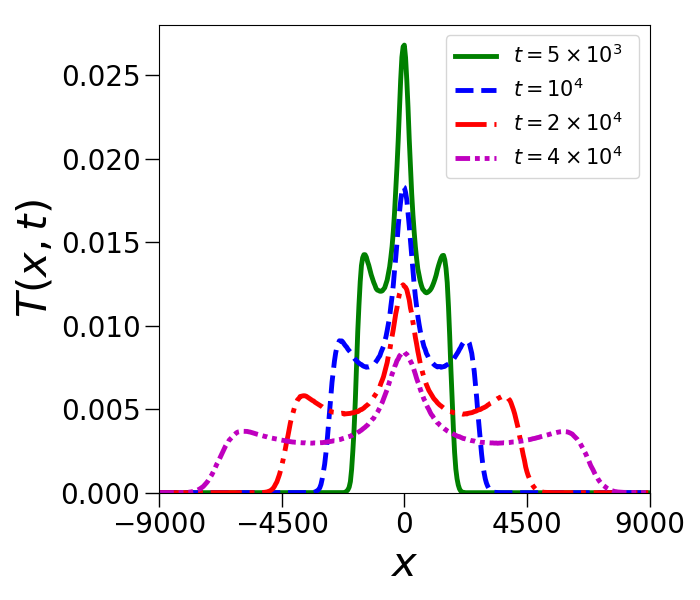}
		\put (-75,130) {$\textbf{(c)}$}
		
		\includegraphics[width=5.5cm,angle=0]{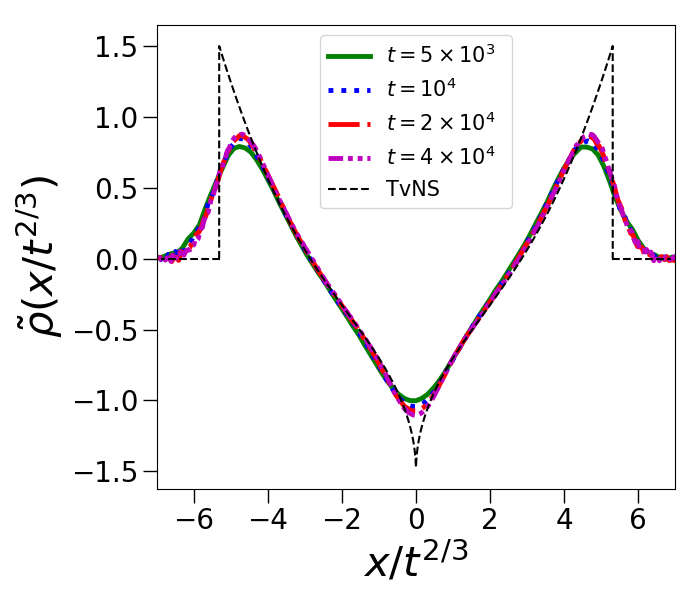}
		\put (-75,130) {$\textbf{(d)}$}
		\includegraphics[width=5.5cm,angle=0]{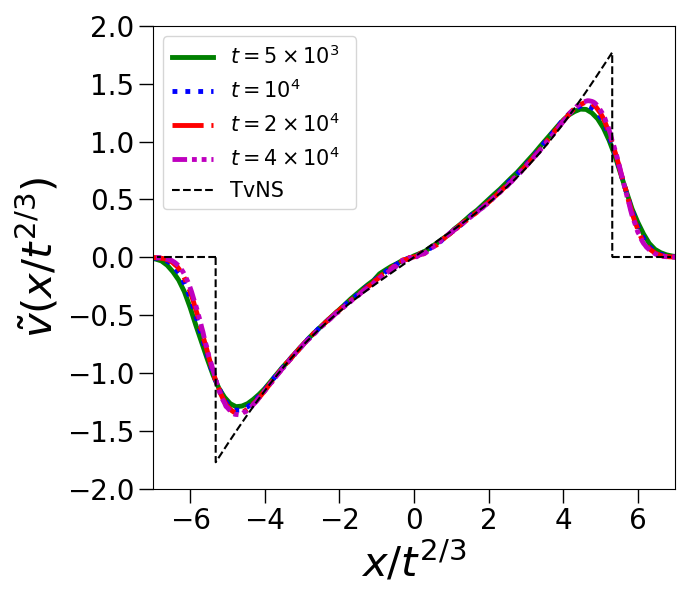}
		\put (-75,130) {$\textbf{(e)}$}
		\includegraphics[width=5.5cm,angle=0]{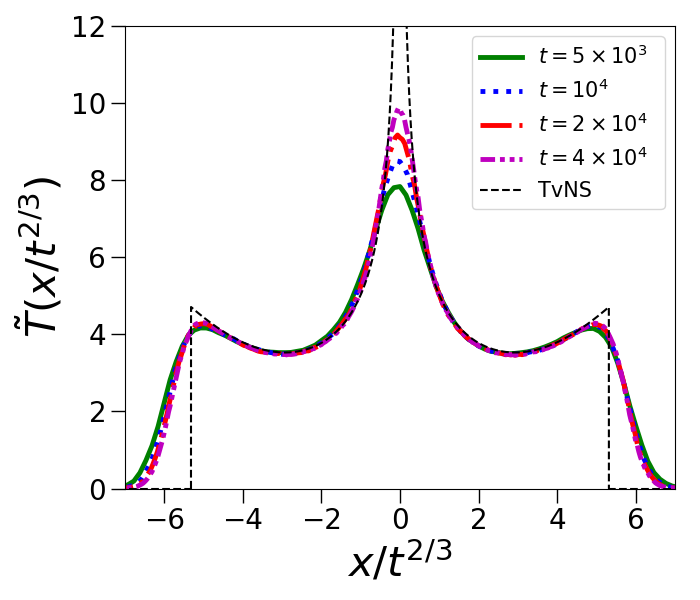}
		\put (-75,130) {$\textbf{(f)}$}
		\caption{{{\bf Initial ensemble with fluctuations in initial total energy and total momentum, with mean energy fixed at $E$ and mean momentum at 0} (a,b,c) Evolution of the density, velocity, and temperature starting from the initial conditions corresponding to a Gaussian initial temperature profile and an average over $10^4$ initial conditions. (d,e,f) The $x/t^{2/3}$ scaling demonstrates data collapse in the long time limit everywhere except at the center. Comparing with Fig.~\ref{longtime} we see that the agreement with the TvNS scaling solution, shown by dashed lines, is now much poorer with significant deviations both at the blast center and near the shock fronts. }}
		\label{longtimecan}
	\end{center}
\end{figure*}

In Fig.~\ref{longtimecan} we repeat the simulations with initial conditions chosen from the ensemble~(A) discussed in Sec.~\ref{sec:ic} where the energy $E $ and total zero initial momentum value are fixed only on average. This means that there are realizations where the energy can be larger or smaller than $E $ while the momentum can have a non-zero value. As a result, this leads to tails developing in the shock front and we lose agreement with the hydrodynamic theory. The correct physically relevant initial condition is the fixed energy, fixed momentum ensemble and so we will not discuss further the case of ensemble~(A).

\subsection{Validity of Local Thermal Equilibrium} 
The local equilibrium (LE) assumption is an important requirement for the validity of the hydrodynamic description. In case of equilibrium, the velocity distribution is Gaussian. Earlier works in two and three dimensions \cite{Joy2021,Joy2021a}  find strong deviations from LE, through observations of the skewness and kurtosis of the local velocity distributions, which indicate non-Gaussianity. These could be a cause for the observed disagreement with the hydrodynamic predictions. Here we show results for these quantities obtained from MD simulations of our 1D system. We define the local skewness $S(x,t)=\frac{1}{n(x,t)}\left \langle \sum [v_i-v(x,t)]^3 \delta(q_i-x) \right \rangle$ and the local kurtosis \newline $\chi(x,t)=\frac{1}{n(x,t)}\left \langle \sum m_i^2 [v_i-v(x,t)]^4 \delta(q_i-x) \right \rangle$, where the number density field $n(x,t)=\left \langle \sum \delta(q_i-x) \right \rangle$. In 1D, the Gaussian distribution of velocity fluctuation implies that the skewness $S(x,t)=0$ and the kurtosis $\chi(x,t) = 3T^2$. 

\begin{figure}
	\includegraphics[width=0.8\textwidth]{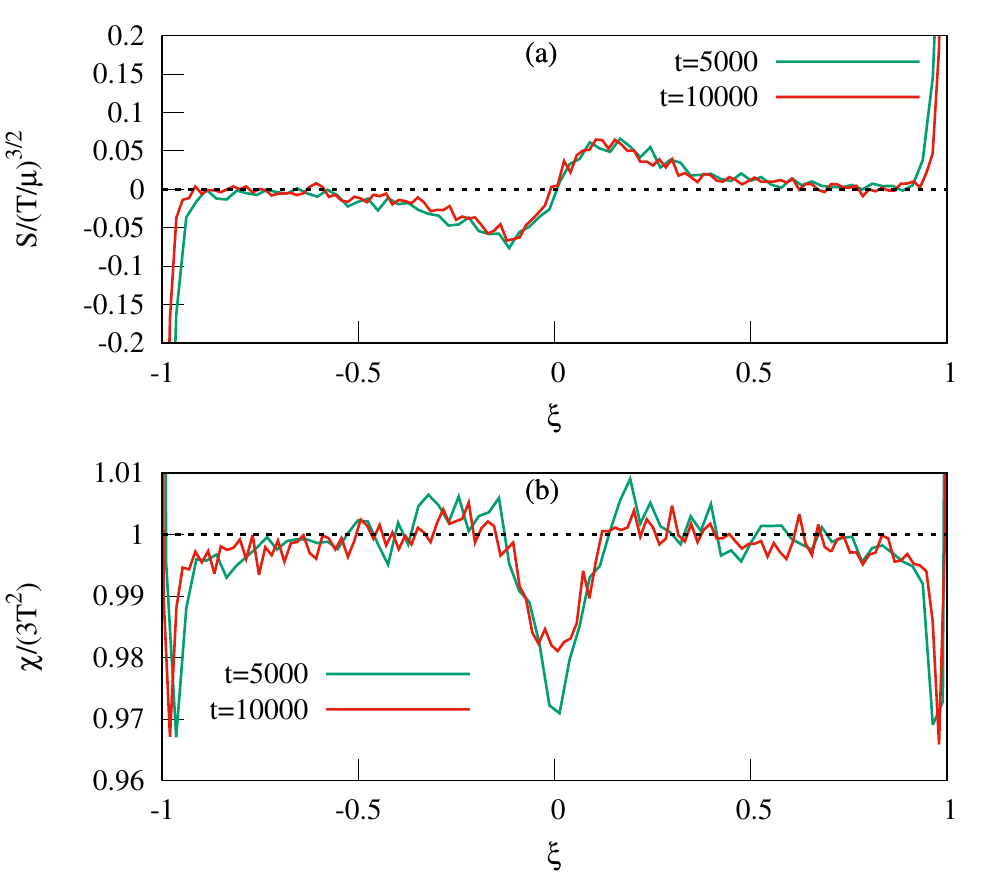}
	\caption{The skewness $S$ and the kurtosis $\chi$, corresponding  to the local velocity distribution at different points inside the blast, plotted as a function of the scaled variable $\xi=x/R(t)$. The skewness shown in (a) is scaled by $[{\rm thermal ~velocity}]^3$ while the kurtosis shown in (b) is scaled by the expected thermal value of $3 T^2(x,t)$. The results at two different times are shown.}
	\label{fig:le}
\end{figure}

In Fig.~\ref{fig:le}, we plot the dimensionless scaled skewness and kurtosis with the scaled variable $\xi=x/R(t)$ at two different times and find that the LE assumption is satisfied quite accurately, and significantly more than what is observed in higher dimensions~\cite{Joy2021,Joy2021a}. We note that there are  small deviations from LE which interestingly, peak in the core and shock regions.  While this could provide a possible explanation for the disagreement between TvNS and simulations in the core region, we note however that we continue to have good agreement between these theories near the shock  front. We now explore the structure of the shock in some more detail.

\subsection{Structure of the Shock} 
\begin{figure}
	\includegraphics[width=0.8\textwidth]{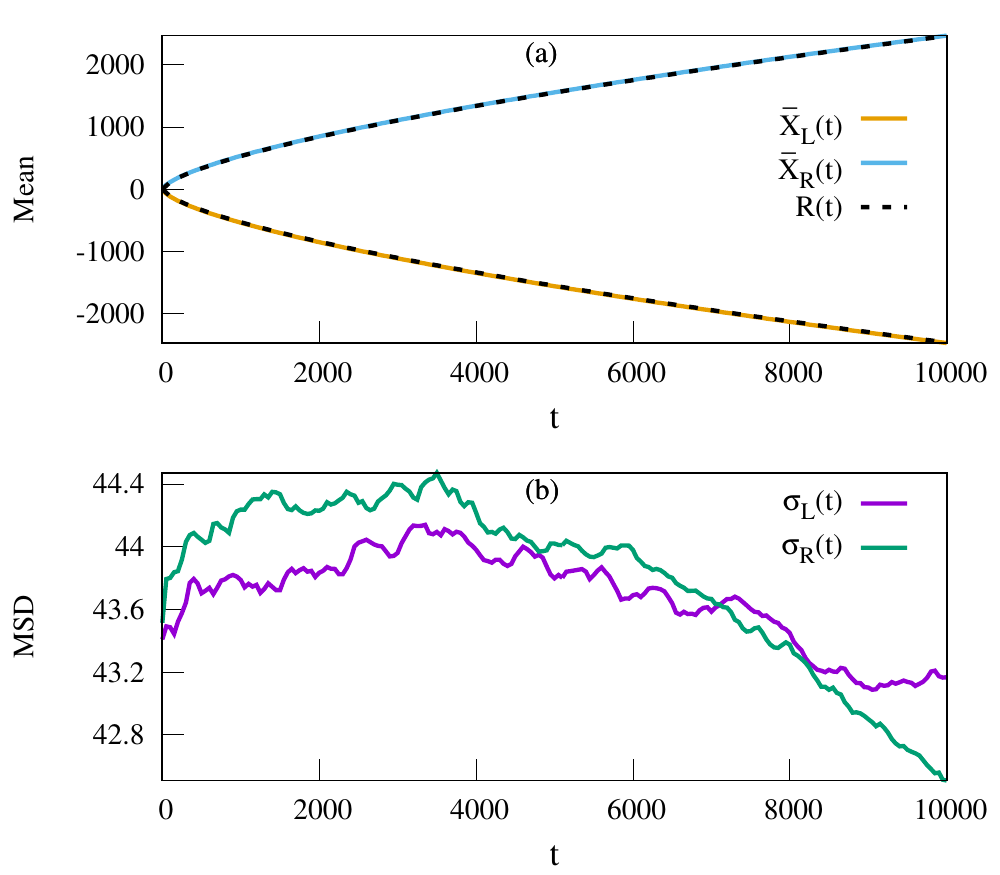}
	\caption{Time evolution of (a) the mean shock positions $\bar{X}_L(t),~\bar{X}_R(t)$, and (b) their fluctuations $\sigma_L(t),~\sigma_R(t)$.  The dashed line $R(t)$ is the TvNS prediction. The MD simulations were done for $N=L=8000$ with an averaging over $10^4$ realizations.}
	\label{fig:front}
\end{figure}

At any moment, we have a finite number of moving particles surrounded by an infinite sea of stationary particles. One can then define the shock positions as that of the extreme moving particles. Denote by $X_L(t)$ and $X_R(t)$ the positions of the left-most and right-most moving particles. These quantities fluctuate between different microscopic realizations. Fluctuations are outside the scope of deterministic hydrodynamics.  We probe the basic features of fluctuations by measuring the mean front positions, $\bar{X}_L(t)=\la X_L(t)\ra$ and $\bar{X}_R(t)=\la X_R(t)\ra$, and the variances, 
$\sigma^2_L(t)=\la X^2_L(t)\ra-\la X_L(t)\ra^2$ and $\sigma^2_R(t)=\la X^2_R(t)\ra-\la X_R(t)\ra^2$. Fig.~\ref{fig:front}a shows that the mean positions $\bar{X}_L(t)$ and $\bar{X}_R(t)$ closely follow the hydrodynamic predictions $\pm R(t)$. Fig.~\ref{fig:front}b shows that the fluctuations remain bounded around the equilibrium value $\sigma_{\rm eq}^2=L^2/(4N)$, which for our parameters gives $\sigma_{\rm eq}^2\approx 45$. Thus we do not observe a broadening of the shock front with increasing time. This behaviour is in sharp contrast to the moving piston case studied in \cite{Hurtado2006} where a diffusive broadening of the shock front was observed. 

The hydrodynamic description assumes local equilibrium (LE). The LE fails inside the shock where the hydrodynamic fields undergo large changes over a small length scale (comparable to the mean free path, i.e., to the gap between adjacent particles in one dimension). The Euler description, where shock fronts are singular surfaces ignores the shock wave structure, while the NSF description is just an uncontrolled approximation. The Boltzmann equation approach provides an adequate framework for analyzing shock waves. Unfortunately, it is notoriously difficult for theoretical analysis even for infinitely strong shock waves \cite{cercignani1999,takata2000,colangeli2013}.

\section{Modelling by the Navier-Stokes-Fourier (NSF) Equations}
\label{sec:hydro}

We now compare microscopic simulations with results obtained by solving the one-dimensional Navier-Stokes-Fourier (NSF) equations: 
\begin{subequations}
	\label{tNS}
	\begin{align}
	&\partial_t \rho + \partial_x (\rho v) = 0, \label{tNS1} \\
	&\partial_t (\rho v) + \partial_x (\rho v^2 +p) = \p_x (\zeta\partial_x v), \label{tNS2} \\
	&\partial_t (\rho e) + \partial_x (\rho ev   + p v)= \p_x (\zeta v\partial_x v + \kappa\partial_x T ),
	\label{tNS3}
	\end{align}
\end{subequations}
supplemented with thermodynamic relations Eqs.~\eqref{therm2}. We aim to show that NSF equations capture the ``essence" of the AHP gas system better than the Euler equations.

The dissipative effects are manifested by the bulk viscosity $\zeta$ and the thermal conductivity $\kappa$ appearing in Eqs.~\eqref{tNS}. These transport coefficients can depend on the fields. The Green-Kubo relations and kinetic theory arguments \cite{Resibois,krapivskybook} lead to the temperature dependence $\zeta \sim T^{1/2}$ and $\kappa \sim T^{1/2}$ characterizing hard-sphere gases in an arbitrary spatial dimension. A recent numerical study \cite{Hurtado2016} suggests the density dependence $\kappa \sim \rho^{1/3}$ specific to one dimension. In our numerical study we have thus used the forms $\zeta =D_1T^{1/2}$ and $\kappa=D_2\rho^{1/3}T^{1/2}$, where $D_1$ and $D_2$ are constants. 
We solve the NSF equations Eqs.~\eqref{tNS} numerically subject to the same initial conditions as in the microscopic simulations: $\rho(x,0)=\rho_\infty, v(x,0)=0$ and $E(x,0)$ given by Eq.~\eqref{Eerf}. We use the MacCormack method~\cite{maccormack1969, maccormack1982} which is second order in both space and time, with discretization $dx=0.1$ and $dt=0.001$. We set the amplitudes in the dissipative terms to unity: $D_1=D_2=1$. In our numerical solution we took system size L=4000 and evolved up to the time ($t \approx 6400$) at which the shock front reached the boundaries.

\begin{figure*}
	\begin{center}
		\leavevmode
		\includegraphics[width=5.5cm,angle=0]{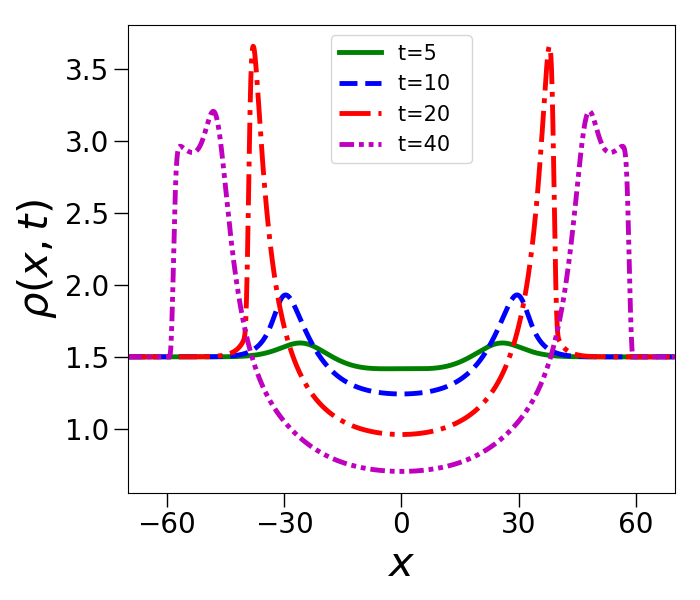}
		\put (-75,130) {$\textbf{(a)}$}
		\includegraphics[width=5.5cm,angle=0]{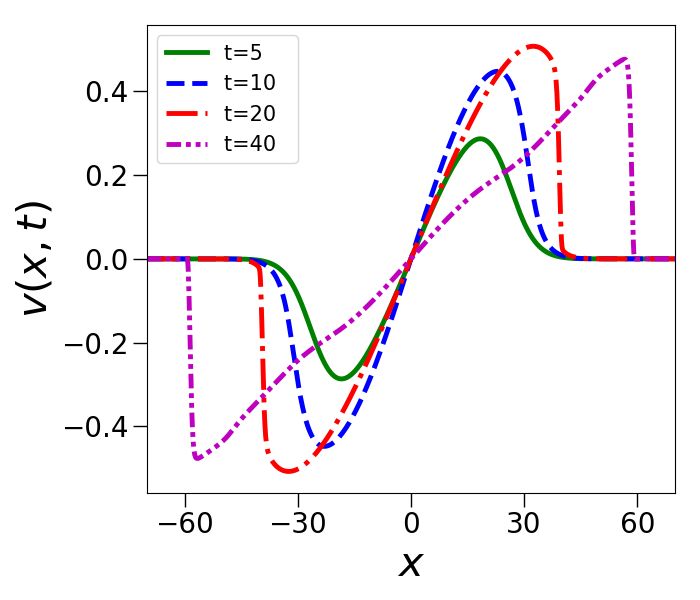}
		\put (-75,130) {$\textbf{(b)}$}
		\includegraphics[width=5.5cm,angle=0]{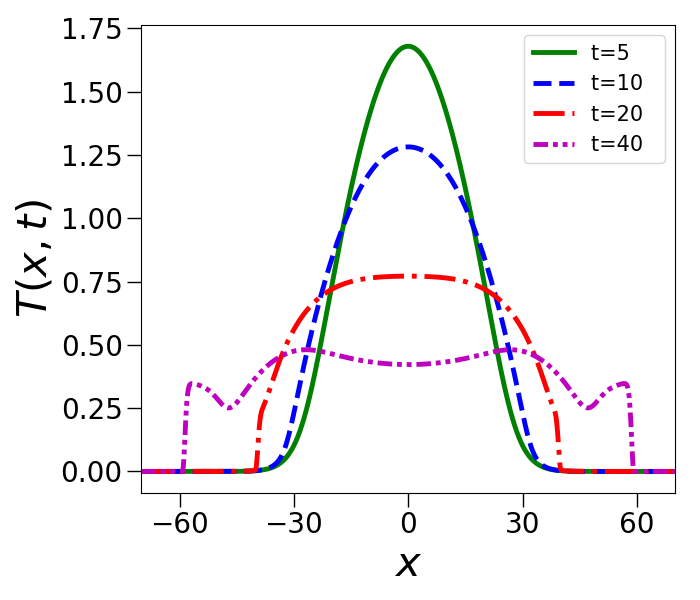}
		\put (-75,130) {$\textbf{(c)}$}
		\caption{Early time evolution of the hydrodynamic fields $\rho(x,t), v(x,t), T(x,t)$ as obtained from a numerical solution of Eqs.~\eqref{tNS}, 
			starting from the  initial conditions with Gaussian temperature profile as in the simulations for Fig.~\ref{longtime}. 
			We see the formation of a sharp front.}
		\label{hydroearly}
	\end{center}
\end{figure*}

\begin{figure*}
	\begin{center}
		\leavevmode
		\includegraphics[width=5.5cm,angle=0]{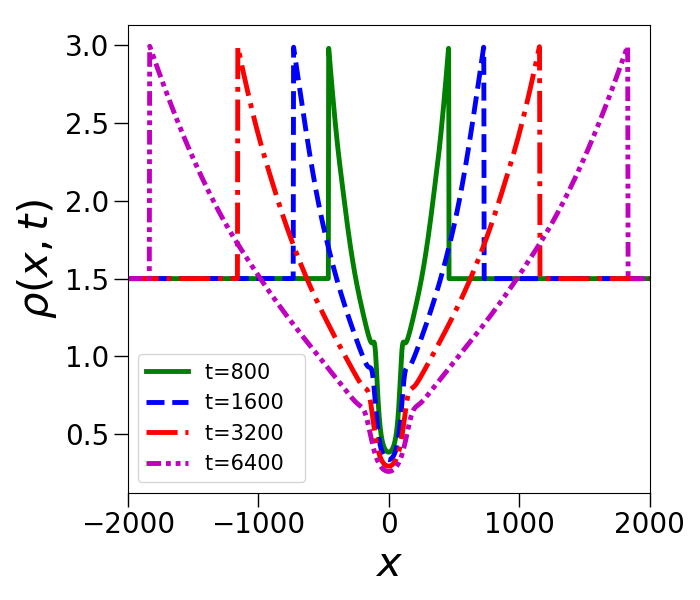}
		\put (-75,130) {$\textbf{(a)}$}
		\includegraphics[width=5.5cm,angle=0]{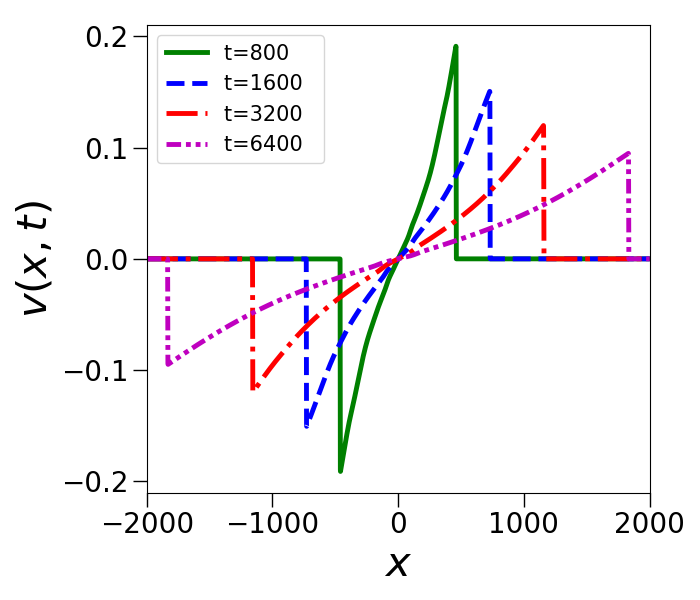}
		\put (-75,130) {$\textbf{(b)}$}
		\includegraphics[width=5.5cm,angle=0]{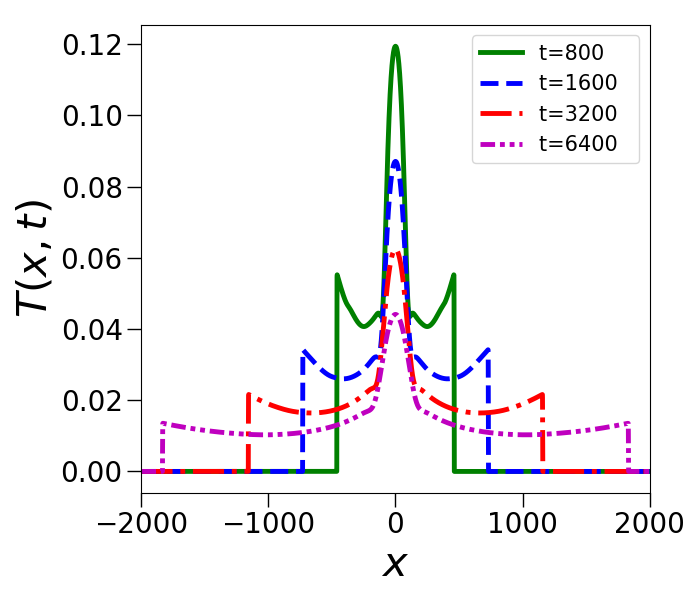}
		\put (-75,130) {$\textbf{(c)}$}
		
		\includegraphics[width=5.5cm,angle=0]{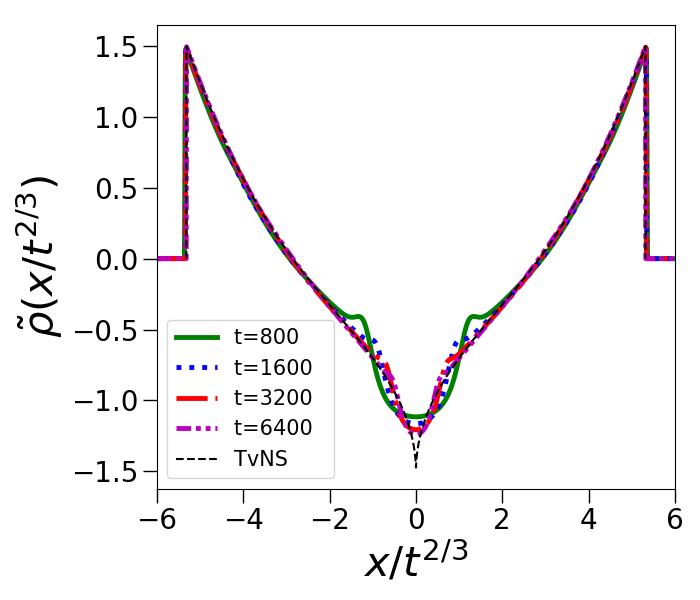}
		\put (-75,130) {$\textbf{(d)}$}
		\includegraphics[width=5.5cm,angle=0]{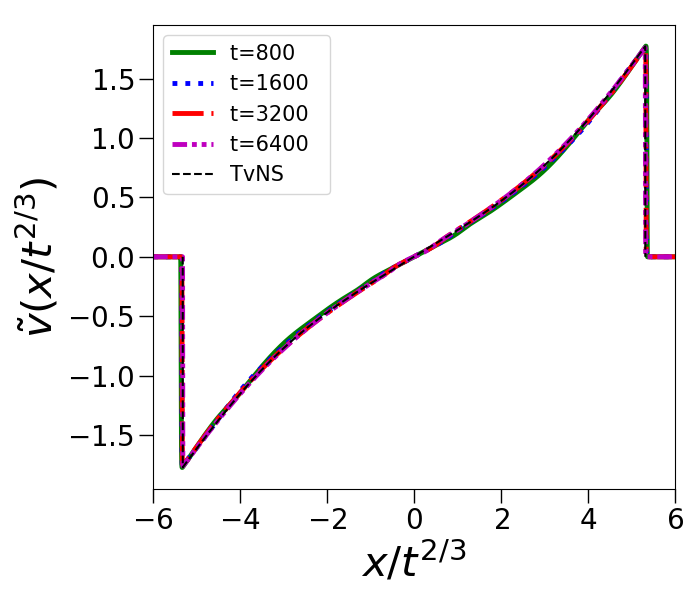}
		\put (-75,130) {$\textbf{(e)}$}
		\includegraphics[width=5.5cm,angle=0]{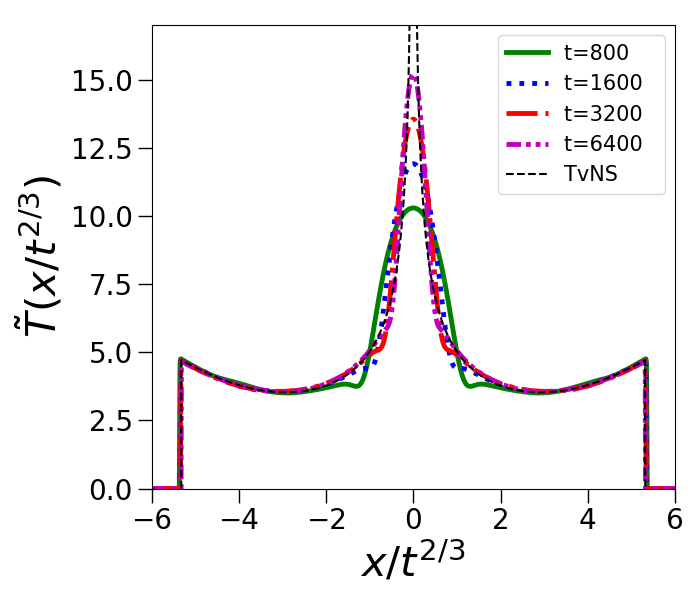}
		\put (-75,130) {$\textbf{(f)}$}
		\caption{(a,b,c) Evolution of the hydrodynamic fields $\rho(x,t), v(x,t), T(x,t)$ as obtained from a numerical solution of Eqs.~\eqref{tNS} 
			starting from the  initial conditions with Gaussian temperature profile used  in the simulations for Fig.~\ref{longtime}. (d,e,f) Plots of the scaling functions $\widetilde{\rho}, \widetilde{v}, \widetilde{T}$ and a comparison with the TvNS solution (black dashed lines).}
		\label{noscaleHydnew}
	\end{center}
\end{figure*}

\begin{figure*}
	\begin{center}
		\leavevmode
		\includegraphics[width=5.5cm,angle=0]{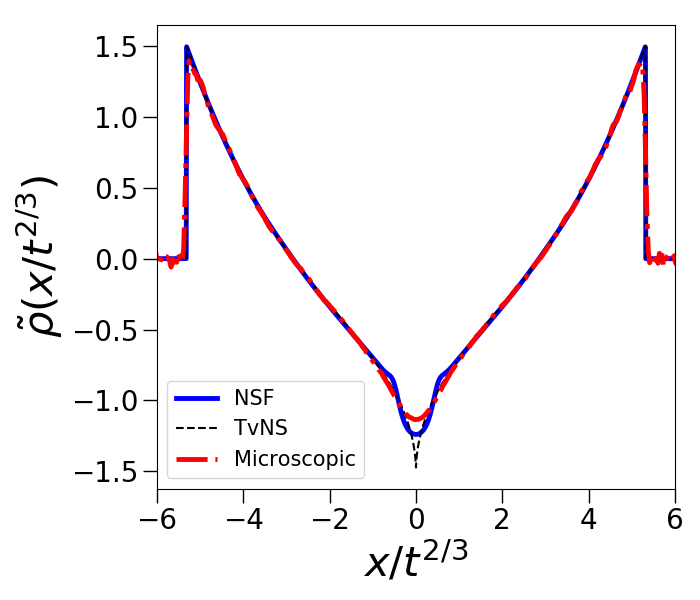}
		\put (-75,130) {$\textbf{(a)}$}
		\includegraphics[width=5.5cm,angle=0]{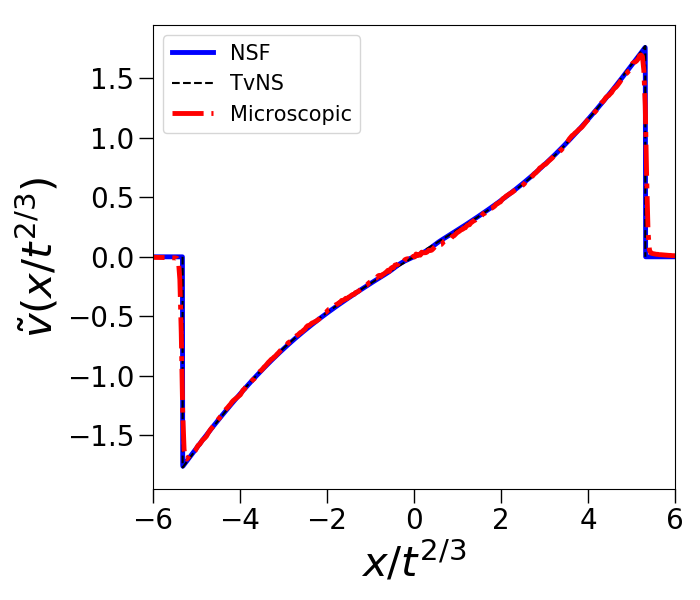}
		\put (-75,130) {$\textbf{(b)}$}
		\includegraphics[width=5.5cm,angle=0]{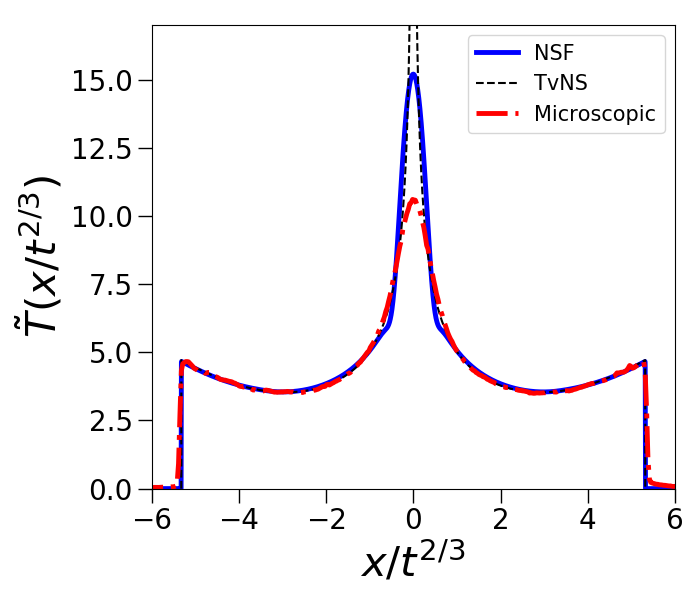}
		\put (-75,130) {$\textbf{(c)}$}
		\caption{Comparison of the scaled fields $\widetilde{\rho},\widetilde{v},\widetilde{T}$ plotted as a function of the scaled variable $x/t^{2/3}$, from microscopic simulation (at time $t=80000$) given by red dash dotted lines, from dissipative hydrodynamics (at $t=6400$) given by blue solid lines and from  the exact TvNS solution of the Euler equations given by black dashed lines.}
		\label{comparisonlong}
	\end{center}
\end{figure*}

The numerical solution at early times plotted in Fig.~\ref{hydroearly} shows that sharp fronts quickly develop for all three fields. The late time numerical results are plotted in  Fig.~\ref{noscaleHydnew} (a,b,c) where we show the profiles of the fields, $\rho(x,t)$, $v(x,t)$ and $T(x,t)$ at different times.  In Fig.~\ref{noscaleHydnew} (d,e,f) we verify the expected scaling forms  $\rho(x,t)= \rho_\infty+ \widetilde \rho(x/t^{2/3})$, $  v(x,t)= t^{-1/3} \widetilde v(x/t^{2/3})$ and $T(x,t)=t^{-2/3}\widetilde T(x/t^{2/3})$, but observe that there is again a core region where we do not get a good data collapse for the density and temperature fields.  For comparison, the exact TvNS scaling solution is  also plotted (dark dashed lines) and we see now that the singularities at the origin have now disappeared and the profiles are closer to those observed in simulations confirming thus the important role of dissipation. In Fig.~\ref{comparisonlong} we show a comparison of the numerical predictions of the dissipative hydrodynamics (at the longest available time) with MD simulations and the exact solution of Euler equations.  Simulation results and the numerical solution of the NSF equations show that the TvNS scaling breaks down in the core region of the blast and we should therefore seek a different solution in this region. In the next section, we analyze the NSF equations to answer this question.

\section{Scaling near the Core of the Blast}
\label{sec:core}
We start with the exact TvNS solution in one dimension, which was derived in the previous chapter. The scaling functions $G(\xi),V(\xi)$ and $Z(\xi)$ are implicitly given by:
\begin{subequations}
	\label{GVZ-xi:2}
	\begin{align} 
	\label{GV:2}
	&G = 2^\frac{16}{5}\, (1-V)^2\, (3V-1)^\frac{1}{5}\, (3-4V)^{-\frac{11}{5}},\\
	\label{xi-V:2}
	&\xi^5 = 2^{-\frac{4}{3}}\,(3V-1)^2\, V^{-\frac{10}{3}}\,(3-4V)^{-\frac{11}{3}},\\
	&\label{ZV:2}
	Z = \frac{3(1-V)V^2}{3V - 1}.
	\end{align} 
\end{subequations}
Near the center of the explosion, the scaled hydrodynamic quantities exhibit the following behaviours: 
\begin{subequations}
	\begin{align}
	\label{G-0}
	G & \simeq B\, |\xi|^\frac{1}{2}\,, ~\quad B =  2^\frac{16}{3}\,3^{-\frac{1}{2}}\,5^{-\frac{11}{6}}, \\
	\label{V-0}
	3V-1&\simeq \beta\, |\xi|^\frac{5}{2}\,, \quad ~~\beta = 2^\frac{2}{3}\,3^{-\frac{7}{2}}\,5^\frac{11}{6}, \\
	\label{Z-0}
	Z & \simeq C\, |\xi|^{-\frac{5}{2}}\,, \quad C =  2^\frac{1}{3}\,3^\frac{3}{2}\,5^{-\frac{11}{6}},
	\end{align}
\end{subequations}
where $\xi=x/R(t)$. Equation \eqref{V-0} follows from Eq.~\eqref{xi-V:2}. Substituting Eq.~\eqref{V-0} into Eq.~\eqref{GV:2} [resp. Eq.~\eqref{ZV:2}] one derives 
Eq.~\eqref{G-0} [resp. Eq.~\eqref{Z-0}].

In one dimension, the scaling form in Eq.~\eqref{scaling} gives
\begin{equation}
\label{vrc}
\rho= \rho_\infty\, G(\xi),~~ v=\frac{2x}{3t}\, V(\xi), ~~T = \frac{4 \mu x^2}{27 t^2}\,  Z(\xi).
\end{equation}
These equations  are valid for all $x\in [-R(t), R(t)]$. The scaled variable is still given by $\xi=|x|/R$ and varies in the $0\leq \xi\leq 1$ range. Equations \eqref{G-0}--\eqref{Z-0} and Eq.~\eqref{vrc} determine the behavior of the original hydrodynamic quantities near the center of explosion. Near the center of explosion, the gas moves with velocity varying linearly with distance from the center of explosion
\begin{equation}
\label{v-core}
v \simeq  \frac{2x}{9t}.
\end{equation}
The velocity vanishes at the origin, while the temperature diverges. Indeed, using Eqs.~\eqref{Z-0} and Eq.~\eqref{vrc} together with $T=\mu c^2/\gamma=\mu c^2/3$ one gets
\begin{equation}
T \simeq  \frac{4 \mu C}{27}\,\frac{x^2}{t^2}\,\xi^{-\frac{5}{2}}.
\end{equation}
Since $\xi\sim |x|/t^{2/3}$, the temperature diverges as
\begin{equation}
\label{outside}
T \sim |x|^{-\frac{1}{2}}\,t^{-\frac{1}{3}}.
\end{equation}
The divergence of the temperature is unphysical. Simulation results (Sec.~\ref{sec:micro}) show that the temperature at the center of the explosion remains finite and decreases with time. The divergence is the consequence of an idealized modeling using the Euler equations that ignores dissipation.  The divergence indicates that the heat conduction term becomes non-negligible near the origin.

Let us now analyze the source of this problem from a slightly different point of view. In the previous chapter in Appendix~\ref{sec:scalings}, we derived the TvNS exponents for the scaling solutions of the Euler equations as:
\begin{equation}
a=-2/3,~b=0,c=-1/3,~ d=-2/3.
\end{equation}
It is easy to verify that if we substitute this scaling in Eq.~\eqref{tNS}, the disspation terms will have a factor of $t^{-1/3}$ and we therefore say that
they do not contribute in the long time limit. Now we say in the previous paragraph that the heat conduction term becomes non-negligible near the core of the blast. That can only happen if the TvNS scaling is not valid near the core of the blast. Not just the scaling functions, but also the values of the exponents $a,b,c$ and $d$. We should therefore expect different values of $a,b,c$ and $d$ which lead to a finite contribution of the conduction term near  the core and also a different scaling solution. Therefore, we have to rectify our analysis.

To rectify the analysis we now take into account the dissipative processes and consider an analysis of the full hydrodynamic equations Eq.~\eqref{tNS}. As shown in Appendix \ref{app:core}, the terms corresponding to viscous dissipation drop off in the scaling limit and we effectively get the following equations for the fields $\rho,v,T$:
\begin{subequations}
	\label{core-eqs}
	\begin{align} 
	\label{cont-eq-1}
	&\partial_t \rho + \partial_x (\rho v) = 0, \\
	\label{NS-zeta}
	&\rho (\partial_t + v \partial_x) v + \mu^{-1}\partial_x (\rho T) = 0, \\
	\label{heat}
	&\rho^3(\partial_t + v \partial_x)\left[\f{T}{\rho^2}\right] =  2 \mu \partial_x \big(D_2T^{1/2}\,\rho^{1/3}  \partial_x T\big). 
	\end{align}
\end{subequations}
These are the basic governing equations for the core region. Using Eq.~\eqref{heat} we deduce an estimate
\begin{equation}
\rho\,\frac{T}{t} \sim \frac{T^{3/2}}{x^2}\, \rho^{1/3},
\end{equation}
which is combined with $\rho\sim \sqrt{\xi}\sim |x|^{1/2}/t^{1/3}$ to give an estimate of the temperature in the core: 
\begin{equation}
\label{inside}
T \sim |x|^\frac{14}{3}\,t^{-\frac{22}{9}},
\end{equation}
where thermal conductivity plays a significant role.  The estimates Eqs.~\eqref{outside} and \eqref{inside} are comparable at $|x|=X$ determined by
\begin{equation*}
X^\frac{14}{3}\,t^{-\frac{22}{9}} \sim X^{-\frac{1}{2}}\,t^{-\frac{1}{3}}\,.
\end{equation*}
Thus the size $X$ of the core grows as 
\begin{equation}
\label{size}
X\sim t^\frac{38}{93}\,. 
\end{equation}
The relative size of the core compared to the size of the excited region approaches to zero as $X\,t^{-\frac{2}{3}} \sim t^{-\frac{8}{31}}$. 

\begin{figure}
	\begin{center}
		\includegraphics[width=6.6cm,angle=0]{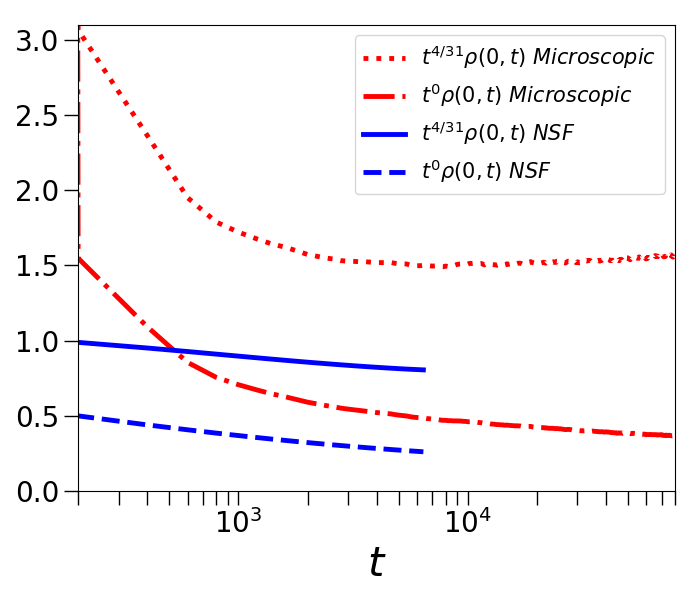}
		\put (-100,158) {$\textbf{(a)}$}
		\includegraphics[width=6.6cm,angle=0]{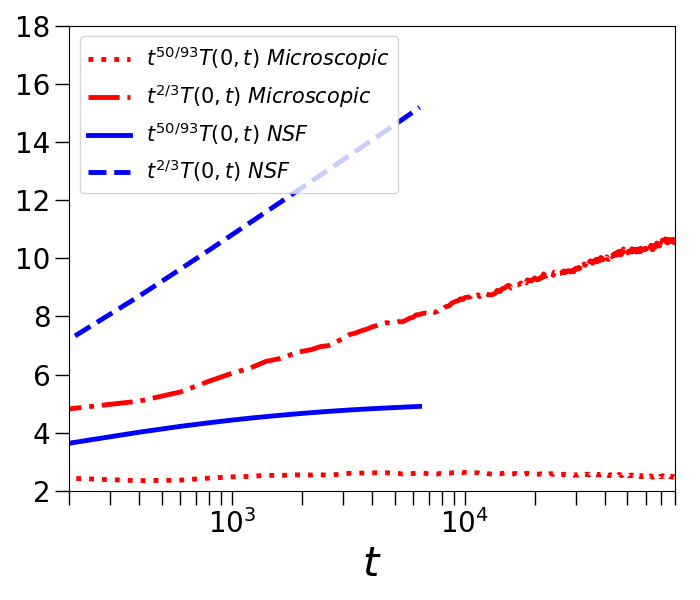}
		\put (-100,158) {$\textbf{(b)}$}
		\caption{Time evolution of  (a) density and (b) temperature at the center of the blast. The data is taken from microscopic simulations 
			and the NSF numerics. The scaling pre-factors correspond to the predictions from TvNS and the core-scaling analysis. For the density, it is difficult to differentiate between the two scaling forms, while for temperature, the scaling prediction $T_0 \sim t^{-50/93}$ in the core describes the data better than the TvNS prediction $T_0 \sim t^{-2/3}$.}
		\label{Fig:centerprofile}
	\end{center}
\end{figure}

The temperature at the center of the explosion is estimated from Eq.~\eqref{outside} after setting $x=X$,  with $X$ given by Eq.~\eqref{size}. Similarly, the density at the center of the explosion is estimated from $\rho_0\sim X^{1/2}/t^{1/3}$ and Eq.~\eqref{size}. We thus arrive at the following 
asymptotic decay laws 
\begin{equation}
\label{T:origin}
T_0 \sim t^{-\frac{50}{93}}\,, \qquad \rho_0 \sim t^{-\frac{4}{31}}.
\end{equation}
Using Eqs.~\eqref{v-core}, \eqref{size} we estimate the velocity near the core: $v_0\sim X/t\sim t^{-\frac{55}{93}}$. The decay laws Eq.~\eqref{T:origin} agree better with microscopic simulations than the TvNS solution as shown in Fig.~\ref{Fig:centerprofile}. 

\begin{figure*}
	\begin{center}
		\includegraphics[width=5.5cm,angle=0]{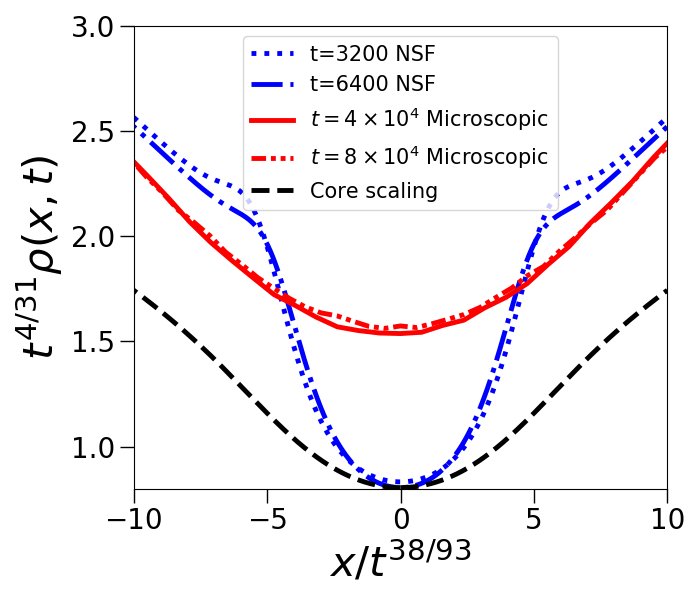}
		\put (-75,130) {$\textbf{(a)}$}
		\includegraphics[width=5.5cm,angle=0]{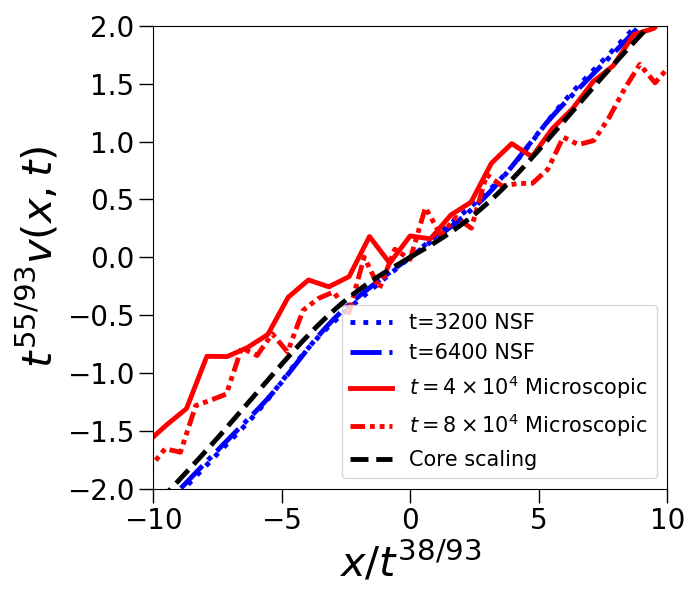}
		\put (-75,130) {$\textbf{(b)}$}
		\includegraphics[width=5.5cm,angle=0]{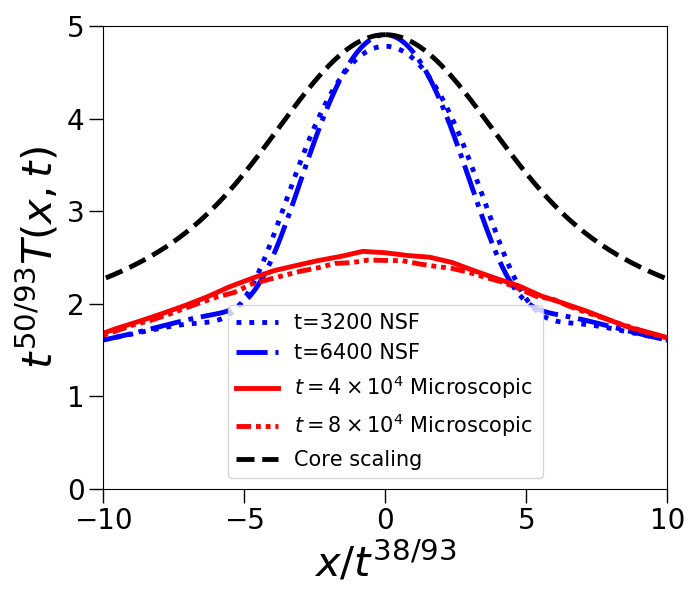}
		\put (-75,130) {$\textbf{(c)}$}
		\caption{A numerical verification of the core scaling forms Eqs.~(\ref{scaling:core})  for data for the fields $\rho,~v$ and  $T$ from both the microscopic simulations and the NSF hydrodynamic numerics. The black dashed lines show the analytic scaling functions obtained from a numerical solution of Eqs.~(\ref{FV-eq}, \ref{F-V}).}
		\label{Fig:coreprofile}
	\end{center}
\end{figure*}

The hydrodynamic variables in the hot core where dissipative effects play the dominant role exhibit scaling behaviors, but instead of $\xi=x/R$ we should use the scaled spatial coordinate $\eta=x/X$. Thus the hydrodynamic variables have a self-similar form
\begin{equation}
\label{scaling:core}
\rho = t^{-\frac{4}{31}} \widetilde{G}(\eta), \quad v = t^{-\frac{55}{93}} \widetilde{V}(\eta), \quad T = t^{-\frac{50}{93}} \widetilde{Z}(\eta)
\end{equation}
in the scaling region
\begin{equation}
\label{scaling:eta}
x\to \infty, \quad  t\to \infty, \quad \eta \sim x\,t^{-\frac{38}{93}} \to\text{finite}.
\end{equation}
The scaling functions $\widetilde{G},\widetilde{V}$ and $\widetilde{Z}$ are now different from $G,V$ and $Z$. The data from the NSF numerics and the microscopic simulations satisfy these scaling forms and give a better collapse than the TvNS scaling, see Fig.~\ref{Fig:coreprofile}. All that is remaining now is to find these scaling functions $\widetilde{G},\widetilde{V}$ and $\widetilde{Z}$ near the core.

We now combine the scaling ansatz Eqs.~\eqref{scaling:core}--\eqref{scaling:eta} with the governing equations \eqref{core-eqs} and we set $\mu=1.5,\ D_2=1$. Eq.~\eqref{NS-zeta} yields the most simple outcome $-$ it becomes $(\widetilde{G}\widetilde{Z})'=0$ in the leading order. (Hereinafter prime denotes differentiating with respect to $\eta$.) Thus $\widetilde{G}=1/\widetilde{Z}$, where for the moment we have set the integration constant to one (we will discuss this point again later). Eq.~\eqref{cont-eq-1} becomes
\begin{subequations}
	\begin{equation}
	\label{GV-eq}
	\widetilde{V}' - \tfrac{4}{31} =  \big(\widetilde{V}-\tfrac{38}{93}\eta\big) (\ln \widetilde{Z})',
	\end{equation}
	while Eq.~\eqref{heat} becomes 
	\begin{equation}
	\label{FV-eq}
	\big(\widetilde{Z}^\frac{1}{6}\,\widetilde{Z}'\big)' + \tfrac{13}{93} =  \tfrac{3}{2}\big(\widetilde{V}-\tfrac{38}{93}\eta\big) (\ln \widetilde{Z})'.
	\end{equation}
\end{subequations}
Comparing Eqs.~\eqref{GV-eq} and \eqref{FV-eq} we obtain 
\begin{equation*}
\big(\widetilde{Z}^\frac{1}{6}\,\widetilde{Z}'\big)' + \tfrac{13}{93} =  \tfrac{3}{2}\big(\widetilde{V}' - \tfrac{4}{31}\big),
\end{equation*}
which is integrated to give
\begin{equation}
\label{F-V}
\widetilde{V} = \tfrac{2}{9}\eta + \tfrac{2}{3} \widetilde{Z}^\frac{1}{6}\,\widetilde{Z}'.
\end{equation}
The integration constant vanishes since $\widetilde{V}(0)=0$ and $\widetilde{Z}'(0)=0$ due to symmetry. Inserting Eq.~\eqref{F-V} into Eq.~\eqref{FV-eq} we obtain
\begin{equation}
\label{F-eq}
\big(\widetilde{Z}^\frac{1}{6}\,\widetilde{Z}'\big)' + \tfrac{13}{93} =  \big(-\tfrac{26}{93}\eta+\widetilde{Z}^\frac{1}{6}\,\widetilde{Z}'\big) (\ln \widetilde{Z})'.
\end{equation}

Solving Eq.~\eqref{F-eq} numerically [equivalently Eqs.~(\ref{FV-eq}, \ref{F-V})], with $\widetilde{Z}(0)$ fixed from the NSF numerical data at $t=6400$ and $\widetilde{Z}'(0)=0$,  we find the scaling functions $\widetilde{G},\widetilde{V},\widetilde{Z}$. In Fig.~\ref{Fig:coreprofile}, we plot these functions and compare with the  data from the NSF. We find qualitative agreement in the form of the scaling functions but not a complete agreement. See Appendix~\ref{app:core} for a more detailed derivation of the equations for $\widetilde{G},\widetilde{V},\widetilde{Z}$, where we also introduce an additional scale factor $\lambda$ to make $\eta$ dimensionless. This gives another fitting parameter  but does not lead to a significant improvement in comparison with data.
The asymptotic behaviors of the velocity field are
\begin{subequations}
	\begin{align}
	\label{V-origin}
	&\widetilde{V} \simeq \tfrac{4}{31} \eta \quad\qquad \qquad\text{when} \quad |\eta|\ll 1, \\
	\label{V-far}
	&\tfrac{2}{9}\eta - \widetilde{V} \sim \eta^{-\frac{19}{12}} \qquad\text{when}\quad \eta\gg 1.
	\end{align}
\end{subequations}
The asymptotic of Eq.~\eqref{V-origin} follows from Eq.~\eqref{GV-eq}. The slope predicted by Eq.~\eqref{V-origin} is a bit smaller than the slope $\frac{2}{9}$ in the blast region.  The asymptotic Eq.~\eqref{V-far} follows from Eq.~\eqref{F-V}. Further,  
\begin{equation}
\label{F-far}
\widetilde{Z}\sim \eta^{-1/2}   \qquad\text{when}\quad \eta\gg 1
\end{equation}
follows from Eq.~\eqref{F-eq}. Indeed, $\widetilde{Z}^\frac{1}{6}\,\widetilde{Z}'$ is asymptotically negligible, so Eq.~\eqref{F-eq} simplifies to $(\ln \widetilde{Z})'=(2\eta)^{-1}$ leading to Eq.~\eqref{F-far}.

{\bf Matching of the inner and outer solutions}: The asymptotic behavior Eq.~\eqref{F-far} yields $\widetilde{G}=\widetilde{Z}^{-1}\sim\eta^{1/2}$ matching the small $\xi$ behavior Eq.~\eqref{G-0}. For the temperature in the outer region we have 
\begin{align}
T \sim \left(\frac{x}{t}\right)^2 Z(\xi) \sim  \left(\frac{x}{t}\right)^2  \left(\frac{x}{t^{2/3}}\right)^{-\frac{5}{2}} \sim x^{-1/2} t^{-1/3}, 
\end{align}
while in the inner region
\begin{align}
T = \f{\widetilde{Z}(\eta)}{t^{50/93}}  \sim t^{-50/93} \left(\f{x}{t^{38/93}}\right)^{-1/2}  
\sim x^{-1/2} t^{-1/3}.
\end{align}
Thus we have perfect matching in the overlap of the inner and outer regions.
Similarly for the density 
\begin{align}
\rho = {G}(\xi)\sim (x/t^{2/3})^{1/2} \sim x^{1/2} t^{-1/3}
\end{align}
in the outer region, while in the inner region 
\begin{align}
\rho = t^{-4/31} \widetilde{G}(\eta) \sim  t^{-4/31} (x/t^{38/93})^{1/2} \sim
x^{1/2} t^{-1/3},
\end{align}
where we used $\widetilde{G}=1/\widetilde{Z}$. Again we get perfect matching. 

\section{Summary}
\label{sec:conclusion}

We analyzed the connection between the atomistic description of matter with particles obeying Newton's equations and the continuum hydrodynamic descriptions relying on the Euler equations (i.e., neglecting dissipative effects) and Navier-Stokes-Fourier equations. Our system is a blast wave in a one-dimensional gas initially at rest (zero temperature). The hydrodynamic description is provided by $\rho(x,t), v(x,t), T(x,t)$ describing the mass density, flow velocity, and temperature. We compared the exact TvNS solution of the hydrodynamic Euler equations with results of direct molecular dynamics simulations of a gas composed of hard point particles with alternating masses. Our main results are:
\begin{enumerate}
	\item[--] We obtained the exact TvNS scaling solution  of the 1D Euler equation for the ideal gas. The position of the shock is $R(t)=[1071 E t^2/(152 \rho_\infty)]^{1/3}$. The hydrodynamic fields are expressed through the scaling variable $\xi=x/R(t)$. 
	\item[--] For the AHP gas, we find a remarkably close agreement between the simulation results and the TvNS predictions.  This includes a matching of the location of the front $R(t)$ and of the scaling functions over most of the range of the scaling variable $\xi$, except for a core region near the origin whose size shrinks (when measured in $\xi$) at larger times. 
	\item[--] We  obtained results both for the ensemble averaged fields and the empirical fields and find agreement between them. We also showed that it is important to use a microcanonical ensemble --- otherwise the energy fluctuations in a canonical ensemble lead to a broadening of the shock front.
	\item[--] The deviations from the TvNS scaling at the core occur due to the contribution of heat conduction that becomes non-negligible in this region. Euler's equations do not take these terms into account.
	\item[--] We thus model the hydrodynamics of our system using the NSF equations, which provided us a detailed understanding of the distinct scaling form in the core region, whose size scales as $X\sim t^{38/93}$. 
	The divergence of the temperature field at the origin and the vanishing of the density field, which are  unphysical predictions of the TvNS solution,  are avoided in the solutions of the  dissipative equations.
\end{enumerate}

\begin{subappendices}
	\section{Derivation of Scaling Functions near the Core}
	\label{app:core}
	
	Consider the heat equation,
	\begin{equation}
	\label{heatapp}
	\rho\partial_t (T/2 \mu) = \partial_x (\kappa  \partial_x T), 
	\end{equation}
	with $\kappa = D_2\rho^{1/3} T^{1/2}$.  Near the origin, the ratio of the right-hand side to the left-hand side of Eq.~\eqref {heatapp} is estimated from the TvNS solution to yield
	\begin{align}
	\f{T^{1/2} t}{\rho^{2/3} x^2} \sim \f{1}{\xi^{5/4} \xi^{1/3} x} \sim \f{t^{19/18}}{x^{31/12}}.  
	\end{align}
	We see that this becomes large in the region $x \lesssim t^{38/93}$. Hence we expect a different scaling solution in the region $ x < t^{-a}$ where $a=-38/93$. 
	As argued in Sec.~\ref{sec:core}, the expected scaling form in the core reads
	\begin{align}
	\rho = t^b\widetilde{G}(xt^a),~~ v= t^c\widetilde{V}(xt^a),~~
	T = t^d\widetilde{Z}(xt^a),
	\end{align}
	with $a=-\f{38}{93},~b=-\f{4}{31},~c=-\f{55}{93}$ and $d=-\f{50}{93}$.  We plug this scaling ansatz into the governing NSF equations:
	
	\begin{align}
	&\partial_t \rho + \partial_x (\rho v) = 0, \\
	&\rho (\p_t + v \p_x) v + \p_x p = \p_x (\eta \p_x v), \\
	& \rho (\p_t + v \p_x) \epsilon+ p\p_x v = \p_x (\eta v \p_x v)+ \p_x  (\kappa\partial_x T),
	\end{align}
	with  $\epsilon = T/(2 \mu)$ and $p=\rho T/\mu$, and arrive at
	\begin{align}
	&b \widetilde{G} + a \eta \widetilde{G}' +  t^{a+c+1} \lambda (\widetilde{G} \widetilde{V})' = 0, \label{eq:sc1} \\
	&t^{c-1} ( c \widetilde{G} \widetilde{V} + a \eta \widetilde{G} \widetilde{V}') + t^{2c +a} \lambda \widetilde{G} \widetilde{V} \widetilde{V}'+ t^{a+d} \lambda \f{(\widetilde{G}\widetilde{Z})'}{\mu} \nn  \\
	& = t^{d/2+c+2 a -b} \lambda^2 (\widetilde{Z}^{1/2} \widetilde{V}')', \label{eq:sc2}\\
	& t^{b+d-2a-1} \,\f{d \widetilde{G}\widetilde{Z} + a \eta \widetilde{G} \widetilde{Z}'}{2 \mu \lambda^2} + t^{b+c+d-a}\,\f{ \lambda \widetilde{G} \big[\widetilde{V} \widetilde{Z}' + 2  \widetilde{Z} \widetilde{V}'\big]}{2 \mu \lambda^2} \nn \\ 
	&= t^{d/2+2c}  (\widetilde{Z}^{1/2}\widetilde{V} \widetilde{V}')' +  t^{b/3+3d/2} (\widetilde{G}^{1/3} \widetilde{Z}^{1/2}  \widetilde{Z}')'. \label{eq:sc3}
	\end{align} 
	Since $c=-1-a=-\f{55}{93}$ and  $b=3d/4+3a+3/2=-\f{4}{31}$, the leading term in Eq.~\eqref{eq:sc2}  gives 
	$(\widetilde{G}\widetilde{Z})'=0$ implying $\widetilde{G}=k/\widetilde{Z}$, where $k=\widetilde{G}(0)\widetilde{Z}(0)$ is a constant. Using this then leads to the following equations for $\widetilde{V}$ and $\widetilde{Z}$:
	\begin{align}
	&  \widetilde{V}'-\f{4}{31} + \left(\f{38\eta  }{93} - \widetilde{V} \right) \f{\widetilde{Z}'}{\widetilde{Z}} = 0, \label{eq:scp1} \\
	&  \widetilde{V}'-\f{25}{93}- \f{1}{2} \left( \f{38 \eta}{93} -   \widetilde{V} \right) \f{ \widetilde{Z}'}{\widetilde{Z}}  
	=  \nu \lambda^2 (\widetilde{Z}^{1/6}  \widetilde{Z}')', \label{eq:scp3}
	\end{align}
	where $\nu = \mu/k^{2/3}$. From these two equations we get
	\begin{align}
	9 \widetilde{V}'-2= 6 \nu  \lambda^2 (\widetilde{Z}^{1/6}  \widetilde{Z}')',
	\end{align}
	which is integrated to give
	\begin{align}
	\widetilde{V}= 2\eta/9+(2\nu \lambda^2 /3) \widetilde{Z}^{1/6}  \widetilde{Z}'.
	\end{align}
	Substituting this into Eq.~\eqref{eq:scp1} yields a non-linear ordinary differential equation for $\widetilde{Z}$:
	\begin{align*}
	\lambda^2 \nu  (\widetilde{Z}^{1/6}  \widetilde{Z}')'+ \left(\f{26 \eta}{93} - \lambda^2 \nu  \widetilde{Z}^{1/6}  \widetilde{Z}' \right) \f{\widetilde{Z}'}{\widetilde{Z}}+\f{13}{93} = 0.
	\end{align*}
	The constants $\widetilde{G}(0),\widetilde{V}(0),\widetilde{Z}(0)$ and $\lambda$ should be fixed by matching the $\eta\gg 1$ behavior with the TvNS solution at $\xi\ll 1$, which however did not lead to a perfect agreement as shown in Fig.~\ref{Fig:coreprofile}.

\end{subappendices}

\chapter{Conclusion}
\label{chap:conc}
\section{Outcomes of the Study}

In this work we studied various aspects of thermalization, chaos and hydrodynamics in one dimensional Hamiltonian systems. Our motivation was two folds. First, to understand what leads a system to thermalization and we studied the role of initial conditions, choice of observables, choice of averaging and the role of chaos. For this, we studied the $\alpha-$FPUT system through the equipartition of energy among the local observables. And second, we studied the evolution of the alternate mass hard particle (AHP) gas from a nonequilibrium initial state and tried to understand the evolution of conserved fields via hydrodynamics.

For the $\alpha-$FPUT system, we found that the equilibration time $\tau_{eq}$ of the local observables scaled with the dimensionless nonlinearity parameter $\epsilon$ as $\epsilon^{-a}$, with the value of $a$ found to be between $4$ and $6$, dependent on the width of the initial distribution $\gamma$ used to compute the ensemble averages. This was in contrast to the equilibration of normal modes, where $a \approx 8$. We found that local observables equilibrate on much shorter time scales when compared to the normal modes. For a given $\epsilon$, we also found that $\tau_{eq}$ depended on the initial ensemble and infact we observe $\tau_{eq} \propto \log \gamma$ with the slope being close to the inverse of the maximal Lyapunov exponent of the system, thus quantifying the relation between thermalization and chaos in the $\alpha-$FPUT system. We also found that the Toda chain thermalizes for a broad enough initial distribution. We tried to explain these differences on the basis of their phase space dynamics. 

We then studied the evolution of the AHP gas from ``blast-wave'' initial condition until the time the energy reaches its boundary. This leads to the formation of shock waves in the system and the evolution of the conserved fields become self-similar in time. The scaling solution of the conserved fields is given by the Taylor, von Neumann and Sedov (TvNS) solution. We find that in our system, the evolution of conserved quantities matches perfectly with the TvNS predictions everywhere except in a small region near the core of the blast whose size shrinks at large times. We explain this deviation as arising due to the non-negligible contribution of the conductivity term near the core of the blast, which is not accounted for by the Euler equations. We thus, model our system by using NSF equations - Euler equations with additional dissipation terms due to viscosity and conductivity.

\section{Limitations of the Study}
	
	For the $\alpha-$FPUT system we found that $\tau_{\rm eq} \propto \log(\gamma)$. The inverse of the proportionality constant $3.3\times 10^{-4}$, is close to the Lyapunov exponent $\Lambda\approx4.1\times 10^{-4}$ of the system. We believe that the properties of the initial conditions, such as its symmetries and its vicinity from breather solutions can affect the exponential growth of perturbations of the initial conditions, which would lead to a retardation of thermalization and higher values of $\tau_{\rm eq}$. This could explain why we did not get a closer agreement. Far away from breathers we can expect there is no such effect. This needs to be investigated further and is beyond the scope of this work. 
	
	For the alternate mass hard particle gas system, our understanding of the core region remains incomplete. We don't get a perfect agreement in the form of scaling solutions near the core of the blast. A comprehensive agreement between microscopic simulations and hydrodynamics is difficult to claim due to the lack of knowledge of the dissipative terms in one dimension. Anomalous transport in one dimensional systems may also have a say in this. This has to be studied further. 
	
	\section{Future Directions}
	\subsection{Thermalization and Chaos}
	
	For the $\alpha-$FPUT system, our results indicate that the equilibration time for local observables $\tau_{eq}$ scales with $\epsilon$ as $\epsilon^{-a}$, with the value of $a$ found to be between $4$ and $6$, in contrast to the equipartition of normal modes, where $a = 8$. One problem is to understand the source of this deviation. Why is this different? One way to study this is to extend the formalism of wave turbulence to the case of space localized initial conditions and observables. Another problem would be to study 
	thermalization in the AHP gas (a non-chaotic system), by studying the evolution of conserved quantities for much longer times and compare results of the dependence of $\tau_{eq}$ on $\gamma$ with the FPUT system, which can give us more clues regarding the role of chaos. 
	
	Thermalization in finite Hamiltonian systems has usually been studied by either considering a time averaging protocol or an ensemble averaging protocol. As we illustrate, they can lead to very different estimates for the time scale of equilibration. In our example, the time averaging protocol gives a thermalization time that is several orders of magnitude larger than that obtained from ensemble averaging. We believe that the ensemble averaging protocol is relevant for understanding aspects of the classical-quantum correspondence in the context of thermalization in finite systems. 
	For quantum systems several studies show, e.g \cite{rigol2008}, that a finite quantum system prepared in a pure initial state and evolving under unitary dynamics can exhibit thermalization (without requiring any
	time averaging). A corresponding statement for the classical system is difficult if (corresponding to the  quantum pure state) one considers a single point in phase space. However if we smear out the point into an initial blob in phase space, as is done in our study, then we can get an equivalent classical statement. The smearing out process can be thought of as arising from the uncertainty principle. As an example we point out recent work \cite{bukov2019} on the quantum-classical correspondence in Floquet systems where such an averaging protocol is essential.

	We note that another important question is that of thermalization of macroscopic systems. There we expect that thermalization requires neither  a temporal nor an ensemble averaging protocol, but arises from the fact that physical observables are macro-variables and their measurement typically involves an averaging over many degrees of freedom \cite{Lebowitz1999,Goldstein2017}. An interesting problem would be to compare the results of the spatial averaging protocol with that of time and ensemble averaging protocols in the thermodynamic limit. This would give us a better understanding thermalization of macroscopic systems. Studying the evolution of conserved quantities in the FPUT system through hydrodynamics is challenging because the equation of state is not known analytically and can only be computed numerically, nevertheless this would be an interesting study. 
	
	\subsection{Hydrodynamics}	
	
	Earlier studies ~\cite{Barbier2016,Joy2021,Joy2021a,Joy2017,Barbier2015,Barbier2015a} of hard-sphere gases in two and three dimensions were unable to obtain a clear and convincing agreement between simulations and the TvNS solution even in the non-core region.  Some possible reasons for this are: (i) for our system, an ideal gas, the equilibrium properties including the equation of state are known exactly and hence one can obtain an exact solution; (ii) simulations in the 1D AHP gas can be done very accurately, the system has good ergodicity properties, and the asymptotic behavior is easy to study. Following our work, an agreement between molecular dynamics simulations and hydrodynamics equations was reported in higher dimensions \cite{Kumar2021}, once dissipative terms are included.

	One of the most interesting extensions  of the present work would be to consider the effect of  potential energy terms in the Hamiltonian, e.g to study systems such as Lennard-Jones gases or lattice models like the Fermi-Pasta-Ulam-Tsingou (FPUT) chain. For the FPUT model, the Euler equations are known~\cite{Mendl2017}, however the equation of state is not known in closed form and this makes a detailed comparison between simulations and hydrodynamic predictions quite nontrivial. Recently, the Euler equations along with Navier-Stokes corrections, for integrable Hamiltonian systems such as the Toda lattice and the hard rod gas have been written and the evolution of domain wall initial conditions investigated~\cite{spohn2021hydrodynamic,doyon2019generalized,mendl2020high}. The effect of integrability would be another interesting direction to explore. For integrable systems, we use generalized hydrodynamics and for non-integrable systems we use the usual hydrodynamics. It will be very interesting to consider what happens with near-integrable systems, and if generalized hydrodynamics is valid in these systems. The evolution of blast-wave initial conditions in these systems, and in particular the nature of self-similar solutions of the conserved fields are fascinating open questions. 
	
	The emergence of the mean-field hydrodynamic behavior is questionable in low dimensional systems \cite{krapivskybook} at low temperatures. Contrary to these generic expectations, our extremal one-dimensional zero-temperature system disturbed by sudden local release of energy demonstrates the hydrodynamic behaviour surprisingly well. Our understanding of the core region remains incomplete. A comprehensive agreement between microscopic simulations and hydrodynamics is difficult to claim due to the lack of knowledge of the dissipative terms in one dimension. The well-known phenomenon of anomalous heat transport in one-dimensional systems (with mass, momentum, and energy conserved) means that the thermal conductivity $\kappa$  is not given by the standard Green-Kubo formula, see \cite{saito2020} for a recent resolution.  
	
	\subsection{Concluding Remarks}
	
	I would like to conclude this thesis by pointing out that thermalization and hydrodynamics in Hamiltonian systems are far from being completely understood. On the one hand, we use statistical physics to understand the emergent behaviour in  macroscopic systems. On the other, there is a lot to understand about finite sized systems from the point of view of phase space dynamics. 
	It is possible that Hamiltonian chaos \cite{Arnold2013, Chirikov1979, Lichtenberg2013} could give us a deeper understanding regarding the transition from finite sized systems to macroscopic systems. However, there are some studies \cite{Lebowitz1993a,Lebowitz1993b,Lebowitz2007,Goldstein2020,Chakraborti2021e} that provide evidence of chaos being misleading, misguided or unnecessary in the context of discussing thermalization of macroscopic systems. It would nevertheless be very interesting to probe this question further and actually evaluate the exact role of chaos. And one wonders if Bob Dylan, an American singer is correct when he said ``I accept chaos, I'm not sure whether it accepts me."

\addcontentsline{toc}{section}{Bibliography}
\bibliographystyle{unsrt}
\bibliography{references}
\end{document}